\documentclass[twocolumn,aps,prl,amsmath,amssymb,groupofadress,showpacs]{revtex4-1}

\usepackage{graphicx}
\usepackage{graphics}
\usepackage{dcolumn}
\usepackage{bm}
\usepackage{color}
\usepackage{soul}
\usepackage{textcomp}
\usepackage{soul}

\usepackage{multirow}
\usepackage{setspace}
\usepackage[mathlines]{lineno}
\usepackage{textcomp}
\usepackage{mathcomp}

\usepackage[final]{pdfpages} 

\usepackage{microtype}
\usepackage[breaklinks=true,hyperindex=true,
pdftitle={Optimal sideband cooling under non-Markovian evolution},
colorlinks=true,pagebackref=false,citecolor=blue,plainpages=false,pdfpagelabels,
linkcolor=blue,urlcolor=blue]{hyperref}

\newcommand{\ie}{i.e., }

\newcommand{\affUdeA}{Grupo de F\'isica At\'omica y Molecular, Instituto de F\'{\i}sica,  
Facultad de Ciencias Exactas y Naturales,\\
Universidad de Antioquia UdeA; Calle 70 No. 52-21, Medell\'in, Colombia.}

\begin{document}

\title{
Ultrafast Optimal Sideband Cooling under Non-Markovian Evolution
}

\author{Johan F. Triana}
\affiliation{\affUdeA}
\author{Andr\'es F. Estrada}
\affiliation{\affUdeA}
\author{Leonardo A. Pach\'on}
\affiliation{\affUdeA}
%

\begin{abstract}
A sideband cooling strategy that incorporates (i) the dynamics induced 
by structured (non-Markovian) environments in the target and auxiliary 
systems and (ii) the optimally-time-modulated interaction between them 
is developed. 
For the context of cavity optomechanics, when non-Markovian dynamics are 
considered in the target system, ground state cooling is reached at much 
faster rates and at much lower phonon occupation number than previously 
reported.
In contrast to similar current strategies, ground state cooling is reached 
here for coupling-strength rates that are experimentally accesible for 
the state-of-the-art implementations.
After the ultrafast optimal-ground-state-cooling protocol is accomplished,
an additional optimal control strategy is considered to maintain the
phonon number as closer as possible to the one obtained in the cooling
procedure.
Contrary to the conventional expectation, when non-Markovian 
dynamics are considered in the auxiliary system, the efficiency of the 
cooling protocol is undermined. 
\end{abstract}

\date{\today}


\maketitle

{\it Introduction.}\textemdash
Fabricating and controlling micro- and nano-devices is of primary 
importance to develop quantum technologies. 
In this effort, optomechanical devices play a central role because they 
may become a building block for hybrid systems \cite{MVT98,MC&07,MC&12,
AKM14}.
To benefit from quantum effects, potential quantum technologies that 
utilize optomechanical systems need to prepare the mechanical component 
in the ground state.
Thus, searching alternative schemes and refining present strategies to 
improve cooling protocols are of great interest. 
However, the highly non-trivial interplay between time-dependent 
external fields and structured environments poses constraints on
the coherent control of these systems \cite{SN&11} that have been barely
explored.

Despite the copious recent literature on optomechanical cooling \cite{MVT98,MC&07,
MC&12,MG09,WN&07,WV&11,CA&11,SC&13,LX&13,LL&14}, a key point of the 
physics at low temperature remains unexplored, namely, low temperature
induces non-Markovian dynamics \cite{Wei12,PB14,PT&14,EP15}.
This very fact and the recent experimental evidence that the dynamics 
of micro-resonators are non-Markovian \cite{GT&13}, certainly, suggest 
that the present understating of cooling processes is incomplete.
Moreover, because these approaches work on very short time-scales, of the 
order of the period of the resonator, it is expected that non-Markovian 
effects, either from low-temperature fluctuations or structured environments, 
dominate the energy and entropy transfer.

The exploration of non-Markovian dynamics has already leaded to the 
enhancement of, e.g., quantum speed limits \cite{DL13}, survival of 
entanglement \cite{TE&09,HRP12,EP15}, improvements in quantum metrology 
\cite{CHP12} and corrections to thermal equilibrium states \cite{PT&14,YA&14}.
In the context of cooling, it was even shown that only under non-Markovian dynamics
the entropy of a parametrically driven resonator can decrease \cite{SN&11}. 
The non-Markovian aspects of the dynamics discussed above are introduced here
to one of the most promising techniques to reach the minimum phonon number, 
namely, finding an optimal coupling function via optimal control theory 
\cite{SN&11,WV&11}.
Based on analytic exact results derived in the context of the Feynman-Vernon 
influence theory, it is found that when the non-Markovian character 
is considered in the target resonator, the phonon number is lower than the 
predicted by Markovian processes; however, in contrast to the conventional 
expectation, non-Markovian dynamics in the auxiliary system deteriorates 
the cooling protocol.

\textit{Model and Theory.}\textemdash
For mechanical systems, sideband cooling consists in coupling a 
resonator to a microwave or optical resonator whose frequency is 
sufficiently high that it effectively sits in its ground state 
at ambient temperature.
The coupling between the mechanical and electromagnetic modes is 
mediated by radiation pressure and is nonlinear.
However, under realistic experimental conditions \cite{TD&11,CA&11,
GB&06,KB06}, the coupling can be assumed to be linear \cite{MC&07,AKM14}.
This common model with linear coupling is considered below
\footnote{Note that in standard cavity optomechanics, the linearization  is realized
using a steady state value for the cavity field, and depends on the damping 
rate of the cavity field. 
Thus, to proceed with no further refinement, it is necessary to verify that the 
spectral used allows for time-independent steady-states.
This occurs for the spectral density used below and, in general, for spectral 
densities with no gaps \cite{XL&15}}. 
Specifically, in the linear approximation, the system consists in two harmonic 
modes that describe the mechanical mode and the optical mode with the 
annhiqulation operators $\hat{a}$ and $\hat{b}$, respectively.
The Hamiltonian reads
\begin{equation}
\hat{H}=\hbar {\omega_{\mathrm{mm}}}\hat{a}^{\dagger}\hat{a} + 
\hbar {\omega_{\mathrm{om}}}\hat{b}^{\dagger}\hat{b} + 
\hbar g(t)(\hat{a}^{\dagger}+\hat{a})(\hat{b}^{\dagger}+\hat{b}) 
\label{eqn:Hamlnt}
\end{equation}
with {$\omega_{\mathrm{mm}}$ and $\omega_{\mathrm{om}}$}, the frequency 
of the mechanical and electromagnetic modes, respectively, and $g(t)$ 
is the arbitrary time-dependent optomechanical coupling function to be 
found via optimal control theory 
\cite{Kir12}. 
For later convenience, the period of the mechanical mode is labeled by 
{$\tau_{\mathrm{mm}} = 2\pi/ \omega_{\mathrm{mm}}$}. 
To describe the interaction with the environment of the resonator
and the losses in the cavity, each mode is coupled to an independent 
thermal bath with arbitrary spectrum and described in the context 
of the Ullersma-Caldeira-Leggett model \cite{Ull66,CL81} [see the Supplementary 
Material for details].

The particular functional form of the Hamiltonian (\ref{eqn:Hamlnt}), 
the environment model and the thermal initial states allow for 
the complete description of the dynamics in terms of the variances of 
the modes' coordinates (see, e.g., Ref.~\cite{EP15}). 
This was utilized to combine sideband cooling and optimal control theory \cite{WV&11}.
Specifically, the dynamics were formulated in terms of an adjoint master 
equation for the quadratures of the modes derived from the Markovian 
version of the Brownian-motion master equation \cite{CD&10,GZ04}.
However, this master equation may violate the density-operator positivity 
(see Sec.~3.6 in Ref.~\cite{BP07}).
Moreover, for time dependent Hamiltonians, an adjoint Lindblad master equation 
can be derived only if the Liouvillian of the non-unitary dynamics commutes with its 
associated-time-ordered propagator (see Sec.~3.2 in Ref.~\cite{BP07}), this 
is not the case for the Hamiltonian (\ref{eqn:Hamlnt}). 

To circumvent the issues raised above, the Feynman-Vernon's influence 
functional theory \cite{FH12} is employed here.
It allows for performing a description of the dynamics 
without any approximation [see Supplementary Material].
and it further obeys the positivity of the density operator and straightforwardly 
incorporates driving fields.
By combining the exact analytic results provided by the influence functional 
approach and an efficient numeral algorithm to solve integro-differential equations, 
the dynamics are solved here for arbitrary bath spectrum and driving field.

\textit{Non-Markovian Optimal Sideband Cooling.}\textemdash
The potential benefits from cooling optomechanical systems have attracted a great deal 
of attention \cite{TD&11,MC&07,GV&08,WN&07,SC&13,RD&11,MVT98} and recently, it has
been powered by optimal control theory \cite{MC&12,WV&11}.
However, a key dynamical feature has been left out of the discussion, namely,
the non-Markovian dynamics induced by low temperature and structured environments
\cite{Wei12,PB14,PT&14,EP15}.
Therefore, the goal here is to calculate and investigate the dependence of the minimum phonon 
number in the mechanical resonator $\langle \hat{n} \rangle(t_{\mathrm{cool}})$, at the shortest 
possible time $t_{\mathrm{cool}}$, on the non-Markovian character of the dynamics.
In terms of the second moments of the position $\langle \hat{q}^{2} \rangle$ and momentum 
$\langle \hat{p}^{2} \rangle$, the phonon number at time $t$ is given by
\begin{equation}
\langle \hat{n} \rangle(t) = \frac{1}{2\hbar\omega}\left[ \frac{\langle \hat{p}^{2} \rangle(t)}{m} 
+ m {\omega_{\mathrm{mm}}}^{2}\langle \hat{q}^{2}\rangle(t)\right] -\frac{1}{2} .
\label{eq:npromnm}
\end{equation}
The optimization procedure under non-Markovian dynamics, described in the Supplementary 
Material, leads to the minimum number phonon occupation at time $t_{\mathrm{cool}}$.
The cooling time $t_{\mathrm{cool}}$ is chosen as short as possible by hand.
The spectrum of the mechanical-resonator's environment is described by the spectral density 
{$J_{\mathrm{mm}}(\omega) = \gamma \omega_{\mathrm{D}}^2/\left( \omega_{\mathrm{D}}^2 
+ \omega^2\right)$} whereas the cavity environment is chosen, for convenience,  as 
{$J_{\mathrm{om}}(\omega) = \kappa \Omega_{\mathrm{D}}^2/\left( \Omega_{\mathrm{D}}^2 
+ \omega^2\right)$}.
{As in Ref.~\cite{WV&11},  it is assumed that the frequency conversion is exact and
set $\omega_{\mathrm{mm}}=\omega_{\mathrm{om}}$.
The corrections to this approximation are of the order of 
$\left(\omega_{\mathrm{mm}}/\omega_{\mathrm{om}}\right)^2$}
This spectral density relates to temperature fluctuations \cite{CR02}. 

\begin{figure}[t]
\centering
\includegraphics[width=0.5\textwidth]{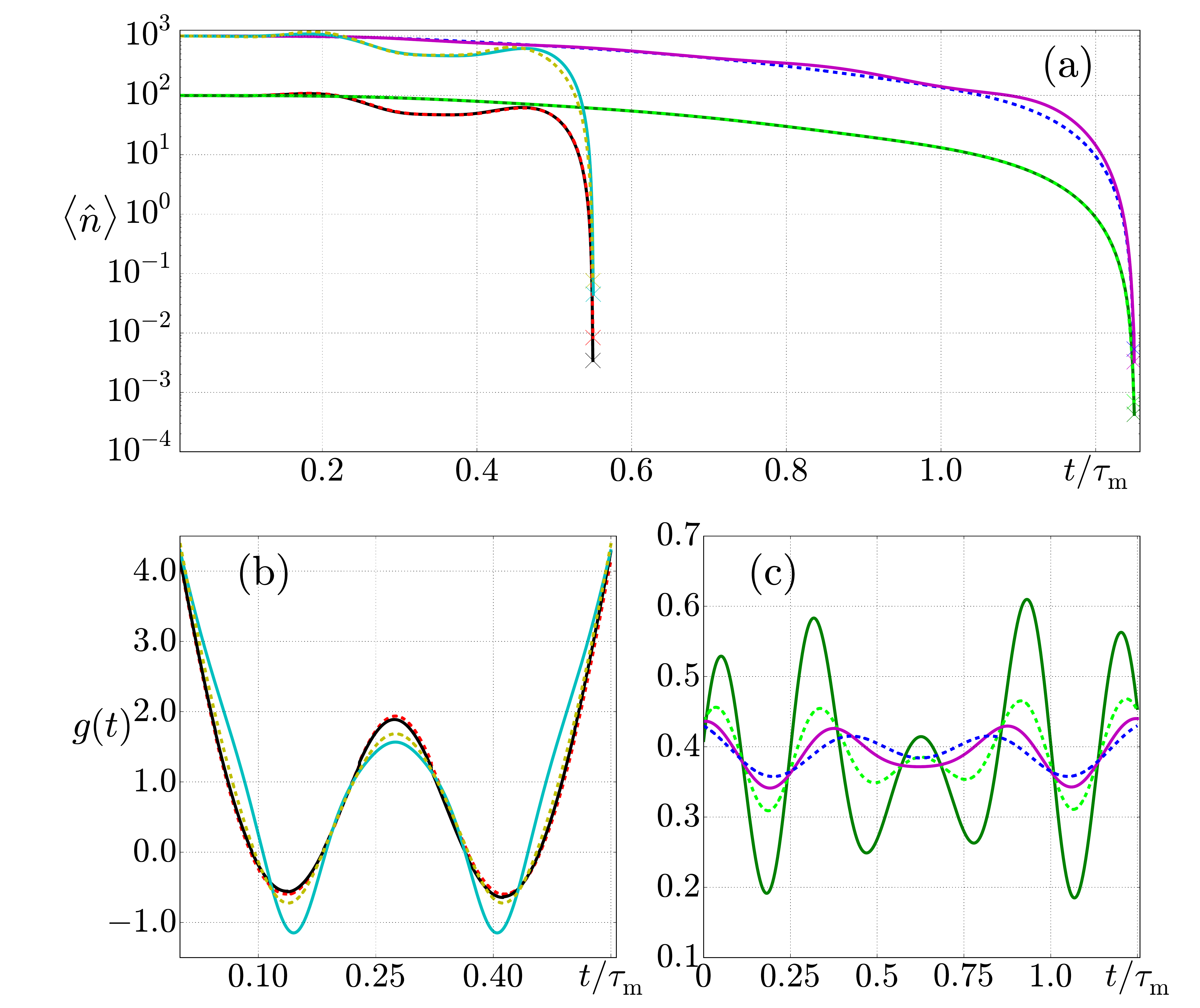}
\caption{(a) Time evolution of the phonon number $\hat{n}(t)$ during the optimal modulation 
of the interaction with $n_{\mathrm{c}}=0$, $\gamma=10^{-6}\omega_{\mathrm{m}}$ 
and $\kappa=2.15\times10^{-4}\omega_{\mathrm{m}}$.
In particular, for $n_{\mathrm{T}}=10^{2}$ ($n_{\mathrm{T}}=10^{3}$) and final time 
$t_{\mathrm{cool}}=0.55 \tau_{\mathrm{m}}$ ($t_{\mathrm{cool}}=1.25 \tau_{\mathrm{m}}$),
the dashed red (blue) curve depicts the dynamics of the phonon number under Markovian 
dynamics ($\omega_{\mathrm{D}}  = \Omega_{\mathrm{D}}  = 100 \omega_{\mathrm{m}}$) 
while the continuos black (magenta) curve depicts the dynamics under non-Markovian dynamics  
($\omega_{\mathrm{D}}  =  \Omega_{\mathrm{D}}  = \omega_{\mathrm{m}}$).
The continuos dark green (cyan) curve depicts the non-Markovian phonon-occupation-number 
dynamics for $n_{\mathrm{T}}=10^{2}$ ($n_{\mathrm{T}}=10^{3}$) and $t_{\mathrm{cool}}=1.25 
\tau_{\mathrm{m}}$ ($t_{\mathrm{cool}}=0.55 \tau_{\mathrm{m}}$) whereas the dashed olive 
(light green) curve stands for the Markovian dynamics.
The $y$-axis is in logarithmic scale. 
(b) Optimal coupling function with $t_{\mathrm{cool}}=0.55\tau_{\mathrm{m}}$ under non-Markovian 
(continuous lines) and Markovian dynamics (dashed line).
(c) Optimal coupling function with $t_{\mathrm{cool}}=1.25\tau_{\mathrm{m}}$ under non-Markovian 
(continuous lines) and Markovian dynamics (dashed line).
}
\label{fig:ncoolnmop}
\end{figure}

For typical values of optomechanical setups, Fig.~\ref{fig:ncoolnmop}\textcolor{blue}{(a)} shows 
the optimal dynamics of the phonon number in the Markovian (dashed lines) and in non-Markovian 
(continuous lines) regime. 
Figure \ref{fig:ncoolnmop} shows two alternatives to decrease the minimum phonon number, 
namely, (i) extending the cooling time or (ii) considering non-Markovian effects in the dynamics. 
By lengthening the cooling process from $t_{\mathrm{cool}}=0.55 \tau_{\mathrm{m}}$ (black 
and cyan lines) to $t_{\mathrm{cool}}=1.25 \tau_{\mathrm{m}}$ (green and magenta lines) and
for {the initials phonon number in the mechanical mode} $n_{\mathrm{T}}=10^{2}$ 
and $n_{\mathrm{T}}=10^{3}$, it is possible to decrease the minimum 
phonon number by an order of magnitude, from ca. $\langle n(t_{\mathrm{cool}}) \rangle=10^{-2}$ to 
$\langle n(t_{\mathrm{cool}}) \rangle=10^{-3}$.
The  Markovian and non-Markovian character of the dynamics does not change considerably the 
functional form of the optomechancal coupling; however, the small difference is enough to decrease 
the minimum phonon-number by around one order of magnitude (see also Table~\ref{tab:cooling}).

\begin{table*}
\begin{tabular}{c||c||c||c||c||c||c||}
\multirow{2}{*}{ } & \multicolumn{5}{c||}{$\langle \hat{n}(t_{\mathrm{cool}}) \rangle$} 
\\
\cline{2-6}
$\gamma/\omega_{\mathrm{m}}$ &
opt.  om$_{\mathrm{M}}+$ mm$_{\mathrm{M}}$& 
om$_{\mathrm{nM}}+$ mm$_{\mathrm{nM}}$ & 
opt. om$_{\mathrm{nM}}+$ mm$_{\mathrm{nM}}$& 
om$_{\mathrm{M}}+$ mm$_{\mathrm{nM}}$& 
opt. om$_{\mathrm{M}}+$ mm$_{\mathrm{nM}}$
 \\
 \hline \hline
$10^{-6}$ & $9.03\times10^{-3}$ & 8.86$\times10^{-3}$  &3.43$\times10^{-3}$ 
& 1.91$\times10^{-3}$  &9.99$\times10^{-5}$ 
\\
\hline
$10^{-5}$ & $1.04\times10^{-2}$ &  1.02$\times10^{-2}$  &4.79$\times10^{-3}$ 
&  2.60$\times10^{-3}$  &7.91$\times10^{-4}$ 
\\
\hline
$10^{-4}$ & $3.28\times10^{-2}$ &  2.40$\times10^{-2}$ &1.99$\times10^{-2}$ 
&  1.63$\times10^{-2}$ &1.52$\times10^{-2}$ 
\\
\hline
$10^{-3}$ & $2.61\times10^{-1}$ &  1.61$\times10^{-1}$  &1.53$\times10^{-1}$ 
&  1.53$\times10^{-1}$  &1.51$\times10^{-1}$ 
\\
\hline
$10^{-2}$ & $2.45$ &  1.52  &1.50 &  1.50  &1.50 
\\
\hline
$10^{-1}$ & $21.12$ &  14.21 & 13.52 &  13.52 & 13.52 
\end{tabular}
\caption{Minimum phonon number in the mechanical resonator at a time 
$t_{\mathrm{cool}}=0.55 \tau_{\mathrm{m}}$ for different values of the dissipation
rate $\gamma/\omega_{\mathrm{m}}$ and for different cooling scenarios (see
text).
Cavity dissipation rate $\kappa=10^{-4}\omega_{\mathrm{m}}$.}
\label{tab:cooling}
\end{table*}

The coupling strength for the ultrafast cooling scenario in 
Fig.~\ref{fig:ncoolnmop}\textcolor{blue}{(b)} is beyond the present experimental capabilities;
however, results for the fast cooling case in Fig.~\ref{fig:ncoolnmop}\textcolor{blue}{(c)}
are encouraging  \cite{TD&11}.
Moreover, in the regime of Fig.~\ref{fig:ncoolnmop}\textcolor{blue}{(c)} cooling from room 
temperature is possible \cite{LL&14}.

The non-trivial interplay between non-Markovian dynamics and driving fields is explored in
Table~\ref{tab:cooling}, it displays the predicted phonon number for a variety of scenarios, 
namely, 
(i:~opt.~om$_{\mathrm{M}}+$ mm$_{\mathrm{M}}$) the minimum phonon number 
obtained from optimization under Markovian dynamics, 
(ii:~om$_{\mathrm{nM}}+$ mm$_{\mathrm{nM}}$) the phonon number obtained under 
non-Markovian dynamics with the optimal coupling found under Markovian dynamics,
(iii:~opt.~om$_{\mathrm{nM}}+$ mm$_{\mathrm{nM}}$) the phonon number obtained 
from optimization under non-Markovian dynamics,
(iv:~om$_{\mathrm{M}}+$ mm$_{\mathrm{nM}}$) the phonon number obtained under 
Markovian dynamics in the optical mode and non-Markovian dynamics in the mechanical 
mode with the optimal coupling found under Markovian dynamics and
(v:~opt.~om$_{\mathrm{M}}+$ mm$_{\mathrm{nM}}$) the phonon number obtained 
from optimization under Markovian dynamics in the optical mode and non-Markovian 
dynamics in the mechanical mode.

Comparison of scenarios (i) and (ii) in Table~\ref{tab:cooling} shows that, for the same 
coupling function, the presence of non-Markovian dynamics reduces the minimum
phonon number and that this decrease is more noticeable when the dissipation rate of 
the mechanical resonator increases.
The third scenario depicts the influence of the optimization process. 
Even though there is a reduction in the phonon number respect to the second scenario, 
in absolute terms, the decrease is tiny.
Besides, if the optical-mode dynamics are considered as Markovian and the dynamics of
the mechanical mode as non-Markovian, fourth scenario, there is a decrease in the number 
of phonons for low values of the decay factor $\gamma/\omega_{\mathrm{m}}$.
Surprisingly, only when Markovian, non-Markovian dynamics and optimal control theory
are combined, fifth scenario,  a substantial reduction in the phonon number is reached
for low mechanical decay rates.
For these low-decay-factor cases, which are relevant under experimental conditions, the strategy 
in the fifth scenario is able to bring the number of phonons more than one order of 
magnitud below the number of phonons reached under the conditions of the first scenario.

This non-trivial and unexpected interplay between Markovian, non-Markovian dynamics and 
optimal control theory is also observed for a variety of values of the cavity decay rate in 
Table~\ref{tab:kappatimecool}.
The advantage of a Markovian-dynamics cavity is more transparent for large values of the
cavity losses rate $\kappa$, e.g., for a non-Markovian cavity ($\Omega_{\mathrm{D}} = 
\omega_{\mathrm{c}}$) with $\kappa=2.15\times10^{-1}\omega_{\mathrm{c}}$, there is 
no ground-state cooling
[$\langle \hat{n}_{\mathrm{nM}}(t_{\mathrm{cool}}) = 2.34$].
\begin{center}
\begin{table}[h]
\begin{tabular}{c|c|c|c|c}
$t_{\mathrm{cool}}/\tau_{\mathrm{m}}$ & $\kappa/\omega_{\mathrm{c}}$ & 
$\langle \hat{n}_{\mathrm{M}}(t_{\mathrm{cool}}) \rangle$ & 
$\langle \hat{n}_{\mathrm{nM}}(t_{\mathrm{cool}}) \rangle$ & 
$\langle \hat{n}_{\mathrm{om_{M}}}(t_{\mathrm{cool}}) \rangle$ \\
 \hline \hline
 0.55 & 1$\times10^{-3}$ & 0.016 & 0.015 & 0.014\\
 0.6 & 1.5$\times10^{-2}$ & 0.025 & 0.019 & 0.019
 \\
 0.8 & 2.5$\times10^{-2}$ & 0.029 & 0.030 & 0.021
 \\
 0.8 & 4.5$\times10^{-2}$ & 0.035 & 0.060 & 0.026
 \\
 0.8 & 5.5$\times10^{-2}$ & 0.037 & 0.086 & 0.032
 \\
 1.0 & 1.25$\times10^{-1}$ & 0.048 & 0.356 & 0.040 
 \\
 1.6 & 2.15$\times10^{-1}$ & 0.056 & 2.34 & 0.044
\end{tabular}
\caption{Minimum phonon number for scenarios
(i:~opt.~om$_{\mathrm{M}}+$ mm$_{\mathrm{M}}$) $\langle \hat{n}_{\mathrm{M}}(t_{\mathrm{cool}})$,
(iii:~opt.~om$_{\mathrm{M}}+$ mm$_{\mathrm{M}}$) $\langle \hat{n}_{\mathrm{nM}}(t_{\mathrm{cool}})$
and (v:~opt.~om$_{\mathrm{M}}+$ mm$_{\mathrm{M}}$) $\langle \hat{n}_{\mathrm{om_{M}}}(t_{\mathrm{cool}})$
for different cooling times and for different parameters of cavity dissipation. 
The initial parameters are $n_{\mathrm{T}} =100$, $\gamma = 10^{-4}\omega_{\mathrm{m}}$.}
\label{tab:kappatimecool}
\end{table}
\end{center}

Because the mechanical-mode initial state is highly 
thermally-populated, the non-Markovian character of the dynamics originates mainly from the 
structure of the thermal bath and not from quantum fluctuations at low temperature.
This explains why, for non-structured environments, Markovian descriptions of the dynamics
may had provided sensible results.

\textit{Maintaining the minimum phonon number over time.}\textemdash
Reaching a very low phonon number in a short period of time is a desirable goal.
However, because the resonator is continuously coupled to its environment, keeping that phonon 
number is a must. 
In doing so, introduce a second optimal control scenario where the objetive is 
to maintain, over a long period of time, the minimum number of phonons in the resonator reached
in the ultrafast cooling in Fig.~\ref{fig:ncoolnmop}.
Define then a composite optimal optomechanical coupling function as 
\begin{equation}
c(t)=\left\{\begin{array}{l c}
g_{\mathrm{c}}(t) & 0 \le t \le t_{\mathrm{cool}},
\\
g_{\mathrm{m}}(t) & t > t_{\mathrm{cool}},
\end{array}\right.
\label{eq:ccf}
\end{equation}
which encompasses the entire cooling process, \ie reaching a minimum phonon 
number by applying $g_{\mathrm{c}}(t)$ and maintaining this number once it is obtained
with $g_{\mathrm{m}}(t)$. 
Because performing the second optimization process under non-Markovian dynamics over 
long times requires an enormous amount of computing time, the second optimization process 
is performed under the assumption that 
for  long times, the Markovian approximation holds so that the optimal coupling function is calculated
for Markovian dynamics.

\begin{figure}[t]
\centering
\includegraphics[width=\columnwidth]{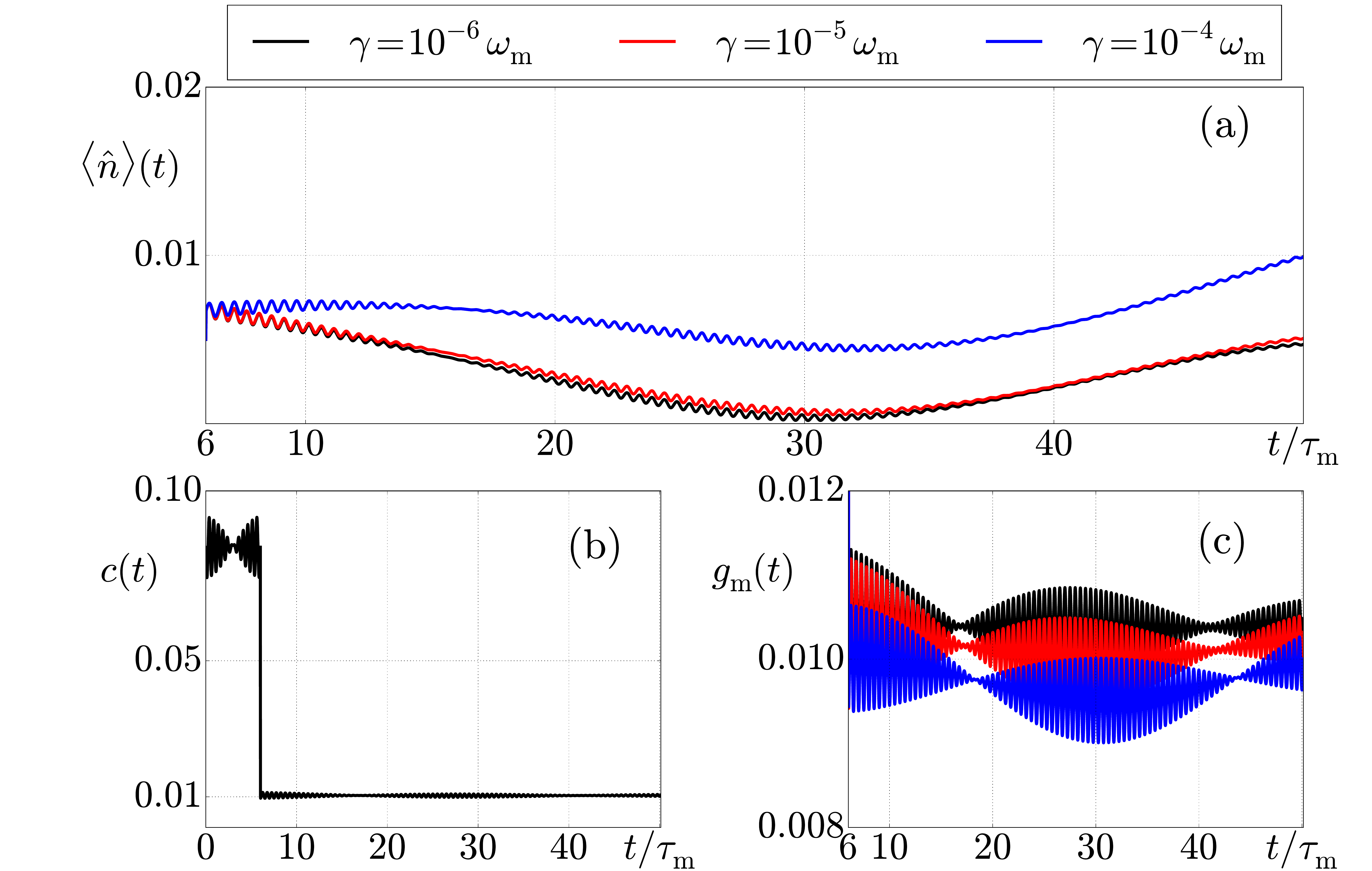}
\caption{(a) Phonon number as a function of time during the optimal control protocol aimed at 
maintaing the minimum phonon number for different values of the dissipation rate. 
The initial parameters are $n_{\mathrm{T}} = \langle n(t_{\mathrm{cool}}) \rangle$, 
$n_{\mathrm{c}} = 0$, $\kappa=2.15\times10^{-4}$. 
(b) The optimal optomechanical coupling function $c(t)$ for $\gamma = 10^{-6}\omega_{\mathrm{m}}$.
(c) The optimal optomechanical coupling function $g_{\mathrm{m}}(t)$.}
\label{fig:nman}
\end{figure}

Figure~\ref{fig:nman}\textcolor{blue}{(a)} depicts the time evolution of the phonon number 
under the action of the optimal coupling function $c(t)$ for a variety of experimentally-relevant 
values of the decay rate $\gamma$.
To obtain an experimentally-accessible-coupling-strength scenario, $|c(t)| < 10^{-1}$, the 
cooling time is set to $t_{\mathrm{cool}} = 6 \tau_{\mathrm{m}}$ and the initial phonon 
number to $n_{\mathrm{T}}=10^2$ [see Fig.~\ref{fig:nman}\textcolor{blue}{(b)}].
For the second control phase, the initial number of phonons is set to be the minimum 
phonon number reached at $t_{\mathrm{cool}}$.
To maintaining the minimum phonon number requires moving from the strong coupling 
regime to a weak coupling regime between the two modes, see 
Figs.~\ref{fig:nman}\textcolor{blue}{(b)} and \ref{fig:nman}\textcolor{blue}{(c)}.
Likewise, with the optimal optomechanical coupling function $c(t)$ the phonon number 
is approximately maintained for 50 periods of the mechanical resonator.

\textit{Discussion.}\textemdash
Analytical solutions and an efficient optimal control protocol showed that 
the presence of non-Markovian dynamics allows for lower phonon numbers than the predicted by 
Markovian-and-RWA-based previous works. 
Surprisingly, significant enhancements are found as the interplay between non-Markovian
dynamics in the mechanical mode, Markovian dynamics in optical mode and optimally-designed
coupling functions.
To understand this effect, note that for well-behaved $J(\omega)$, non-Markovian dynamics define 
an effective coupling to the thermal bath \cite{PT&14,EP15} that, in general, is weaker than the 
Markovian one.
Thus, when non-Markovian dynamics are considered in the cavity dynamics, the rate at which 
the electromagnetic mode releases the  entropy, that absorbes from the mechanical mode, into 
its environment diminishes and therefore, the number of phonons in the mechanical mode does 
not largely decrease compared to the case of Markovian-dynamics cavity.
Furthermore, because the mechanical mode is constantly coupled to its environment, the optimal 
cooling protocol was extended [see Eq.~(\ref{eq:ccf})] to preserve the phonon number as low as 
possible after cooling is reached.
For this second protocol, the coupling amplitude is between reach of present technology and 
can be combined with already-experimentally-implemented cooling protocols 
\cite{CA&11,TD&11,GB&06,KB06}.

The optimization protocol developed here relies on the optimization over the Green functions of the 
trajectories that minimize the influence functional (see Supplementary Material) and can be readily 
implemented in semiclassical formulations of quantum mechanics in phase space \cite{PID10,DGP10}.
Thus, non-linear systems can be addressed and the influence of quantum fluctuations in the 
design of optimal pulses of coupling functions can be analyzed.
This  enables the present proposal, e.g., in the context of optimal control theory of molecular 
processes \cite{SB12}.

\begin{acknowledgments}
This work was supported by the \textit{Comit\'e para el Desarrollo de la Investigaci\'on}
--CODI-- of Universidad de Antioquia, Colombia under contract number E01651 
and under the \textit{Estrategia de Sostenibilidad 2015-2016}, by the \textit{Departamento Administrativo 
de Ciencia, Tecnolog\'ia e Innovaci\'on} --COLCIENCIAS-- of Colombia under the grant number 
111556934912.
\end{acknowledgments}

\bibliography{oscprl}

\begin{thebibliography}{37}%
\makeatletter
\providecommand \@ifxundefined [1]{%
 \@ifx{#1\undefined}
}%
\providecommand \@ifnum [1]{%
 \ifnum #1\expandafter \@firstoftwo
 \else \expandafter \@secondoftwo
 \fi
}%
\providecommand \@ifx [1]{%
 \ifx #1\expandafter \@firstoftwo
 \else \expandafter \@secondoftwo
 \fi
}%
\providecommand \natexlab [1]{#1}%
\providecommand \enquote  [1]{``#1''}%
\providecommand \bibnamefont  [1]{#1}%
\providecommand \bibfnamefont [1]{#1}%
\providecommand \citenamefont [1]{#1}%
\providecommand \href@noop [0]{\@secondoftwo}%
\providecommand \href [0]{\begingroup \@sanitize@url \@href}%
\providecommand \@href[1]{\@@startlink{#1}\@@href}%
\providecommand \@@href[1]{\endgroup#1\@@endlink}%
\providecommand \@sanitize@url [0]{\catcode `\\12\catcode `\$12\catcode
  `\&12\catcode `\#12\catcode `\^12\catcode `\_12\catcode `\%12\relax}%
\providecommand \@@startlink[1]{}%
\providecommand \@@endlink[0]{}%
\providecommand \url  [0]{\begingroup\@sanitize@url \@url }%
\providecommand \@url [1]{\endgroup\@href {#1}{\urlprefix }}%
\providecommand \urlprefix  [0]{URL }%
\providecommand \Eprint [0]{\href }%
\providecommand \doibase [0]{http://dx.doi.org/}%
\providecommand \selectlanguage [0]{\@gobble}%
\providecommand \bibinfo  [0]{\@secondoftwo}%
\providecommand \bibfield  [0]{\@secondoftwo}%
\providecommand \translation [1]{[#1]}%
\providecommand \BibitemOpen [0]{}%
\providecommand \bibitemStop [0]{}%
\providecommand \bibitemNoStop [0]{.\EOS\space}%
\providecommand \EOS [0]{\spacefactor3000\relax}%
\providecommand \BibitemShut  [1]{\csname bibitem#1\endcsname}%
\let\auto@bib@innerbib\@empty
\bibitem [{\citenamefont {Mancini}\ \emph {et~al.}(1998)\citenamefont
  {Mancini}, \citenamefont {Vitali},\ and\ \citenamefont {Tombesi}}]{MVT98}%
  \BibitemOpen
  \bibfield  {author} {\bibinfo {author} {\bibfnamefont {S.}~\bibnamefont
  {Mancini}}, \bibinfo {author} {\bibfnamefont {D.}~\bibnamefont {Vitali}}, \
  and\ \bibinfo {author} {\bibfnamefont {P.}~\bibnamefont {Tombesi}},\ }\href
  {\doibase 10.1103/PhysRevLett.80.688} {\bibfield  {journal} {\bibinfo
  {journal} {Phys. Rev. Lett.}\ }\textbf {\bibinfo {volume} {80}},\ \bibinfo
  {pages} {688} (\bibinfo {year} {1998})}\BibitemShut {NoStop}%
\bibitem [{\citenamefont {Marquardt}\ \emph {et~al.}(2007)\citenamefont
  {Marquardt}, \citenamefont {Chen}, \citenamefont {Clerk},\ and\ \citenamefont
  {Girvin}}]{MC&07}%
  \BibitemOpen
  \bibfield  {author} {\bibinfo {author} {\bibfnamefont {F.}~\bibnamefont
  {Marquardt}}, \bibinfo {author} {\bibfnamefont {J.~P.}\ \bibnamefont {Chen}},
  \bibinfo {author} {\bibfnamefont {A.~A.}\ \bibnamefont {Clerk}}, \ and\
  \bibinfo {author} {\bibfnamefont {S.~M.}\ \bibnamefont {Girvin}},\ }\href
  {\doibase 10.1103/PhysRevLett.99.093902} {\bibfield  {journal} {\bibinfo
  {journal} {Phys. Rev. Lett.}\ }\textbf {\bibinfo {volume} {99}},\ \bibinfo
  {pages} {093902} (\bibinfo {year} {2007})}\BibitemShut {NoStop}%
\bibitem [{\citenamefont {Machnes}\ \emph {et~al.}(2012)\citenamefont
  {Machnes}, \citenamefont {Cerrillo}, \citenamefont {Aspelmeyer},
  \citenamefont {Wieczorek}, \citenamefont {Plenio},\ and\ \citenamefont
  {Retzker}}]{MC&12}%
  \BibitemOpen
  \bibfield  {author} {\bibinfo {author} {\bibfnamefont {S.}~\bibnamefont
  {Machnes}}, \bibinfo {author} {\bibfnamefont {J.}~\bibnamefont {Cerrillo}},
  \bibinfo {author} {\bibfnamefont {M.}~\bibnamefont {Aspelmeyer}}, \bibinfo
  {author} {\bibfnamefont {W.}~\bibnamefont {Wieczorek}}, \bibinfo {author}
  {\bibfnamefont {M.~B.}\ \bibnamefont {Plenio}}, \ and\ \bibinfo {author}
  {\bibfnamefont {A.}~\bibnamefont {Retzker}},\ }\href {\doibase
  10.1103/PhysRevLett.108.153601} {\bibfield  {journal} {\bibinfo  {journal}
  {Phys. Rev. Lett.}\ }\textbf {\bibinfo {volume} {108}},\ \bibinfo {pages}
  {153601} (\bibinfo {year} {2012})}\BibitemShut {NoStop}%
\bibitem [{\citenamefont {Aspelmeyer}\ \emph {et~al.}(2014)\citenamefont
  {Aspelmeyer}, \citenamefont {Kippenberg},\ and\ \citenamefont
  {Marquardt}}]{AKM14}%
  \BibitemOpen
  \bibfield  {author} {\bibinfo {author} {\bibfnamefont {M.}~\bibnamefont
  {Aspelmeyer}}, \bibinfo {author} {\bibfnamefont {T.~J.}\ \bibnamefont
  {Kippenberg}}, \ and\ \bibinfo {author} {\bibfnamefont {F.}~\bibnamefont
  {Marquardt}},\ }\href {\doibase 10.1103/RevModPhys.86.1391} {\bibfield
  {journal} {\bibinfo  {journal} {Rev. Mod. Phys.}\ }\textbf {\bibinfo {volume}
  {86}},\ \bibinfo {pages} {1391} (\bibinfo {year} {2014})}\BibitemShut
  {NoStop}%
\bibitem [{\citenamefont {Schmidt}\ \emph {et~al.}(2011)\citenamefont
  {Schmidt}, \citenamefont {Negretti}, \citenamefont {Ankerhold}, \citenamefont
  {Calarco},\ and\ \citenamefont {Stockburger}}]{SN&11}%
  \BibitemOpen
  \bibfield  {author} {\bibinfo {author} {\bibfnamefont {R.}~\bibnamefont
  {Schmidt}}, \bibinfo {author} {\bibfnamefont {A.}~\bibnamefont {Negretti}},
  \bibinfo {author} {\bibfnamefont {J.}~\bibnamefont {Ankerhold}}, \bibinfo
  {author} {\bibfnamefont {T.}~\bibnamefont {Calarco}}, \ and\ \bibinfo
  {author} {\bibfnamefont {J.~T.}\ \bibnamefont {Stockburger}},\ }\href
  {\doibase 10.1103/PhysRevLett.107.130404} {\bibfield  {journal} {\bibinfo
  {journal} {Phys. Rev. Lett.}\ }\textbf {\bibinfo {volume} {107}},\ \bibinfo
  {pages} {130404} (\bibinfo {year} {2011})}\BibitemShut {NoStop}%
\bibitem [{\citenamefont {Marquardt}\ and\ \citenamefont
  {Girvin}(2009)}]{MG09}%
  \BibitemOpen
  \bibfield  {author} {\bibinfo {author} {\bibfnamefont {F.}~\bibnamefont
  {Marquardt}}\ and\ \bibinfo {author} {\bibfnamefont {S.~M.}\ \bibnamefont
  {Girvin}},\ }\href {\doibase 10.1103/Physics.2.40} {\bibfield  {journal}
  {\bibinfo  {journal} {Physics}\ }\textbf {\bibinfo {volume} {2}},\ \bibinfo
  {pages} {40} (\bibinfo {year} {2009})}\BibitemShut {NoStop}%
\bibitem [{\citenamefont {Wilson-Rae}\ \emph {et~al.}(2007)\citenamefont
  {Wilson-Rae}, \citenamefont {Nooshi}, \citenamefont {Zwerger},\ and\
  \citenamefont {Kippenberg}}]{WN&07}%
  \BibitemOpen
  \bibfield  {author} {\bibinfo {author} {\bibfnamefont {I.}~\bibnamefont
  {Wilson-Rae}}, \bibinfo {author} {\bibfnamefont {N.}~\bibnamefont {Nooshi}},
  \bibinfo {author} {\bibfnamefont {W.}~\bibnamefont {Zwerger}}, \ and\
  \bibinfo {author} {\bibfnamefont {T.~J.}\ \bibnamefont {Kippenberg}},\ }\href
  {\doibase 10.1103/PhysRevLett.99.093901} {\bibfield  {journal} {\bibinfo
  {journal} {Phys. Rev. Lett.}\ }\textbf {\bibinfo {volume} {99}},\ \bibinfo
  {pages} {093901} (\bibinfo {year} {2007})}\BibitemShut {NoStop}%
\bibitem [{\citenamefont {Wang}\ \emph {et~al.}(2011)\citenamefont {Wang},
  \citenamefont {Vinjanampathy}, \citenamefont {Strauch},\ and\ \citenamefont
  {Jacobs}}]{WV&11}%
  \BibitemOpen
  \bibfield  {author} {\bibinfo {author} {\bibfnamefont {X.}~\bibnamefont
  {Wang}}, \bibinfo {author} {\bibfnamefont {S.}~\bibnamefont {Vinjanampathy}},
  \bibinfo {author} {\bibfnamefont {F.~W.}\ \bibnamefont {Strauch}}, \ and\
  \bibinfo {author} {\bibfnamefont {K.}~\bibnamefont {Jacobs}},\ }\href
  {\doibase 10.1103/PhysRevLett.107.177204} {\bibfield  {journal} {\bibinfo
  {journal} {Phys. Rev. Lett.}\ }\textbf {\bibinfo {volume} {107}},\ \bibinfo
  {pages} {177204} (\bibinfo {year} {2011})}\BibitemShut {NoStop}%
\bibitem [{\citenamefont {{Chan}}\ \emph {et~al.}(2011)\citenamefont {{Chan}},
  \citenamefont {{Alegre}}, \citenamefont {{Safavi-Naeini}}, \citenamefont
  {{Hill}}, \citenamefont {{Krause}}, \citenamefont {{Gr{\"o}blacher}},
  \citenamefont {{Aspelmeyer}},\ and\ \citenamefont {{Painter}}}]{CA&11}%
  \BibitemOpen
  \bibfield  {author} {\bibinfo {author} {\bibfnamefont {J.}~\bibnamefont
  {{Chan}}}, \bibinfo {author} {\bibfnamefont {T.~P.~M.}\ \bibnamefont
  {{Alegre}}}, \bibinfo {author} {\bibfnamefont {A.~H.}\ \bibnamefont
  {{Safavi-Naeini}}}, \bibinfo {author} {\bibfnamefont {J.~T.}\ \bibnamefont
  {{Hill}}}, \bibinfo {author} {\bibfnamefont {A.}~\bibnamefont {{Krause}}},
  \bibinfo {author} {\bibfnamefont {S.}~\bibnamefont {{Gr{\"o}blacher}}},
  \bibinfo {author} {\bibfnamefont {M.}~\bibnamefont {{Aspelmeyer}}}, \ and\
  \bibinfo {author} {\bibfnamefont {O.}~\bibnamefont {{Painter}}},\ }\href
  {\doibase 10.1038/nature10461} {\bibfield  {journal} {\bibinfo  {journal}
  {Nature}\ }\textbf {\bibinfo {volume} {478}},\ \bibinfo {pages} {89}
  (\bibinfo {year} {2011})}\BibitemShut {NoStop}%
\bibitem [{\citenamefont {Safavi-Naeini}\ \emph {et~al.}(2013)\citenamefont
  {Safavi-Naeini}, \citenamefont {Chan}, \citenamefont {Hill}, \citenamefont
  {Gröblacher}, \citenamefont {Miao}, \citenamefont {Chen}, \citenamefont
  {Aspelmeyer},\ and\ \citenamefont {Painter}}]{SC&13}%
  \BibitemOpen
  \bibfield  {author} {\bibinfo {author} {\bibfnamefont {A.~H.}\ \bibnamefont
  {Safavi-Naeini}}, \bibinfo {author} {\bibfnamefont {J.}~\bibnamefont {Chan}},
  \bibinfo {author} {\bibfnamefont {J.~T.}\ \bibnamefont {Hill}}, \bibinfo
  {author} {\bibfnamefont {S.}~\bibnamefont {Gröblacher}}, \bibinfo {author}
  {\bibfnamefont {H.}~\bibnamefont {Miao}}, \bibinfo {author} {\bibfnamefont
  {Y.}~\bibnamefont {Chen}}, \bibinfo {author} {\bibfnamefont {M.}~\bibnamefont
  {Aspelmeyer}}, \ and\ \bibinfo {author} {\bibfnamefont {O.}~\bibnamefont
  {Painter}},\ }\href {http://stacks.iop.org/1367-2630/15/i=3/a=035007}
  {\bibfield  {journal} {\bibinfo  {journal} {New Journal of Physics}\ }\textbf
  {\bibinfo {volume} {15}},\ \bibinfo {pages} {035007} (\bibinfo {year}
  {2013})}\BibitemShut {NoStop}%
\bibitem [{\citenamefont {Liu}\ \emph {et~al.}(2013)\citenamefont {Liu},
  \citenamefont {Xiao}, \citenamefont {Luan},\ and\ \citenamefont
  {Wong}}]{LX&13}%
  \BibitemOpen
  \bibfield  {author} {\bibinfo {author} {\bibfnamefont {Y.-C.}\ \bibnamefont
  {Liu}}, \bibinfo {author} {\bibfnamefont {Y.-F.}\ \bibnamefont {Xiao}},
  \bibinfo {author} {\bibfnamefont {X.}~\bibnamefont {Luan}}, \ and\ \bibinfo
  {author} {\bibfnamefont {C.~W.}\ \bibnamefont {Wong}},\ }\href {\doibase
  10.1103/PhysRevLett.110.153606} {\bibfield  {journal} {\bibinfo  {journal}
  {Phys. Rev. Lett.}\ }\textbf {\bibinfo {volume} {110}},\ \bibinfo {pages}
  {153606} (\bibinfo {year} {2013})}\BibitemShut {NoStop}%
\bibitem [{\citenamefont {Liu}\ \emph {et~al.}(2015)\citenamefont {Liu},
  \citenamefont {Liu}, \citenamefont {Dong}, \citenamefont {Li}, \citenamefont
  {Gong},\ and\ \citenamefont {Xiao}}]{LL&14}%
  \BibitemOpen
  \bibfield  {author} {\bibinfo {author} {\bibfnamefont {Y.-C.}\ \bibnamefont
  {Liu}}, \bibinfo {author} {\bibfnamefont {R.-S.}\ \bibnamefont {Liu}},
  \bibinfo {author} {\bibfnamefont {C.-H.}\ \bibnamefont {Dong}}, \bibinfo
  {author} {\bibfnamefont {Y.}~\bibnamefont {Li}}, \bibinfo {author}
  {\bibfnamefont {Q.}~\bibnamefont {Gong}}, \ and\ \bibinfo {author}
  {\bibfnamefont {Y.-F.}\ \bibnamefont {Xiao}},\ }\href {\doibase
  10.1103/PhysRevA.91.013824} {\bibfield  {journal} {\bibinfo  {journal} {Phys.
  Rev. A}\ }\textbf {\bibinfo {volume} {91}},\ \bibinfo {pages} {013824}
  (\bibinfo {year} {2015})}\BibitemShut {NoStop}%
\bibitem [{\citenamefont {Weiss}(2012)}]{Wei12}%
  \BibitemOpen
  \bibfield  {author} {\bibinfo {author} {\bibfnamefont {U.}~\bibnamefont
  {Weiss}},\ }\href {https://books.google.com.co/books?id=qgfuFZxvGKQC} {\emph
  {\bibinfo {title} {Quantum Dissipative Systems}}},\ Series in modern
  condensed matter physics\ (\bibinfo  {publisher} {World Scientific},\
  \bibinfo {year} {2012})\BibitemShut {NoStop}%
\bibitem [{\citenamefont {Pach\'on}\ and\ \citenamefont {Brumer}(2014)}]{PB14}%
  \BibitemOpen
  \bibfield  {author} {\bibinfo {author} {\bibfnamefont {L.~A.}\ \bibnamefont
  {Pach\'on}}\ and\ \bibinfo {author} {\bibfnamefont {P.}~\bibnamefont
  {Brumer}},\ }\href {\doibase 10.1063/1.4858915} {\bibfield  {journal}
  {\bibinfo  {journal} {J. Math. Phys.}\ }\textbf {\bibinfo {volume} {55}},\
  \bibinfo {eid} {012103} (\bibinfo {year} {2014}),\
  10.1063/1.4858915}\BibitemShut {NoStop}%
\bibitem [{\citenamefont {Pach\'on}\ \emph {et~al.}(2014)\citenamefont
  {Pach\'on}, \citenamefont {Triana}, \citenamefont {Zueco},\ and\
  \citenamefont {Brumer}}]{PT&14}%
  \BibitemOpen
  \bibfield  {author} {\bibinfo {author} {\bibfnamefont {L.~A.}\ \bibnamefont
  {Pach\'on}}, \bibinfo {author} {\bibfnamefont {J.~F.}\ \bibnamefont
  {Triana}}, \bibinfo {author} {\bibfnamefont {D.}~\bibnamefont {Zueco}}, \
  and\ \bibinfo {author} {\bibfnamefont {P.}~\bibnamefont {Brumer}},\
  }\href@noop {} {\bibfield  {journal} {\bibinfo  {journal} {arXiv}\ }\textbf
  {\bibinfo {volume} {1401.1418}} (\bibinfo {year} {2014})},\ \Eprint
  {http://arxiv.org/abs/arXiv:1401.1418} {arXiv:1401.1418} \BibitemShut
  {NoStop}%
\bibitem [{\citenamefont {Estrada}\ and\ \citenamefont {Pachon}(2015)}]{EP15}%
  \BibitemOpen
  \bibfield  {author} {\bibinfo {author} {\bibfnamefont {A.~F.}\ \bibnamefont
  {Estrada}}\ and\ \bibinfo {author} {\bibfnamefont {L.~A.}\ \bibnamefont
  {Pachon}},\ }\href {http://stacks.iop.org/1367-2630/17/i=3/a=033038}
  {\bibfield  {journal} {\bibinfo  {journal} {New J. Phys.}\ }\textbf {\bibinfo
  {volume} {17}},\ \bibinfo {pages} {033038} (\bibinfo {year}
  {2015})}\BibitemShut {NoStop}%
\bibitem [{\citenamefont {Groblacher}\ \emph {et~al.}(2015)\citenamefont
  {Groblacher}, \citenamefont {Trubarov}, \citenamefont {Prigge}, \citenamefont
  {Cole}, \citenamefont {Aspelmeyer},\ and\ \citenamefont {Eisert}}]{GT&13}%
  \BibitemOpen
  \bibfield  {author} {\bibinfo {author} {\bibfnamefont {S.}~\bibnamefont
  {Groblacher}}, \bibinfo {author} {\bibfnamefont {A.}~\bibnamefont
  {Trubarov}}, \bibinfo {author} {\bibfnamefont {N.}~\bibnamefont {Prigge}},
  \bibinfo {author} {\bibfnamefont {G.~D.}\ \bibnamefont {Cole}}, \bibinfo
  {author} {\bibfnamefont {M.}~\bibnamefont {Aspelmeyer}}, \ and\ \bibinfo
  {author} {\bibfnamefont {J.}~\bibnamefont {Eisert}},\ }\href
  {http://dx.doi.org/10.1038/ncomms8606} {\bibfield  {journal} {\bibinfo
  {journal} {Nat Commun}\ }\textbf {\bibinfo {volume} {6}},\ \bibinfo {pages}
  {7606} (\bibinfo {year} {2015})}\BibitemShut {NoStop}%
\bibitem [{\citenamefont {Deffner}\ and\ \citenamefont {Lutz}(2013)}]{DL13}%
  \BibitemOpen
  \bibfield  {author} {\bibinfo {author} {\bibfnamefont {S.}~\bibnamefont
  {Deffner}}\ and\ \bibinfo {author} {\bibfnamefont {E.}~\bibnamefont {Lutz}},\
  }\href {\doibase 10.1103/PhysRevLett.111.010402} {\bibfield  {journal}
  {\bibinfo  {journal} {Phys. Rev. Lett.}\ }\textbf {\bibinfo {volume} {111}},\
  \bibinfo {pages} {010402} (\bibinfo {year} {2013})}\BibitemShut {NoStop}%
\bibitem [{\citenamefont {Thorwart}\ \emph {et~al.}(2009)\citenamefont
  {Thorwart}, \citenamefont {Eckel}, \citenamefont {Reina}, \citenamefont
  {Nalbach},\ and\ \citenamefont {Weiss}}]{TE&09}%
  \BibitemOpen
  \bibfield  {author} {\bibinfo {author} {\bibfnamefont {M.}~\bibnamefont
  {Thorwart}}, \bibinfo {author} {\bibfnamefont {J.}~\bibnamefont {Eckel}},
  \bibinfo {author} {\bibfnamefont {J.}~\bibnamefont {Reina}}, \bibinfo
  {author} {\bibfnamefont {P.}~\bibnamefont {Nalbach}}, \ and\ \bibinfo
  {author} {\bibfnamefont {S.}~\bibnamefont {Weiss}},\ }\href {\doibase
  http://dx.doi.org/10.1016/j.cplett.2009.07.053} {\bibfield  {journal}
  {\bibinfo  {journal} {Chemical Physics Letters}\ }\textbf {\bibinfo {volume}
  {478}},\ \bibinfo {pages} {234 } (\bibinfo {year} {2009})}\BibitemShut
  {NoStop}%
\bibitem [{\citenamefont {Huelga}\ \emph {et~al.}(2012)\citenamefont {Huelga},
  \citenamefont {Rivas},\ and\ \citenamefont {Plenio}}]{HRP12}%
  \BibitemOpen
  \bibfield  {author} {\bibinfo {author} {\bibfnamefont {S.~F.}\ \bibnamefont
  {Huelga}}, \bibinfo {author} {\bibfnamefont {A.}~\bibnamefont {Rivas}}, \
  and\ \bibinfo {author} {\bibfnamefont {M.~B.}\ \bibnamefont {Plenio}},\
  }\href {\doibase 10.1103/PhysRevLett.108.160402} {\bibfield  {journal}
  {\bibinfo  {journal} {Phys. Rev. Lett.}\ }\textbf {\bibinfo {volume} {108}},\
  \bibinfo {pages} {160402} (\bibinfo {year} {2012})}\BibitemShut {NoStop}%
\bibitem [{\citenamefont {Chin}\ \emph {et~al.}(2012)\citenamefont {Chin},
  \citenamefont {Huelga},\ and\ \citenamefont {Plenio}}]{CHP12}%
  \BibitemOpen
  \bibfield  {author} {\bibinfo {author} {\bibfnamefont {A.~W.}\ \bibnamefont
  {Chin}}, \bibinfo {author} {\bibfnamefont {S.~F.}\ \bibnamefont {Huelga}}, \
  and\ \bibinfo {author} {\bibfnamefont {M.~B.}\ \bibnamefont {Plenio}},\
  }\href {\doibase 10.1103/PhysRevLett.109.233601} {\bibfield  {journal}
  {\bibinfo  {journal} {Phys. Rev. Lett.}\ }\textbf {\bibinfo {volume} {109}},\
  \bibinfo {pages} {233601} (\bibinfo {year} {2012})}\BibitemShut {NoStop}%
\bibitem [{\citenamefont {Yang}\ \emph {et~al.}(2014)\citenamefont {Yang},
  \citenamefont {An}, \citenamefont {Luo}, \citenamefont {Li},\ and\
  \citenamefont {Oh}}]{YA&14}%
  \BibitemOpen
  \bibfield  {author} {\bibinfo {author} {\bibfnamefont {C.-J.}\ \bibnamefont
  {Yang}}, \bibinfo {author} {\bibfnamefont {J.-H.}\ \bibnamefont {An}},
  \bibinfo {author} {\bibfnamefont {H.-G.}\ \bibnamefont {Luo}}, \bibinfo
  {author} {\bibfnamefont {Y.}~\bibnamefont {Li}}, \ and\ \bibinfo {author}
  {\bibfnamefont {C.~H.}\ \bibnamefont {Oh}},\ }\href {\doibase
  10.1103/PhysRevE.90.022122} {\bibfield  {journal} {\bibinfo  {journal} {Phys.
  Rev. E}\ }\textbf {\bibinfo {volume} {90}},\ \bibinfo {pages} {022122}
  (\bibinfo {year} {2014})}\BibitemShut {NoStop}%
\bibitem [{\citenamefont {Teufel}\ \emph {et~al.}(2011)\citenamefont {Teufel},
  \citenamefont {Donner}, \citenamefont {Li}, \citenamefont {Harlow},
  \citenamefont {Allman}, \citenamefont {Cicak}, \citenamefont {Sirois},
  \citenamefont {Whittaker}, \citenamefont {Lehnert},\ and\ \citenamefont
  {Simmonds}}]{TD&11}%
  \BibitemOpen
  \bibfield  {author} {\bibinfo {author} {\bibfnamefont {J.~D.}\ \bibnamefont
  {Teufel}}, \bibinfo {author} {\bibfnamefont {T.}~\bibnamefont {Donner}},
  \bibinfo {author} {\bibfnamefont {D.}~\bibnamefont {Li}}, \bibinfo {author}
  {\bibfnamefont {J.~W.}\ \bibnamefont {Harlow}}, \bibinfo {author}
  {\bibfnamefont {M.~S.}\ \bibnamefont {Allman}}, \bibinfo {author}
  {\bibfnamefont {K.}~\bibnamefont {Cicak}}, \bibinfo {author} {\bibfnamefont
  {A.~J.}\ \bibnamefont {Sirois}}, \bibinfo {author} {\bibfnamefont {J.~D.}\
  \bibnamefont {Whittaker}}, \bibinfo {author} {\bibfnamefont {K.~W.}\
  \bibnamefont {Lehnert}}, \ and\ \bibinfo {author} {\bibfnamefont {R.~W.}\
  \bibnamefont {Simmonds}},\ }\href {\doibase 10.1038/nature10261} {\bibfield
  {journal} {\bibinfo  {journal} {Nature}\ }\textbf {\bibinfo {volume} {475}},\
  \bibinfo {pages} {359} (\bibinfo {year} {2011})}\BibitemShut {NoStop}%
\bibitem [{\citenamefont {Gigan}\ \emph {et~al.}(2006)\citenamefont {Gigan},
  \citenamefont {Bohm}, \citenamefont {Paternostro}, \citenamefont {Blaser},
  \citenamefont {Langer}, \citenamefont {Hertzberg}, \citenamefont {Schwab},
  \citenamefont {Bauerle}, \citenamefont {Aspelmeyer},\ and\ \citenamefont
  {Zeilinger}}]{GB&06}%
  \BibitemOpen
  \bibfield  {author} {\bibinfo {author} {\bibfnamefont {S.}~\bibnamefont
  {Gigan}}, \bibinfo {author} {\bibfnamefont {H.~R.}\ \bibnamefont {Bohm}},
  \bibinfo {author} {\bibfnamefont {M.}~\bibnamefont {Paternostro}}, \bibinfo
  {author} {\bibfnamefont {F.}~\bibnamefont {Blaser}}, \bibinfo {author}
  {\bibfnamefont {G.}~\bibnamefont {Langer}}, \bibinfo {author} {\bibfnamefont
  {J.~B.}\ \bibnamefont {Hertzberg}}, \bibinfo {author} {\bibfnamefont {K.~C.}\
  \bibnamefont {Schwab}}, \bibinfo {author} {\bibfnamefont {D.}~\bibnamefont
  {Bauerle}}, \bibinfo {author} {\bibfnamefont {M.}~\bibnamefont {Aspelmeyer}},
  \ and\ \bibinfo {author} {\bibfnamefont {A.}~\bibnamefont {Zeilinger}},\
  }\href {http://dx.doi.org/10.1038/nature05273} {\bibfield  {journal}
  {\bibinfo  {journal} {Nature}\ }\textbf {\bibinfo {volume} {444}},\ \bibinfo
  {pages} {67} (\bibinfo {year} {2006})}\BibitemShut {NoStop}%
\bibitem [{\citenamefont {{Kleckner}}\ and\ \citenamefont
  {{Bouwmeester}}(2006)}]{KB06}%
  \BibitemOpen
  \bibfield  {author} {\bibinfo {author} {\bibfnamefont {D.}~\bibnamefont
  {{Kleckner}}}\ and\ \bibinfo {author} {\bibfnamefont {D.}~\bibnamefont
  {{Bouwmeester}}},\ }\href {\doibase 10.1038/nature05231} {\bibfield
  {journal} {\bibinfo  {journal} {Nature}\ }\textbf {\bibinfo {volume} {444}},\
  \bibinfo {pages} {75} (\bibinfo {year} {2006})}\BibitemShut {NoStop}%
\bibitem [{\citenamefont {Kirk}(2012)}]{Kir12}%
  \BibitemOpen
  \bibfield  {author} {\bibinfo {author} {\bibfnamefont {D.}~\bibnamefont
  {Kirk}},\ }\href {http://books.google.es/books?id=onuH0PnZwV4C} {\emph
  {\bibinfo {title} {Optimal Control Theory: An Introduction}}},\ Dover Books
  on Electrical Engineering\ (\bibinfo  {publisher} {Dover Publications},\
  \bibinfo {year} {2012})\BibitemShut {NoStop}%
\bibitem [{\citenamefont {Ullersma}(1966)}]{Ull66}%
  \BibitemOpen
  \bibfield  {author} {\bibinfo {author} {\bibfnamefont {P.}~\bibnamefont
  {Ullersma}},\ }\href@noop {} {\bibfield  {journal} {\bibinfo  {journal}
  {Physica}\ }\textbf {\bibinfo {volume} {32}},\ \bibinfo {pages} {27}
  (\bibinfo {year} {1966})}\BibitemShut {NoStop}%
\bibitem [{\citenamefont {Caldeira}\ and\ \citenamefont
  {Leggett}(1981)}]{CL81}%
  \BibitemOpen
  \bibfield  {author} {\bibinfo {author} {\bibfnamefont {A.~O.}\ \bibnamefont
  {Caldeira}}\ and\ \bibinfo {author} {\bibfnamefont {A.~J.}\ \bibnamefont
  {Leggett}},\ }\href {\doibase 10.1103/PhysRevLett.46.211} {\bibfield
  {journal} {\bibinfo  {journal} {Phys. Rev. Lett.}\ }\textbf {\bibinfo
  {volume} {46}},\ \bibinfo {pages} {211} (\bibinfo {year} {1981})}\BibitemShut
  {NoStop}%
\bibitem [{\citenamefont {Clerk}\ \emph {et~al.}(2010)\citenamefont {Clerk},
  \citenamefont {Devoret}, \citenamefont {Girvin}, \citenamefont {Marquardt},\
  and\ \citenamefont {Schoelkopf}}]{CD&10}%
  \BibitemOpen
  \bibfield  {author} {\bibinfo {author} {\bibfnamefont {A.~A.}\ \bibnamefont
  {Clerk}}, \bibinfo {author} {\bibfnamefont {M.~H.}\ \bibnamefont {Devoret}},
  \bibinfo {author} {\bibfnamefont {S.~M.}\ \bibnamefont {Girvin}}, \bibinfo
  {author} {\bibfnamefont {F.}~\bibnamefont {Marquardt}}, \ and\ \bibinfo
  {author} {\bibfnamefont {R.~J.}\ \bibnamefont {Schoelkopf}},\ }\href
  {\doibase 10.1103/RevModPhys.82.1155} {\bibfield  {journal} {\bibinfo
  {journal} {Rev. Mod. Phys.}\ }\textbf {\bibinfo {volume} {82}},\ \bibinfo
  {pages} {1155} (\bibinfo {year} {2010})}\BibitemShut {NoStop}%
\bibitem [{\citenamefont {Gardiner}\ and\ \citenamefont {Zoller}(2004)}]{GZ04}%
  \BibitemOpen
  \bibfield  {author} {\bibinfo {author} {\bibfnamefont {C.}~\bibnamefont
  {Gardiner}}\ and\ \bibinfo {author} {\bibfnamefont {P.}~\bibnamefont
  {Zoller}},\ }\href {http://books.google.co.uk/books?id=a\_xsT8oGhdgC} {\emph
  {\bibinfo {title} {Quantum Noise: A Handbook of {M}arkovian and
  Non-{M}arkovian Quantum Stochastic Methods with Applications to Quantum
  Optics}}},\ Springer Series in Synergetics\ (\bibinfo  {publisher}
  {Springer},\ \bibinfo {year} {2004})\BibitemShut {NoStop}%
\bibitem [{\citenamefont {Breuer}\ and\ \citenamefont
  {Petruccione}(2007)}]{BP07}%
  \BibitemOpen
  \bibfield  {author} {\bibinfo {author} {\bibfnamefont {H.}~\bibnamefont
  {Breuer}}\ and\ \bibinfo {author} {\bibfnamefont {F.}~\bibnamefont
  {Petruccione}},\ }\href {http://books.google.co.uk/books?id=DkcJPwAACAAJ}
  {\emph {\bibinfo {title} {The Theory of Open Quantum Systems}}}\ (\bibinfo
  {publisher} {OUP Oxford},\ \bibinfo {year} {2007})\BibitemShut {NoStop}%
\bibitem [{\citenamefont {Feynman}\ and\ \citenamefont {Hibbs}(2012)}]{FH12}%
  \BibitemOpen
  \bibfield  {author} {\bibinfo {author} {\bibfnamefont {R.}~\bibnamefont
  {Feynman}}\ and\ \bibinfo {author} {\bibfnamefont {A.}~\bibnamefont
  {Hibbs}},\ }\href {http://books.google.com.co/books?id=-jjvxfwcleYC} {\emph
  {\bibinfo {title} {Quantum Mechanics and Path Integrals: Emended Edition}}}\
  (\bibinfo  {publisher} {Dover Publications, Incorporated},\ \bibinfo {year}
  {2012})\BibitemShut {NoStop}%
\bibitem [{\citenamefont {Genes}\ \emph {et~al.}(2008)\citenamefont {Genes},
  \citenamefont {Vitali}, \citenamefont {Tombesi}, \citenamefont {Gigan},\ and\
  \citenamefont {Aspelmeyer}}]{GV&08}%
  \BibitemOpen
  \bibfield  {author} {\bibinfo {author} {\bibfnamefont {C.}~\bibnamefont
  {Genes}}, \bibinfo {author} {\bibfnamefont {D.}~\bibnamefont {Vitali}},
  \bibinfo {author} {\bibfnamefont {P.}~\bibnamefont {Tombesi}}, \bibinfo
  {author} {\bibfnamefont {S.}~\bibnamefont {Gigan}}, \ and\ \bibinfo {author}
  {\bibfnamefont {M.}~\bibnamefont {Aspelmeyer}},\ }\href {\doibase
  10.1103/PhysRevA.77.033804} {\bibfield  {journal} {\bibinfo  {journal} {Phys.
  Rev. A}\ }\textbf {\bibinfo {volume} {77}},\ \bibinfo {pages} {033804}
  (\bibinfo {year} {2008})}\BibitemShut {NoStop}%
\bibitem [{\citenamefont {Rivi\`ere}\ \emph {et~al.}(2011)\citenamefont
  {Rivi\`ere}, \citenamefont {Del\'eglise}, \citenamefont {Weis}, \citenamefont
  {Gavartin}, \citenamefont {Arcizet}, \citenamefont {Schliesser},\ and\
  \citenamefont {Kippenberg}}]{RD&11}%
  \BibitemOpen
  \bibfield  {author} {\bibinfo {author} {\bibfnamefont {R.}~\bibnamefont
  {Rivi\`ere}}, \bibinfo {author} {\bibfnamefont {S.}~\bibnamefont
  {Del\'eglise}}, \bibinfo {author} {\bibfnamefont {S.}~\bibnamefont {Weis}},
  \bibinfo {author} {\bibfnamefont {E.}~\bibnamefont {Gavartin}}, \bibinfo
  {author} {\bibfnamefont {O.}~\bibnamefont {Arcizet}}, \bibinfo {author}
  {\bibfnamefont {A.}~\bibnamefont {Schliesser}}, \ and\ \bibinfo {author}
  {\bibfnamefont {T.~J.}\ \bibnamefont {Kippenberg}},\ }\href {\doibase
  10.1103/PhysRevA.83.063835} {\bibfield  {journal} {\bibinfo  {journal} {Phys.
  Rev. A}\ }\textbf {\bibinfo {volume} {83}},\ \bibinfo {pages} {063835}
  (\bibinfo {year} {2011})}\BibitemShut {NoStop}%
\bibitem [{\citenamefont {Pach\'on}\ \emph {et~al.}(2010)\citenamefont
  {Pach\'on}, \citenamefont {Ingold},\ and\ \citenamefont {Dittrich}}]{PID10}%
  \BibitemOpen
  \bibfield  {author} {\bibinfo {author} {\bibfnamefont {L.}~\bibnamefont
  {Pach\'on}}, \bibinfo {author} {\bibfnamefont {G.-L.}\ \bibnamefont
  {Ingold}}, \ and\ \bibinfo {author} {\bibfnamefont {T.}~\bibnamefont
  {Dittrich}},\ }\href {\doibase
  http://dx.doi.org/10.1016/j.chemphys.2010.05.024} {\bibfield  {journal}
  {\bibinfo  {journal} {Chemical Physics}\ }\textbf {\bibinfo {volume} {375}},\
  \bibinfo {pages} {209 } (\bibinfo {year} {2010})}\BibitemShut {NoStop}%
\bibitem [{\citenamefont {Dittrich}\ \emph {et~al.}(2010)\citenamefont
  {Dittrich}, \citenamefont {G\'omez},\ and\ \citenamefont {Pach\'on}}]{DGP10}%
  \BibitemOpen
  \bibfield  {author} {\bibinfo {author} {\bibfnamefont {T.}~\bibnamefont
  {Dittrich}}, \bibinfo {author} {\bibfnamefont {E.~A.}\ \bibnamefont
  {G\'omez}}, \ and\ \bibinfo {author} {\bibfnamefont {L.~A.}\ \bibnamefont
  {Pach\'on}},\ }\href {\doibase http://dx.doi.org/10.1063/1.3425881}
  {\bibfield  {journal} {\bibinfo  {journal} {The Journal of Chemical Physics}\
  }\textbf {\bibinfo {volume} {132}},\ \bibinfo {eid} {214102} (\bibinfo {year}
  {2010})}\BibitemShut {NoStop}%
\bibitem [{\citenamefont {Shapiro}\ and\ \citenamefont {Brumer}(2012)}]{SB12}%
  \BibitemOpen
  \bibfield  {author} {\bibinfo {author} {\bibfnamefont {M.}~\bibnamefont
  {Shapiro}}\ and\ \bibinfo {author} {\bibfnamefont {P.}~\bibnamefont
  {Brumer}},\ }\href@noop {} {\emph {\bibinfo {title} {Quantum Control of
  Molecular Processes}}},\ \bibinfo {edition} {2nd}\ ed.\ (\bibinfo
  {publisher} {Wiley-VCH},\ \bibinfo {address} {Weinheim},\ \bibinfo {year}
  {2012})\BibitemShut {NoStop}%
\end{thebibliography}%

\newpage
\widetext{


\section*{Supplementary material}

# Supplementary Material for Ultrafast Optimal Sideband Cooling under Non-Markovian Evolution

Johan F. Triana, Andrés F. Estrada, and Leonardo A. Pachón

*Grupo de Física Atómica y Molecular, Instituto de*

*Física, Facultad de Ciencias Exactas y Naturales,*

*Universidad de Antioquia UdeA; Calle 70 No. 52-21, Medellín, Colombia.*

**INFLUENCE FUNCTIONAL**

The state of the mechanical mode and the optical mode $\tilde{\rho}(q''_{1+}, q''_{2+}, q''_{1-}, q''_2, t)$ can be determined from

$$\tilde{\rho}(q''_{1+}, q''_{2+}, q''_{1-}, q''_2, t) = \int_{-\infty}^{\infty} \mathrm{d}q'_{1+} \mathrm{d}q'_{2+} \mathrm{d}q'_{1-} \mathrm{d}q'_{2-} J(q''_{1+}, q''_{2+}, q''_{1-}, q''_{2-}, t; q'_{1+}, q'_{2+}, q'_{1-}, q'_{2-}, 0)$$
$$\times \rho_S(q'_{1+}, q'_{2+}, q'_{1-}, q'_{2-}, 0) \tag{1}$$

where $\rho_S(q'_{1+}, q'_{2+}, q'_{1-}, q'_{2-}, 0)$ accounts for the initial state. $J(q''_{1+}, q''_{2+}, q''_{1-}, q''_{2-}, t; q'_{1+}, q'_{2+}, q'_{1-}, q'_{2-}, 0)$ is the propagating function to be obtained from the Feynman-Vernon influence functional approach. After assuming that each mode is coupled to its own Ullersma-Caldeira-Leggett-type bath at temperature $T_{1,2}$, the propagating function reads

$$J(q''_{1+}, q''_{2+}, q''_{1-}, q''_{2-}, t; q'_{1+}, q'_{2+}, q'_{1-}, q'_{2-}, 0) =$$
$$\frac{1}{N(t)} \exp\left\{ \frac{i}{\hbar} \int_0^t \mathrm{d}s \left[ \sum_{\alpha=1}^{2} \left( \frac{1}{2} m_\alpha \dot{q}_{\alpha+}^2(s) - \frac{1}{2} m_\alpha \omega_\alpha^2 q_{\alpha+}^2(s) \right) - c(s) q_{1+}(s) q_{2+}(s) \right] \right\}$$
$$\times \exp\left\{ -\frac{i}{\hbar} \int_0^t \mathrm{d}s \left[ \sum_{\alpha=1}^{2} \left( \frac{1}{2} m_\alpha \dot{q}_{\alpha-}^2(s) - \frac{1}{2} m_\alpha \omega_\alpha^2 q_{\alpha-}^2(s) \right) - c(s) q_{1-}(s) q_{2-}(s) \right] \right\}$$
$$\times \mathcal{F}[q_{1+}, q_{2+}, q_{1-}, q_{2-}] \tag{2}$$

with

$$\mathcal{F}[q_{1+}, q_{2+}, q_{1-}, q_{2-}] = \prod_{\alpha=1}^{2} \exp\left( -\frac{im_\alpha}{2\hbar} \left\{ (q'_{\alpha+} + q'_{\alpha-}) \int_0^t \mathrm{d}s\, \gamma_\alpha(s) [q_{\alpha+}(s) - q_{\alpha-}(s)] \right. \right.$$
$$\left. + \int_0^t \mathrm{d}s \int_0^s \mathrm{d}u\, \gamma_\alpha(s-u) [\dot{q}_{\alpha+}(u) + \dot{q}_{\alpha-}(u)][q_{\alpha+}(s) - q_{\alpha-}(s)] \right\} \right)$$
$$\times \exp\left\{ -\frac{1}{\hbar} \int_0^t \mathrm{d}s \int_0^s \mathrm{d}u\, K_\alpha(u-s) [q_{\alpha+}(s) - q_{\alpha-}(s)][q_{\alpha+}(u) - q_{\alpha-}(u)] \right\}, \tag{3}$$

where

$$\gamma_\alpha(s) = \frac{2}{m_\alpha} \int_0^\infty \frac{\mathrm{d}\omega_\alpha}{\pi} \frac{J_\alpha(\omega_\alpha)}{\omega_\alpha} \cos(\omega_\alpha s), \tag{4}$$

$$K_\alpha(s) = \int_0^\infty \frac{\mathrm{d}\omega_\alpha}{\pi} J_\alpha(\omega_\alpha) \coth\left(\frac{\hbar\beta_\alpha\omega_\alpha}{2}\right) \cos(\omega_\alpha s), \tag{5}$$

being $J_\alpha(\omega_\alpha)$ the spectral density. In the main text, $J_\alpha(\omega_\alpha)$ was chosen such that is accounts for Ohmic dissipation with a cutoff frequency $\omega_{\alpha D}$. The functions $q_{\alpha\pm}(s)$ obey the following relations (1 for mechanical mode and 2 for optical mode)

$$\ddot{q}_{1+}(s) + \omega_1^2 q_{1+}(s) + \frac{c(s)}{m_1} q_{2+}(s) + \frac{1}{2}\frac{\mathrm{d}}{\mathrm{d}s}\left[\int_0^t \mathrm{d}u\, \gamma_1(s-u)(q_{1+}+q_{1-}) - \int_s^t \mathrm{d}u\, \gamma_1(u-s)(q_{1+}-q_{1-})\right] = 0,$$

$$\ddot{q}_{1-}(s) + \omega_1^2 q_{1-}(s) + \frac{c(s)}{m_1} q_{2-}(s) + \frac{1}{2}\frac{\mathrm{d}}{\mathrm{d}s}\left[\int_0^t \mathrm{d}u\, \gamma_1(s-u)(q_{1+}+q_{1-}) + \int_s^t \mathrm{d}u\, \gamma_1(u-s)(q_{1+}-q_{1-})\right] = 0,$$

$$\ddot{q}_{2+}(s) + \omega_1^2 q_{2+}(s) + \frac{c(s)}{m_2} q_{1+}(s) + \frac{1}{2}\frac{\mathrm{d}}{\mathrm{d}s}\left[\int_0^t \mathrm{d}u\, \gamma_2(s-u)(q_{2+}+q_{2-}) - \int_s^t \mathrm{d}u\, \gamma_2(u-s)(q_{2+}-q_{2-})\right] = 0,$$

$$\ddot{q}_{2-}(s) + \omega_1^2 q_{2-}(s) + \frac{c(s)}{m_2} q_{1-}(s) + \frac{1}{2}\frac{\mathrm{d}}{\mathrm{d}s}\left[\int_0^t \mathrm{d}u\, \gamma_2(s-u)(q_{2+}+q_{2-}) + \int_s^t \mathrm{d}u\, \gamma_2(u-s)(q_{2+}-q_{2-})\right] = 0. \tag{6}$$

To simplify the above expressions and subsequent calculations, introduce the half sum and difference coordinates as

$$Q_\alpha = \frac{1}{2}(q_{\alpha+} + q_{\alpha-}), \qquad q_\alpha = q_{\alpha+} - q_{\alpha-} \tag{7}$$

where the Jacobian of the coordinates transformation is equal to one.

The interest is to find the minimum phonon number in the mechanical mode, which in terms of the second moments is given by

$$\langle \hat{n}\rangle(t) = \frac{1}{2\hbar\omega}\left[\frac{\langle \hat{p}^2\rangle(t)}{m} + m\omega^2\langle \hat{q}^2\rangle(t)\right] - \frac{1}{2}. \tag{8}$$

Following the technical details in Ref. [1], the second moments read

$$\left\langle q_1^2\right\rangle(t) = \int_{-\infty}^\infty \mathrm{d}Q_1''\, Q_1''\, \tilde{\rho}_1(Q_1'', q_1''=0, t), \tag{9}$$

$$\left\langle p_1^2\right\rangle(t) = -\hbar^2 \int_{-\infty}^\infty \mathrm{d}Q_1''\, \frac{\mathrm{d}^2}{\mathrm{d}q_1''^2}\tilde{\rho}_1(Q_1'', q_1'', t)\bigg|_{q_1''=0}, \tag{10}$$

3

with

$$\tilde{\rho}_1(q''_{1+}, q''_{1-}, t) = \langle q''_{1+}|\text{tr}_2\tilde{\rho}(t)|q''_{1-}\rangle = \int_{-\infty}^{\infty} dQ''_2 \tilde{\rho}(q''_{1+}, Q''_2, q''_{1-}, q''_2 = 0, t). \tag{11}$$

## EQUATIONS OF MOTION AND ITS SOLUTION

The differential equations of motion for the coordinates $Q_i$ and $q_i$ take the form

$$\ddot{Q}_{1,2}(s) + \omega_{1,2}^2 Q_{1,2}(s) + \frac{c(s)}{m_{1,2}} Q_{2,1}(s) + \frac{d}{ds} \int_0^s du\, \gamma_{1,2}(s-u) Q_{1,2}(u) = 0,$$

$$\ddot{q}_{1,2}(s) + \omega_{1,2}^2 q_{1,2}(s) + \frac{c(s)}{m_{1,2}} q_{2,1}(s) - \frac{d}{ds} \int_s^t du\, \gamma_{1,2}(u-s) q_{1,2}(u) = 0, \tag{12}$$

and have the following boundary conditions

$$Q_i(s) = \begin{cases} Q'_i, & s = 0 \\ Q''_i, & s = t \end{cases}, \qquad q_i(s) = \begin{cases} q'_i, & s = 0 \\ q''_i, & s = t \end{cases}. \tag{13}$$

The solution to Eqs. (12) can be expresed as

$$Q_1(t,s) = U_1(t,s)Q'_1 + U_2(t,s)Q''_1 + U_3(t,s)Q'_2 + U_4(t,s)Q''_2,$$

$$Q_2(t,s) = V_1(t,s)Q'_2 + V_2(t,s)Q''_2 + V_3(t,s)Q'_1 + V_4(t,s)Q''_1,$$

$$q_1(t,s) = u_1(t,s)q'_1 + u_2(t,s)q''_1 + u_3(t,s)q'_2 + u_4(t,s)q''_2,$$

$$q_2(t,s) = v_1(t,s)q'_2 + v_2(t,s)q''_2 + v_3(t,s)q'_1 + v_4(t,s)q''_1, \tag{14}$$

where the boundary conditions that satisfied the functions $U_i$, $V_i$, $u_i$ and $v_i$ in Eqs. (14) are given by

$$U_1(t,0) = 1, \quad U_1(t,t) = 0, \quad U_2(t,0) = 0, \quad U_2(t,t) = 1, \quad U_i(t,0) = U_i(t,t) = 0, \tag{15}$$

$$V_1(t,0) = 1, \quad V_1(t,t) = 0, \quad V_2(t,0) = 0, \quad V_2(t,t) = 1, \quad V_i(t,0) = V_i(t,t) = 0, \tag{16}$$

$$u_1(t,0) = 1, \quad u_1(t,t) = 0, \quad u_2(t,0) = 0, \quad u_2(t,t) = 1, \quad u_i(t,0) = u_i(t,t) = 0, \tag{17}$$

$$v_1(t,0) = 1, \quad v_1(t,t) = 0, \quad v_2(t,0) = 0, \quad v_2(t,t) = 1, \quad v_i(t,0) = v_i(t,t) = 0, \tag{18}$$

for $i = 3, 4$.

One form to solve the problem is solving in terms of the initial conditions

$$Q_i(0) = q_i(0) = 1, \quad \dot{Q}_i(0) = \dot{q}_i(0) = 1, \quad Q_i(0) = q_i(0) = 0, \quad \text{and} \quad \dot{Q}_i(0) = \dot{q}_i(0) = 0,$$

where $U_i(t,s)$, $V_i(t,s)$, $u_i(t,s)$, $v_i(t,s)$ in terms of the solutions $\varphi_i(s)$, $\phi_i(s)$, $\nu_i(s)$ and $\vartheta_i(s)$ (solution functions of the problem initial condition) are given by

$$U_1(t,s) = \varphi_1(s) - \frac{\varphi_1(t)\varphi_2(s)\phi_2(t)}{\varphi_2(t)\phi_2(t) - \varphi_4(t)\phi_4(t)} - \frac{\varphi_2(t)\phi_3(t)\varphi_4(s)}{\varphi_2(t)\phi_2(t) - \varphi_4(t)\phi_4(t)} + \frac{\varphi_2(s)\phi_3(t)\varphi_4(t)}{\varphi_2(t)\phi_2(t) - \varphi_4(t)\phi_4(t)} + \frac{\varphi_1(t)\varphi_4(s)\phi_4(t)}{\varphi_2(t)\phi_2(t) - \varphi_4(t)\phi_4(t)}, \tag{19}$$

$$U_2(t,s) = \frac{\varphi_2(s)\phi_2(t)}{\varphi_2(t)\phi_2(t) - \varphi_4(t)\phi_4(t)} - \frac{\varphi_4(s)\phi_4(t)}{\varphi_2(t)\phi_2(t) - \varphi_4(t)\phi_4(t)}, \tag{20}$$

$$U_3(t,s) = \varphi_3(s) - \frac{\varphi_2(s)\phi_2(t)\varphi_3(t)}{\varphi_2(t)\phi_2(t) - \varphi_4(t)\phi_4(t)} - \frac{\phi_1(t)\varphi_2(t)\varphi_4(s)}{\varphi_2(t)\phi_2(t) - \varphi_4(t)\phi_4(t)} + \frac{\phi_1(t)\varphi_2(s)\varphi_4(t)}{\varphi_2(t)\phi_2(t) - \varphi_4(t)\phi_4(t)} + \frac{\varphi_3(t)\varphi_4(s)\phi_4(t)}{\varphi_2(t)\phi_2(t) - \varphi_4(t)\phi_4(t)}, \tag{21}$$

$$U_4(t,s) = \frac{\varphi_2(t)\varphi_4(s)}{\varphi_2(t)\phi_2(t) - \varphi_4(t)\phi_4(t)} - \frac{\varphi_2(s)\phi_4(t)}{\varphi_2(t)\phi_2(t) - \varphi_4(t)\phi_4(t)}, \tag{22}$$

$$V_1(t,s) = \phi_1(s) - \frac{\phi_1(t)\varphi_2(t)\phi_2(s)}{\varphi_2(t)\phi_2(t) - \varphi_4(t)\phi_4(t)} - \frac{\phi_2(t)\varphi_3(t)\phi_4(s)}{\varphi_2(t)\phi_2(t) - \varphi_4(t)\phi_4(t)} + \frac{\phi_1(t)\varphi_4(t)\phi_4(s)}{\varphi_2(t)\phi_2(t) - \varphi_4(t)\phi_4(t)} + \frac{\phi_2(s)\varphi_3(t)\phi_4(t)}{\varphi_2(t)\phi_2(t) - \varphi_4(t)\phi_4(t)}, \tag{23}$$

$$V_2(t,s) = \frac{\varphi_2(t)\phi_2(s)}{\varphi_2(t)\phi_2(t) - \varphi_4(t)\phi_4(t)} - \frac{\varphi_4(t)\phi_4(s)}{\varphi_2(t)\phi_2(t) - \varphi_4(t)\phi_4(t)}, \tag{24}$$

$$V_3(t,s) = \phi_3(s) - \frac{\varphi_2(t)\phi_2(s)\varphi_3(t)}{\varphi_2(t)\phi_2(t) - \varphi_4(t)\phi_4(t)} - \frac{\varphi_1(t)\phi_2(t)\phi_4(s)}{\varphi_2(t)\phi_2(t) - \varphi_4(t)\phi_4(t)}$$
$$+ \frac{\phi_3(s)\varphi_4(t)\phi_4(s)}{\varphi_2(t)\phi_2(t) - \varphi_4(t)\phi_4(t)} + \frac{\varphi_1(t)\phi_2(s)\phi_4(t)}{\varphi_2(t)\phi_2(t) - \varphi_4(t)\phi_4(t)}, \tag{25}$$

$$V_4(t,s) = \frac{\varphi_2(t)\phi_2(s)}{\varphi_2(t)\phi_2(t) - \varphi_4(t)\phi_4(t)} - \frac{\varphi_4(t)\phi_4(s)}{\varphi_2(t)\phi_2(t) - \varphi_4(t)\phi_4(t)}, \tag{26}$$

$$u_1(t,s) = \nu_1(s) - \frac{\nu_1(t)\nu_2(s)\vartheta_2(t)}{\nu_2(t)\vartheta_2(t) - \nu_4(t)\vartheta_4(t)} - \frac{\nu_2(t)\vartheta_3(t)\nu_4(s)}{\nu_2(t)\vartheta_2(t) - \nu_4(t)\vartheta_4(t)}$$
$$+ \frac{\nu_2(s)\vartheta_3(t)\nu_4(t)}{\nu_2(t)\vartheta_2(t) - \nu_4(t)\vartheta_4(t)} + \frac{\nu_1(t)\nu_4(s)\vartheta_4(t)}{\nu_2(t)\vartheta_2(t) - \nu_4(t)\vartheta_4(t)}, \tag{27}$$

$$u_2(t,s) = \frac{\nu_2(s)\vartheta_2(t)}{\nu_2(t)\vartheta_2(t) - \nu_4(t)\vartheta_4(t)} - \frac{\nu_4(s)\vartheta_4(t)}{\nu_2(t)\vartheta_2(t) - \nu_4(t)\vartheta_4(t)}, \tag{28}$$

$$u_3(t,s) = \nu_3(s) - \frac{\nu_2(s)\vartheta_2(t)\nu_3(t)}{\nu_2(t)\vartheta_2(t) - \nu_4(t)\vartheta_4(t)} - \frac{\vartheta_1(t)\nu_2(t)\nu_4(s)}{\nu_2(t)\vartheta_2(t) - \nu_4(t)\vartheta_4(t)}$$
$$+ \frac{\vartheta_1(t)\nu_2(s)\nu_4(t)}{\nu_2(t)\vartheta_2(t) - \nu_4(t)\vartheta_4(t)} + \frac{\nu_3(t)\nu_4(s)\vartheta_4(t)}{\nu_2(t)\vartheta_2(t) - \nu_4(t)\vartheta_4(t)}, \tag{29}$$

$$u_4(t,s) = \frac{\nu_2(t)\nu_4(s)}{\nu_2(t)\vartheta_2(t) - \nu_4(t)\vartheta_4(t)} - \frac{\nu_2(s)\vartheta_4(t)}{\nu_2(t)\vartheta_2(t) - \nu_4(t)\vartheta_4(t)}, \tag{30}$$

$$v_1(t,s) = \vartheta_1(s) - \frac{\vartheta_1(t)\nu_2(t)\vartheta_2(s)}{\nu_2(t)\vartheta_2(t) - \nu_4(t)\vartheta_4(t)} - \frac{\vartheta_2(t)\nu_3(t)\vartheta_4(s)}{\nu_2(t)\vartheta_2(t) - \nu_4(t)\vartheta_4(t)}$$
$$+ \frac{\vartheta_1(t)\nu_4(t)\vartheta_4(s)}{\nu_2(t)\vartheta_2(t) - \nu_4(t)\vartheta_4(t)} + \frac{\vartheta_2(s)\nu_3(t)\vartheta_4(t)}{\nu_2(t)\vartheta_2(t) - \nu_4(t)\vartheta_4(t)}, \tag{31}$$

$$v_2(t,s) = \frac{\nu_2(t)\vartheta_2(s)}{\nu_2(t)\vartheta_2(t) - \nu_4(t)\vartheta_4(t)} - \frac{\nu_4(t)\vartheta_4(s)}{\nu_2(t)\vartheta_2(t) - \nu_4(t)\vartheta_4(t)}, \tag{32}$$

$$v_3(t,s) = \vartheta_3(s) - \frac{\nu_2(t)\vartheta_2(s)\nu_3(t)}{\nu_2(t)\vartheta_2(t) - \nu_4(t)\vartheta_4(t)} - \frac{\nu_1(t)\vartheta_2(t)\vartheta_4(s)}{\nu_2(t)\vartheta_2(t) - \nu_4(t)\vartheta_4(t)}$$
$$+ \frac{\vartheta_3(s)\nu_4(t)\vartheta_4(s)}{\nu_2(t)\vartheta_2(t) - \nu_4(t)\vartheta_4(t)} + \frac{\nu_1(t)\vartheta_2(s)\vartheta_4(t)}{\nu_2(t)\vartheta_2(t) - \nu_4(t)\vartheta_4(t)}, \tag{33}$$

$$v_4(t,s) = \frac{\nu_2(t)\vartheta_2(s)}{\nu_2(t)\vartheta_2(t) - \nu_4(t)\vartheta_4(t)} - \frac{\nu_4(t)\vartheta_4(s)}{\nu_2(t)\vartheta_2(t) - \nu_4(t)\vartheta_4(t)}. \tag{34}$$

Therefore, in terms of the solutions of the initial condition problem with $\varphi_i$, $\phi_i$, $\nu_i$ and $\vartheta_i$, the equations of motion are transformed as

$$\ddot{\varphi}_{1,3}(s) + \omega_1^2 \varphi_{1,3}(s) + \frac{c(s)}{m_1}\phi_{3,1}(s) + \frac{\mathrm{d}}{\mathrm{d}s}\int_0^s \mathrm{d}u\,\gamma_1(s-u)\varphi_{1,3}(u) = 0,$$
$$\ddot{\varphi}_{2,4}(s) + \omega_1^2 \varphi_{2,4}(s) + \frac{c(s)}{m_1}\phi_{4,2}(s) + \frac{\mathrm{d}}{\mathrm{d}s}\int_0^s \mathrm{d}u\,\gamma_1(s-u)\varphi_{2,4}(u) = 0,$$
$$\ddot{\phi}_{1,3}(s) + \omega_2^2 \phi_{1,3}(s) + \frac{c(s)}{m_2}\varphi_{3,1}(s) + \frac{\mathrm{d}}{\mathrm{d}s}\int_0^s \mathrm{d}u\,\gamma_2(s-u)\phi_{1,3}(u) = 0,$$
$$\ddot{\phi}_{2,4}(s) + \omega_2^2 \phi_{2,4}(s) + \frac{c(s)}{m_2}\varphi_{4,2}(s) + \frac{\mathrm{d}}{\mathrm{d}s}\int_0^s \mathrm{d}u\,\gamma_2(s-u)\phi_{2,4}(u) = 0,$$

$$\tag{35}$$

$$\ddot{\nu}_{1,3}(s) + \omega_1^2 \nu_{1,3}(s) + \frac{c(s)}{m_1}\vartheta_{3,1}(s) - \frac{\mathrm{d}}{\mathrm{d}s}\int_s^t \mathrm{d}u\,\gamma_1(u-s)\nu_{1,3}(u) = 0,$$
$$\ddot{\nu}_{2,4}(s) + \omega_1^2 \nu_{2,4}(s) + \frac{c(s)}{m_1}\vartheta_{4,2}(s) - \frac{\mathrm{d}}{\mathrm{d}s}\int_s^t \mathrm{d}u\,\gamma_1(u-s)\nu_{2,4}(u) = 0,$$
$$\ddot{\vartheta}_{1,3}(s) + \omega_2^2 \vartheta_{1,3}(s) + \frac{c(s)}{m_2}\nu_{3,1}(s) - \frac{\mathrm{d}}{\mathrm{d}s}\int_s^t \mathrm{d}u\,\gamma_2(u-s)\vartheta_{1,3}(u) = 0,$$
$$\ddot{\vartheta}_{2,4}(s) + \omega_2^2 \vartheta_{2,4}(s) + \frac{c(s)}{m_2}\nu_{4,2}(s) - \frac{\mathrm{d}}{\mathrm{d}s}\int_s^t \mathrm{d}u\,\gamma_2(u-s)\vartheta_{2,4}(u) = 0.$$

$$\tag{36}$$

In this work, the dissipation kernels in Eqs. (35-36) for the mechanical and electromagnetic mode are given by

$$\gamma_1(s) = \gamma_1 \omega_{1D} e^{-\omega_{1D}s}, \qquad \gamma_2(s) = \gamma_2 \omega_{2D} e^{-\omega_{2D}s}. \tag{37}$$

The equations of motion to solve and optimize are obtained after replacing these dissipation

kernels in Eqs. (35-36) and are given by

$$\ddot{\varphi}_{1,3}(s) + \omega_1^2 \varphi_{1,3}(s) + \frac{c(s)}{m_1}\phi_{3,1}(s) + \dot{J}_{1,2}(s) = 0,$$

$$\ddot{\phi}_{1,3}(s) + \omega_2^2 \phi_{1,3}(s) + \frac{c(s)}{m_2}\varphi_{3,1}(s) + \dot{J}_{3,4}(s) = 0,$$

$$\dot{J}_1(s) + \omega_{1D} J_1(s) - \gamma_1 \omega_{1D} \varphi_1(s) = 0,$$

$$\dot{J}_2(s) + \omega_{1D} J_2(s) - \gamma_1 \omega_{1D} \varphi_3(s) = 0,$$   (38)

$$\dot{J}_3(s) + \omega_{2D} J_3(s) - \gamma_2 \omega_{2D} \phi_1(s) = 0,$$

$$\dot{J}_4(s) + \omega_{2D} J_4(s) - \gamma_2 \omega_{2D} \phi_3(s) = 0,$$

$$\ddot{\varphi}_{2,4}(s) + \omega_1^2 \varphi_{2,4}(s) + \frac{c(s)}{m_1}\phi_{4,2}(s) + \dot{J}_{5,6}(s) = 0,$$

$$\ddot{\phi}_{2,4}(s) + \omega_2^2 \phi_{2,4}(s) + \frac{c(s)}{m_2}\varphi_{4,2}(s) + \dot{J}_{7,8}(s) = 0,$$

$$\dot{J}_5(s) + \omega_{1D} J_5(s) - \gamma_1 \omega_{1D} \varphi_2(s) = 0,$$

$$\dot{J}_6(s) + \omega_{1D} J_6(s) - \gamma_1 \omega_{1D} \varphi_4(s) = 0,$$   (39)

$$\dot{J}_7(s) + \omega_{2D} J_7(s) - \gamma_2 \omega_{2D} \phi_2(s) = 0,$$

$$\dot{J}_8(s) + \omega_{2D} J_8(s) - \gamma_2 \omega_{2D} \phi_4(s) = 0,$$

$$\ddot{\nu}_{1,3}(s) + \omega_1^2 \nu_{1,3}(s) + \frac{c(s)}{m_1}\vartheta_{3,1}(s) - \dot{j}_{1,2}(s) = 0,$$

$$\ddot{\vartheta}_{1,3}(s) + \omega_2^2 \vartheta_{1,3}(s) + \frac{c(s)}{m_2}\nu_{3,1}(s) - \dot{j}_{3,4}(s) = 0,$$

$$\dot{j}_1(s) - \omega_{1D} j_1(s) + \gamma_1 \omega_{1D} \nu_1(s) = 0,$$

$$\dot{j}_2(s) - \omega_{1D} j_2(s) + \gamma_1 \omega_{1D} \nu_3(s) = 0,$$   (40)

$$\dot{j}_3(s) - \omega_{2D} j_3(s) + \gamma_2 \omega_{2D} \vartheta_1(s) = 0,$$

$$\dot{j}_4(s) - \omega_{2D} j_4(s) + \gamma_2 \omega_{2D} \vartheta_3(s) = 0,$$

$$\ddot{\nu}_{2,4}(s) + \omega_1^2 \nu_{2,4}(s) + \frac{c(s)}{m_1} \vartheta_{4,2}(s) - \dot{j}_{5,6}(s) = 0,$$

$$\ddot{\vartheta}_{2,4}(s) + \omega_2^2 \vartheta_{2,4}(s) + \frac{c(s)}{m_2} \nu_{4,2}(s) - \dot{j}_{7,8}(s) = 0,$$

$$\dot{j}_5(s) - \omega_{1D} j_5(s) + \gamma_1 \omega_{1D} \nu_2(s) = 0,$$

$$\dot{j}_6(s) - \omega_{1D} j_6(s) + \gamma_1 \omega_{1D} \nu_4(s) = 0,$$

$$\dot{j}_7(s) - \omega_{2D} j_7(s) + \gamma_2 \omega_{2D} \vartheta_2(s) = 0,$$

$$\dot{j}_8(s) - \omega_{2D} j_8(s) + \gamma_2 \omega_{2D} \vartheta_4(s) = 0,$$

$$(41)$$

with

$$J_{1,2}(s) = \int_s^t \mathrm{d}u \, \gamma_1 \omega_{1D} e^{-\omega_{1D}(u-s)} \varphi_{1,3}(u),$$

$$J_{3,4}(s) = \int_s^t \mathrm{d}u \, \gamma_2 \omega_{2D} e^{-\omega_{2D}(u-s)} \phi_{1,3}(u),$$

$$J_{5,6}(s) = \int_s^t \mathrm{d}u \, \gamma_1 \omega_{1D} e^{-\omega_{1D}(u-s)} \varphi_{2,4}(u),$$

$$J_{7,8}(s) = \int_s^t \mathrm{d}u \, \gamma_2 \omega_{2D} e^{-\omega_{2D}(u-s)} \phi_{2,4}(u),$$

$$j_{1,2}(s) = \int_s^t \mathrm{d}u \, \gamma_1 \omega_{1D} e^{-\omega_{1D}(u-s)} \nu_{1,3}(u),$$

$$j_{3,4}(s) = \int_s^t \mathrm{d}u \, \gamma_2 \omega_{2D} e^{-\omega_{2D}(u-s)} \vartheta_{1,3}(u),$$

$$j_{5,6}(s) = \int_s^t \mathrm{d}u \, \gamma_1 \omega_{1D} e^{-\omega_{1D}(u-s)} \nu_{2,4}(u),$$

$$j_{7,8}(s) = \int_s^t \mathrm{d}u \, \gamma_2 \omega_{2D} e^{-\omega_{2D}(u-s)} \vartheta_{2,4}(u),$$

$$(42)$$

where the Eqs. (38-41) are solved by a Runge-Kutta algorithm. A similar process can be followed for arbitrary spectral densities, where the Eqs. (35-36) have to be solve by a numerical method for integro-differential equations. In order to give an idea, the final

equations to solve for any spectral density are

$$\ddot{\varphi}_{1,3}(s) + [\omega_1^2 + \gamma_1(0)]\varphi_{1,3}(s) + \frac{c(s)}{m_1}\phi_{3,1}(s) + \int_0^s du\, \gamma_1'(s-u)\varphi_{1,3}(u) = 0,$$

$$\ddot{\varphi}_{2,4}(s) + [\omega_1^2 + \gamma_1(0)]\varphi_{2,4}(s) + \frac{c(s)}{m_1}\phi_{4,2}(s) + \int_0^s du\, \gamma_1'(s-u)\varphi_{2,4}(u) = 0,$$

$$\ddot{\phi}_{1,3}(s) + [\omega_2^2 + \gamma_2(0)]\phi_{1,3}(s) + \frac{c(s)}{m_2}\varphi_{3,1}(s) + \int_0^s du\, \gamma_2'(s-u)\phi_{1,3}(u) = 0,$$

$$\ddot{\phi}_{2,4}(s) + [\omega_2^2 + \gamma_2(0)]\phi_{2,4}(s) + \frac{c(s)}{m_2}\varphi_{4,2}(s) + \int_0^s du\, \gamma_2'(s-u)\phi_{2,4}(u) = 0,$$

$$(43)$$

$$\ddot{\nu}_{1,3}(s) + [\omega_1^2 + \gamma_1(0)]\nu_{1,3}(s) + \frac{c(s)}{m_1}\vartheta_{3,1}(s) + \int_s^t du\, \gamma_1'(u-s)\nu_{1,3}(u) = 0,$$

$$\ddot{\nu}_{2,4}(s) + [\omega_1^2 + \gamma_1(0)]\nu_{2,4}(s) + \frac{c(s)}{m_1}\vartheta_{4,2}(s) + \int_s^t du\, \gamma_1'(u-s)\nu_{2,4}(u) = 0,$$

$$\ddot{\vartheta}_{1,3}(s) + [\omega_2^2 + \gamma_2(0)]\vartheta_{1,3}(s) + \frac{c(s)}{m_2}\nu_{3,1}(s) + \int_s^t du\, \gamma_2'(u-s)\vartheta_{1,3}(u) = 0,$$

$$\ddot{\vartheta}_{2,4}(s) + [\omega_2^2 + \gamma_2(0)]\vartheta_{2,4}(s) + \frac{c(s)}{m_2}\nu_{4,2}(s) + \int_s^t du\, \gamma_2'(u-s)\vartheta_{2,4}(u) = 0.$$

$$(44)$$

## SQUEEZING GENERATION BY NON-MARKOVIAN DYNAMICS

In general, for non-Markovian dynamics the system does not always thermalize and part of its quantum coherence could be preserved in the steady state'. In our Ref. [15] arXiv:1401.1418, we showed that even at thermal equilibrium, non-Markovian dynamics allow for (i) generating squeezing in a single harmonic mode that induces deviations from the canonical thermal state and (ii) the presence of entanglement between two harmonic modes. For measuring cooling based on the mean phonon number this may be problematic because the final state may not be thermal. However, the presence of quantum correlations assisted by non-Markovian dynamics is possible at equilibrium only when $k_B T/\hbar\omega_m \ll 1$.

For the case at hand, interest is in the the mechanical mode that is assumed to be in a thermal state at $t = 0$ with $k_B T/\hbar\omega_m \gg 1$; therefore, there is no non-Markovianally-induced squeezing initially. During the subsequent ultrafast dynamics, of the order of the mechanical period oscillation, the time-dependent character of the coupling may squeeze the normal modes of the optomechanical systems and, certainly, take the system into a non-thermal state. Note that this time-depedent-coupling-induced squeezing is also present in the Markovian case considered in Ref. [8] Phys. Rev. Lett. 107, 177204 (2011) and in one of the original

proposal of sideband cooling [see Refs., 7 and 22 in the manuscript].

A possible way to characterize the deviations from a thermal state is the $g^2(t)$ correlation function; however, due to the Gaussian character of the state of the mechanical and optical states, and for the present purposes, an equivalent calculation to the $g^{(2)}$ function is the direct calculation of the squeezing parameter $r(t)$ defined as

$$r = \frac{1}{2} \log \left( \frac{\langle \Delta q \rangle^2}{\langle \Delta p \rangle^2} \right), \tag{45}$$

being $\langle \Delta q \rangle^2$ and $\langle \Delta p \rangle^2$ the dispersion of the position and momentum, respectively.

For the parameters used to generate the mean phonon dynamics depicted by the black curve in Fig. 1 of the manuscript, Fig. 1 depicts the time dependence of the squeezing parameter of the mechanical mode state $r(t)$. Because the initial state is thermal at high temperature and no initial system-bath correlations are considered, we have that $r(0) = 0$. The subsequent time-modulation of the coupling induces squeezing that goes very close to zero at the end of the cooling protocol. Specifically, $r_{\mathrm{NM}}(t_{\mathrm{cool}}) = 7.193 \times 10^{-2}$ and $r_{\mathrm{M}}(t_{\mathrm{cool}}) = 3.572 \times 10^{-2}$.

Therefore, non-Markovian dynamics generate more squeezing than the Markovian one; however, due to the parameter regime, the excess of squeezing is very small $\Delta r(t_{\mathrm{cool}}) = 3.621 \times 10^{-2}$. Thus, it is safe to consider the mean phonon number as an measure of cooling.

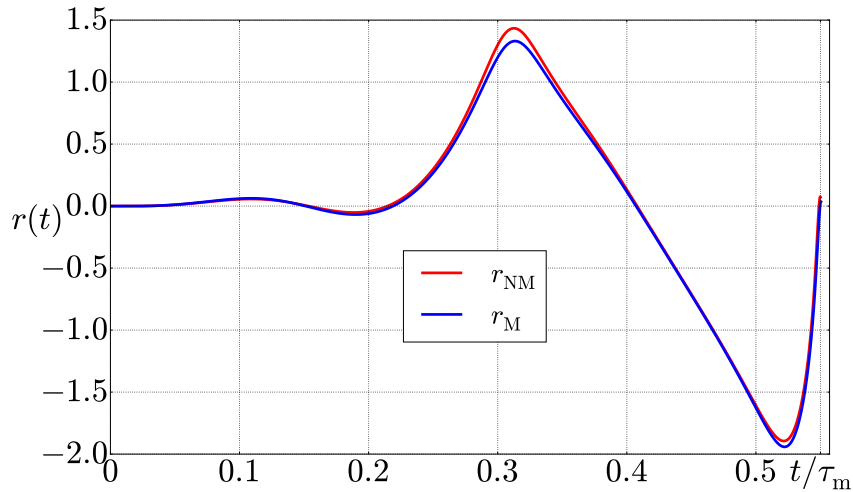

FIG. 1. Squeezing parameter $r(t)$ for the cooling protocol depicted by the black curve in Fig. 1 of the manuscript under Markovian (blue curve) and non-Markovian (red curve) dynamics. The parameters are $\gamma = 10^{-6}\omega_{\mathrm{m}}$, $\kappa = 2.15 \times 10^{-4}\omega_{\mathrm{c}}$, $n_{\mathrm{T}} = 100$ and $n_{\mathrm{cav}} = 0$.

**ENTROPY TRANSFER**

Interest concerns how to reduce the phonon number in the mechanical mode to achieve, as soon as possible, a state near to the ground state. An alternative cooling witness is the entropy at the mechanical mode. If the entropy of the mechanical mode decreases as the optical mode entropy increases, the effective temperature of the mechanical mode decreases, i.e., we are cooling the mechanical mode although we cannot give an absolute value for the temperature. Figure 2 depicts the von Newman entropy [2]

$$S(x) = \left(x + \frac{1}{2}\right)\log\left(x + \frac{1}{2}\right) - \left(x - \frac{1}{2}\right)\log\left(x - \frac{1}{2}\right), \tag{46}$$

$$x = \sqrt{\langle q^2 \rangle \langle p^2 \rangle - \langle pq + qp \rangle^2 / 4}, \tag{47}$$

for the mechanical and the optical mode as a function of time.

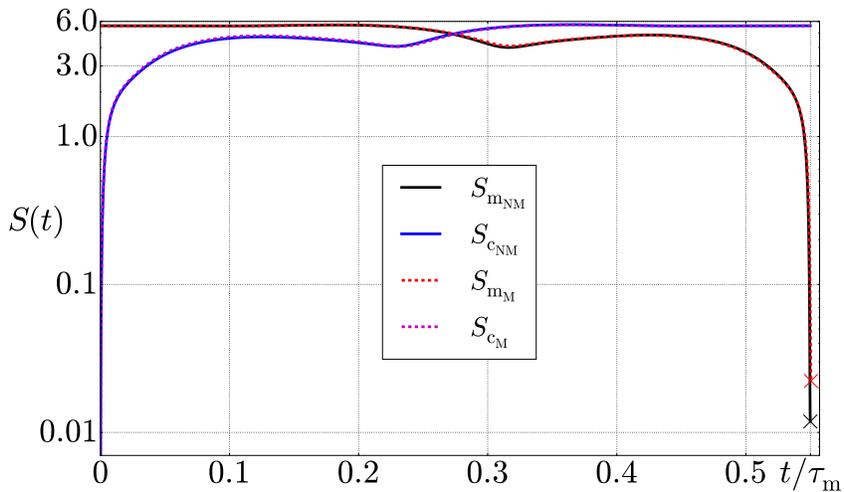

FIG. 2. von Neumann entropy of the both mechanical ($S_{\rm m}$) and optical mode ($S_{\rm c}$). Parameters as in Fig. 1. Continuous lines for non-Markovian dynamics and dashed lines for Markovian dynamics.

Cooling at this level is guaranteed by the fact that the entropy in the mechanical mode decreases; specifically, it decreases from $S_{\rm m}(0) = 5.6101$ to $S_{\rm m}(t_{\rm cool}) = 2.23 \times 10^{-2}$ in the Markovian case and to $S_{\rm m}(t_{\rm cool}) = 2.19 \times 10^{-2}$ in the non-Markovian case. Thus, the systems sits very close to its ground state (zero entropy). For the cavity mode, it varies from $S_{\rm c}(0) = 0$ to $S_{\rm c}(t_{\rm cool}) = 5.601$ in the Markovian case and to $S_{\rm c}(t_{\rm cool}) = 5.607$ in the non-Markovian. Although for this set of parameters the entropy difference is not substantial, from the analysis in the manuscript, it is clear that if the dissipative rate $\gamma$ increases, then the entropy difference

12

does so. Note that non-Markovian dynamics favours entropy flow as discussed in Ref. [5].

## OPTIMIZATION ALGORITHM

To find the minimum phonon number in the mechanical mode, Eqs. (38-41) are optimized to minimize the phonon number in the mechanical mode. In this algorithm the condition to minimize is then

$$J = h\left[\mathbf{x}(t), t\right] + \int_{t_0}^{t_f} V\left[\mathbf{x}(t), \mathbf{g}(t), t\right] \, \mathrm{d}t = \langle \hat{n}\left[\mathbf{x}(t), t\right]\rangle \tag{48}$$

where $h\left[\mathbf{x}(t), t\right]$ and $V\left[\mathbf{x}(t), \mathbf{g}(t), t\right]$ are arbitrary functions, $\mathbf{x}(t)$ and $\mathbf{g}(t)$ are the vector of the equations of motion and the vector of the optimal functions, respectively. In this work, from Eq. (48), $V\left[\mathbf{x}(t), \mathbf{g}(t), t\right] = 0$, $h\left[\mathbf{x}(t), t\right] = \langle \hat{n}\left[\mathbf{x}(t), t\right]\rangle$, $\mathbf{g}(t) = c(t)$ being $c(t)$ the coupling function and

$$H = V\left[\mathbf{x}(t), \mathbf{g}(t), t\right] + \mathbf{p} \cdot \mathbf{x} = \mathbf{p} \cdot \mathbf{x}, \tag{49}$$

where $\mathbf{p}$ is the vector of the costate equations.

However, the condition to minimize, the phonon number in the mechanical mode, is given in terms of the second moments [see Eq. (8)] which in turn are given in terms of the functions $\varphi_i$, $\phi_i$, $\nu_i$ and $\vartheta_i$. This makes quite tedious the calculation of the "initial" conditions for the costate equations (see below) .

The algorithm used to solve the optimization problem was "*the method of steepest descent*" [3] whose algorithm consists of 4 steps:

1. Subdivide the interval $[t_0, t_f]$ into $N$ equal subintervals and assume an initial piecewise-constant control $g^{(0)}(t) = g(0)(t_k)$, $t \in [t_k, t_{k+1}]$ $k = 0, 1, ..., N-1$. In this case, the control function is the coupling between the two resonators $c(s)$ [see Eqs. (38-41)], the mechanical and the electromagnetic mode and the vector of the optimal functions is composed by one function.

2. Apply the assumed control $g^{(i)}$ to integrate the state equations from $t_0$ to $t_f$ with initial conditions $\mathbf{x}(t_0) = \mathbf{x}_0$ and store the state trajectory $\mathbf{x}^{(i)}$. In this case, Eqs. (38-41) are solved and the state trajectory is stored to modify the control function later.

3. Apply $g^{(i)}$ and $\mathbf{x}^{(i)}$ to integrate costate equations backward, i.e., from $[t_f, t_0]$. The "initial value" $\mathbf{p}^{(i)}(t_f)$ is obtained by:

$$\mathbf{p}^{(i)}(t_f) = \frac{\partial h\left[\mathbf{x}^{(i)}(t_f)\right]}{\partial \mathbf{x}} = \frac{\partial \left\langle \hat{n}\left[\mathbf{x}^{(i)}(t_f)\right]\right\rangle}{\partial \mathbf{x}}. \tag{50}$$

Costate equations and their "initial conditions" are given below in the section "Costate equations and their initial conditions in the optimization algorithm (Third Step)".

Evaluate $\partial H^{(i)}(t)/\partial g$, $t \in [t_0, t_f]$ and store this vector.

4. If

$$\left|\left|\frac{\partial H^{(i)}}{\partial g}\right|\right| \leq \epsilon, \tag{51}$$

$$\left|\left|\frac{\partial H^{(i)}}{\partial g}\right|\right|^2 \equiv \int_{t_0}^{t_f} \left[\left|\left|\frac{\partial H^{(i)}}{\partial g}\right|\right|\right]^T \left[\left|\left|\frac{\partial H^{(i)}}{\partial g}\right|\right|\right] \, \mathrm{d}t \tag{52}$$

then stop the iterative procedure. Here $\epsilon$ is a preselected small positive constant used as a tolerance. If Eq. (51) is not satisfied, adjust the piecewise-constant control function by:

$$g^{(i+1)}(t_k) = g^{(i)} - \tau \frac{\partial H^{(i)}}{\partial g}(t_k), \qquad k = 0, 1, ..., N-1. \tag{53}$$

Replace $g^{(i)}$ by $g^{(i+1)}$ and return to step 2. Here, $\tau$ is the step size.

## COSTATE EQUATIONS AND THEIR INITIAL CONDITIONS IN THE OPTIMIZATION ALGORITHM (THIRD STEP)

The costate equations are calculated by

$$\dot{\mathbf{p}} = -\frac{\partial H}{\partial \mathbf{x}}. \tag{54}$$

The "initial" (final) conditions for the costate equations in Eq. (50), in terms of the

seconds moments, are given by

$$p_{\varphi_1}(t_f) = \frac{1}{2}\left(\frac{\partial\langle\hat{p}^2\rangle}{\partial\varphi_1} + \frac{\partial\langle\hat{q}^2\rangle}{\partial\varphi_1}\right), \qquad p_{\varphi_2}(t_f) = \frac{1}{2}\left(\frac{\partial\langle\hat{p}^2\rangle}{\partial\varphi_2} + \frac{\partial\langle\hat{q}^2\rangle}{\partial\varphi_2}\right),$$

$$p_{\varphi_3}(t_f) = \frac{1}{2}\left(\frac{\partial\langle\hat{p}^2\rangle}{\partial\varphi_3} + \frac{\partial\langle\hat{q}^2\rangle}{\partial\varphi_3}\right), \qquad p_{\varphi_4}(t_f) = \frac{1}{2}\left(\frac{\partial\langle\hat{p}^2\rangle}{\partial\varphi_4} + \frac{\partial\langle\hat{q}^2\rangle}{\partial\varphi_4}\right),$$

$$p_{\phi_1}(t_f) = \frac{1}{2}\left(\frac{\partial\langle\hat{p}^2\rangle}{\partial\phi_1} + \frac{\partial\langle\hat{q}^2\rangle}{\partial\phi_1}\right), \qquad p_{\phi_2}(t_f) = \frac{1}{2}\left(\frac{\partial\langle\hat{p}^2\rangle}{\partial\phi_2} + \frac{\partial\langle\hat{q}^2\rangle}{\partial\phi_2}\right),$$

$$p_{\phi_3}(t_f) = \frac{1}{2}\left(\frac{\partial\langle\hat{p}^2\rangle}{\partial\phi_3} + \frac{\partial\langle\hat{q}^2\rangle}{\partial\phi_3}\right), \qquad p_{\phi_4}(t_f) = \frac{1}{2}\left(\frac{\partial\langle\hat{p}^2\rangle}{\partial\phi_4} + \frac{\partial\langle\hat{q}^2\rangle}{\partial\phi_4}\right),$$

$$p_{\varphi_1 p}(t_f) = \frac{1}{2}\left(\frac{\partial\langle\hat{p}^2\rangle}{\partial\varphi_1 p} + \frac{\partial\langle\hat{q}^2\rangle}{\partial\varphi_1 p}\right), \qquad p_{\varphi_2 p}(t_f) = \frac{1}{2}\left(\frac{\partial\langle\hat{p}^2\rangle}{\partial\varphi_2 p} + \frac{\partial\langle\hat{q}^2\rangle}{\partial\varphi_2 p}\right),$$

$$p_{\varphi_3 p}(t_f) = \frac{1}{2}\left(\frac{\partial\langle\hat{p}^2\rangle}{\partial\varphi_3 p} + \frac{\partial\langle\hat{q}^2\rangle}{\partial\varphi_3 p}\right), \qquad p_{\varphi_4 p}(t_f) = \frac{1}{2}\left(\frac{\partial\langle\hat{p}^2\rangle}{\partial\varphi_4 p} + \frac{\partial\langle\hat{q}^2\rangle}{\partial\varphi_4 p}\right),$$

$$p_{\phi_1 p}(t_f) = \frac{1}{2}\left(\frac{\partial\langle\hat{p}^2\rangle}{\partial\phi_1 p} + \frac{\partial\langle\hat{q}^2\rangle}{\partial\phi_1 p}\right), \qquad p_{\phi_2 p}(t_f) = \frac{1}{2}\left(\frac{\partial\langle\hat{p}^2\rangle}{\partial\phi_2 p} + \frac{\partial\langle\hat{q}^2\rangle}{\partial\phi_2 p}\right),$$

$$p_{\phi_3 p}(t_f) = \frac{1}{2}\left(\frac{\partial\langle\hat{p}^2\rangle}{\partial\phi_3 p} + \frac{\partial\langle\hat{q}^2\rangle}{\partial\phi_3 p}\right), \qquad p_{\phi_4 p}(t_f) = \frac{1}{2}\left(\frac{\partial\langle\hat{p}^2\rangle}{\partial\phi_4 p} + \frac{\partial\langle\hat{q}^2\rangle}{\partial\phi_4 p}\right),$$

$$\tag{55}$$

$$p_{\nu_1}(t_f) = \frac{1}{2}\left[\frac{\partial\langle\hat{p}^2\rangle}{\partial\nu_1} + \frac{\partial\langle\hat{q}^2\rangle}{\partial\nu_1} + \sum_{i=1}^{4}\left(\frac{\partial\langle\hat{p}^2\rangle}{\partial\nu_1 i} + \frac{\partial\langle\hat{q}^2\rangle}{\partial\nu_1 i}\right)\nu i_l + \sum_{i=1,i\neq 1}^{4}\left(\frac{\partial\langle\hat{p}^2\rangle}{\partial\nu i 1} + \frac{\partial\langle\hat{q}^2\rangle}{\partial\nu i 1}\right)\nu i_r\right],$$

$$\tag{56}$$

$$p_{\nu_2}(t_f) = \frac{1}{2}\left[\frac{\partial\langle\hat{p}^2\rangle}{\partial\nu_2} + \frac{\partial\langle\hat{q}^2\rangle}{\partial\nu_2} + \sum_{i=1}^{4}\left(\frac{\partial\langle\hat{p}^2\rangle}{\partial\nu_2 i} + \frac{\partial\langle\hat{q}^2\rangle}{\partial\nu_2 i}\right)\nu i_l + \sum_{i=1,i\neq 2}^{4}\left(\frac{\partial\langle\hat{p}^2\rangle}{\partial\nu i 2} + \frac{\partial\langle\hat{q}^2\rangle}{\partial\nu i 2}\right)\nu i_r\right],$$

$$\tag{57}$$

$$p_{\nu_3}(t_f) = \frac{1}{2}\left[\frac{\partial\langle\hat{p}^2\rangle}{\partial\nu_3} + \frac{\partial\langle\hat{q}^2\rangle}{\partial\nu_3} + \sum_{i=1}^{4}\left(\frac{\partial\langle\hat{p}^2\rangle}{\partial\nu_3 i} + \frac{\partial\langle\hat{q}^2\rangle}{\partial\nu_3 i}\right)\nu i_l + \sum_{i=1,i\neq 3}^{4}\left(\frac{\partial\langle\hat{p}^2\rangle}{\partial\nu i 3} + \frac{\partial\langle\hat{q}^2\rangle}{\partial\nu i 3}\right)\nu i_r\right],$$

$$\tag{58}$$

$$p_{\nu_4}(t_f) = \frac{1}{2}\left[\frac{\partial\langle\hat{p}^2\rangle}{\partial\nu_4} + \frac{\partial\langle\hat{q}^2\rangle}{\partial\nu_4} + \sum_{i=1}^{4}\left(\frac{\partial\langle\hat{p}^2\rangle}{\partial\nu_4 i} + \frac{\partial\langle\hat{q}^2\rangle}{\partial\nu_4 i}\right)\nu i_l + \sum_{i=1,i\neq 4}^{4}\left(\frac{\partial\langle\hat{p}^2\rangle}{\partial\nu i 4} + \frac{\partial\langle\hat{q}^2\rangle}{\partial\nu i 4}\right)\nu i_r\right],$$

$$\tag{59}$$

$$p_{\vartheta_1}(t_f) = \frac{1}{2}\left[\frac{\partial\langle\hat{p}^2\rangle}{\partial\vartheta_1} + \frac{\partial\langle\hat{q}^2\rangle}{\partial\vartheta_1} + \sum_{i=1}^{4}\left(\frac{\partial\langle\hat{p}^2\rangle}{\partial\vartheta_1 i} + \frac{\partial\langle\hat{q}^2\rangle}{\partial\vartheta_1 i}\right)\vartheta i_l + \sum_{i=1,i\neq 1}^{4}\left(\frac{\partial\langle\hat{p}^2\rangle}{\partial\vartheta i1} + \frac{\partial\langle\hat{q}^2\rangle}{\partial\vartheta i1}\right)\vartheta i_r\right],$$
(60)

$$p_{\vartheta_2}(t_f) = \frac{1}{2}\left[\frac{\partial\langle\hat{p}^2\rangle}{\partial\vartheta_2} + \frac{\partial\langle\hat{q}^2\rangle}{\partial\vartheta_2} + \sum_{i=1}^{4}\left(\frac{\partial\langle\hat{p}^2\rangle}{\partial\vartheta_2 i} + \frac{\partial\langle\hat{q}^2\rangle}{\partial\vartheta_2 i}\right)\vartheta i_l + \sum_{i=1,i\neq 2}^{4}\left(\frac{\partial\langle\hat{p}^2\rangle}{\partial\vartheta i2} + \frac{\partial\langle\hat{q}^2\rangle}{\partial\vartheta i2}\right)\vartheta i_r\right],$$
(61)

$$p_{\vartheta_3}(t_f) = \frac{1}{2}\left[\frac{\partial\langle\hat{p}^2\rangle}{\partial\vartheta_3} + \frac{\partial\langle\hat{q}^2\rangle}{\partial\vartheta_3} + \sum_{i=1}^{4}\left(\frac{\partial\langle\hat{p}^2\rangle}{\partial\vartheta_3 i} + \frac{\partial\langle\hat{q}^2\rangle}{\partial\vartheta_3 i}\right)\vartheta i_l + \sum_{i=1,i\neq 3}^{4}\left(\frac{\partial\langle\hat{p}^2\rangle}{\partial\vartheta i3} + \frac{\partial\langle\hat{q}^2\rangle}{\partial\vartheta i3}\right)\vartheta i_r\right],$$
(62)

$$p_{\vartheta_4}(t_f) = \frac{1}{2}\left[\frac{\partial\langle\hat{p}^2\rangle}{\partial\vartheta_4} + \frac{\partial\langle\hat{q}^2\rangle}{\partial\vartheta_4} + \sum_{i=1}^{4}\left(\frac{\partial\langle\hat{p}^2\rangle}{\partial\vartheta_4 i} + \frac{\partial\langle\hat{q}^2\rangle}{\partial\vartheta_4 i}\right)\vartheta i_l + \sum_{i=1,i\neq 4}^{4}\left(\frac{\partial\langle\hat{p}^2\rangle}{\partial\vartheta i4} + \frac{\partial\langle\hat{q}^2\rangle}{\partial\vartheta i4}\right)\vartheta i_r\right],$$
(63)

with $\nu i_{l,r}$ and $\vartheta i_{l,r}$ given by

$$\begin{aligned}
\nu i_r &= \int_0^t \mathrm{d}s \int_0^s \mathrm{d}u\, K_1(u-s)u_i(s),\\
\vartheta i_r &= \int_0^t \mathrm{d}s \int_0^s \mathrm{d}u\, K_2(u-s)v_i(s),\\
\nu i_l &= \int_0^t \mathrm{d}s \int_0^s \mathrm{d}u\, K_1(u-s)u_i(u),\\
\vartheta i_l &= \int_0^t \mathrm{d}s \int_0^s \mathrm{d}u\, K_2(u-s)v_i(u),
\end{aligned}$$
(64)

where $K_i(s)$ is the noise kernel generated by the dissipation kernels in Eqs. (37) [1].

**Variances**

The initial conditions for the thermal states of the mechanical (1) and optical (2) modes

16

are

$$\langle q_1 \rangle(0) = 0, \qquad \langle q_2 \rangle(0) = 0,$$

$$\langle p_1 \rangle(0) = 0, \qquad \langle p_2 \rangle(0) = 0,$$

$$\langle q_1 q_1 \rangle(0) = \frac{1}{2m_1\omega_1}\coth\left(\frac{1}{2}\beta_1\omega_1\right), \qquad \langle q_2 q_2 \rangle(0) = \frac{1}{2m_2\omega_2}\coth\left(\frac{1}{2}\beta_2\omega_2\right), \quad (65)$$

$$\langle p_1 p_1 \rangle(0) = \frac{m_1\omega_1}{2}\coth\left(\frac{1}{2}\beta_1\omega_1\right), \qquad \langle p_2 p_2 \rangle(0) = \frac{m_2\omega_2}{2}\coth\left(\frac{1}{2}\beta_2\omega_2\right),$$

$$\langle q_1 p_1 \rangle(0) = 0, \qquad \langle q_2 p_2 \rangle(0) = 0,$$

where $\beta_i$ are the inverse temperatures of the baths, $\omega_i$ are the frequencies of the mechanical and electromagnetic modes respectively, and $m_i$ the masses of the resonators.

The second moments of the mechanical resonator $\langle q_1 q_1 \rangle(t) = \texttt{sq1q1[t]}$ and $\langle p_1 p_1 \rangle(t) = \texttt{sp1p1[t]}$ in terms of the functions $\varphi_i = \texttt{Ui}$, $\phi_i = \texttt{Vi}$, $\nu_i = \texttt{ui}$ and $\vartheta_i = \texttt{vi}$ are given by

sq1q1[t_]:=(sp2p2i (U2p[0] U4[t]-U2[t] U4p[0])^2)/(4 (2 U2[t] V2[t]-2 U4[t]
V4[t])^2 (((-U2p[0] V2[t]+U4p[0] V4[t]) (-m U2[t] V2p[0]+m U4[t] V4p[0]))/(2
U2[t] V2[t]-2 U4[t] V4[t])^2-(m (U2p[0] U4[t]-U2[t] U4p[0]) (V2p[0] V4[t]-V2[t]
V4p[0]))/(2 U2[t] V2[t]-2 U4[t] V4[t])^2)^2)+(vep2i^2 (U2p[0] U4[t]-U2[t]
U4p[0])^2)/(4 (2 U2[t] V2[t]-2 U4[t] V4[t])^2 (((-U2p[0] V2[t]+U4p[0] V4[t])
(-m U2[t] V2p[0]+m U4[t] V4p[0]))/(2 U2[t] V2[t]-2 U4[t] V4[t])^2-(m (U2p[0]
U4[t]-U2[t] U4p[0]) (V2p[0] V4[t]-V2[t] V4p[0]))/(2 U2[t] V2[t]-2 U4[t] V4[t])^2)^2)
+ ((U2p[0] U4[t]-U2[t] U4p[0])^2 (u22[t] (u4[t] v1[t]-u3[t] v2[t])^2+u2[t]^2
(u44[t] v1[t]^2-v2[t] (-(u33[t]+v11[t]) v2[t]+v1[t] (u34[t]+u43[t]+v12[t]+v21[t]))
+ v1[t]^2 v22[t])+u24[t] u3[t] u4[t] v1[t] v4[t]-u23[t] u4[t]^2 v1[t] v4[t]-u32[t]
u4[t]^2 v1[t] v4[t]+u3[t] u4[t] u42[t] v1[t] v4[t]-u4[t]^2 v1[t] v14[t] v4[t]-u24[t]
u3[t]^2 v2[t] v4[t]+u23[t] u3[t] u4[t] v2[t] v4[t]+u3[t] u32[t] u4[t] v2[t]
v4[t]-u3[t]^2 u42[t] v2[t] v4[t]+u3[t] u4[t] v14[t] v2[t] v4[t]+u3[t] u4[t]
v1[t] v24[t] v4[t]-u3[t]^2 v2[t] v24[t] v4[t]-u3[t] u34[t] u4[t] v4[t]^2+u33[t]
u4[t]^2 v4[t]^2-u3[t] u4[t] u43[t] v4[t]^2+u3[t]^2 u44[t] v4[t]^2+u4[t]^2
v11[t] v4[t]^2-u3[t] u4[t] v12[t] v4[t]^2-u3[t] u4[t] v21[t] v4[t]^2+u3[t]^2
v22[t] v4[t]^2-u4[t]^2 v1[t] v4[t] v41[t]+u3[t] u4[t] v2[t] v4[t] v41[t]+u3[t]
u4[t] v1[t] v4[t] v42[t]-u3[t]^2 v2[t] v4[t] v42[t]+u2[t] (u24[t] v1[t] (-u4[t]
v1[t]+u3[t] v2[t])+u4[t] (-u42[t] v1[t]^2+u23[t] v1[t] v2[t]+u32[t] v1[t]

17

v2[t]+v1[t] v14[t] v2[t]-v1[t]^2 v24[t]+u34[t] v1[t] v4[t]+u43[t] v1[t] v4[t]+v1[t]
v12[t] v4[t]-2 u33[t] v2[t] v4[t]-2 v11[t] v2[t] v4[t]+v1[t] v21[t] v4[t]+v1[t]
v2[t] v41[t]-v1[t]^2 v42[t])+u3[t] (u42[t] v1[t] v2[t]-u23[t] v2[t]^2-u32[t]
v2[t]^2-v14[t] v2[t]^2+v1[t] v2[t] v24[t]-2 u44[t] v1[t] v4[t]+u34[t] v2[t]
v4[t]+u43[t] v2[t] v4[t]+v12[t] v2[t] v4[t]+v2[t] v21[t] v4[t]-2 v1[t] v22[t]
v4[t]-v2[t]^2 v41[t]+v1[t] v2[t] v42[t]))+(u4[t] v1[t]-u3[t] v2[t])^2 v44[t]))/(2
(u2[t] v2[t]-u4[t] v4[t])^2 (2 U2[t] V2[t]-2 U4[t] V4[t])^2 (((-U2p[0] V2[t]+U4p[0]
V4[t]) (-m U2[t] V2p[0]+m U4[t] V4p[0]))/(2 U2[t] V2[t]-2 U4[t] V4[t])^2-(m
(U2p[0] U4[t]-U2[t] U4p[0]) (V2p[0] V4[t]-V2[t] V4p[0]))/(2 U2[t] V2[t]-2
U4[t] V4[t])^2)^2) - vep1i vep2i (U2p[0] U4[t]-U2[t] U4p[0]) (-m U2[t] V2p[0]+m
U4[t] V4p[0]))/(2 (2 U2[t] V2[t]-2 U4[t] V4[t])^2 (((-U2p[0] V2[t]+U4p[0]
V4[t]) (-m U2[t] V2p[0]+m U4[t] V4p[0]))/(2 U2[t] V2[t]-2 U4[t] V4[t])^2-(m
(U2p[0] U4[t]-U2[t] U4p[0]) (V2p[0] V4[t]-V2[t] V4p[0]))/(2 U2[t] V2[t]-2
U4[t] V4[t])^2)^2)-(vep2i veq1i (U2p[0] U4[t]-U2[t] U4p[0]) (U2[t] (-U1p[0]
V2[t]+U4p[0] V3[t])+U4[t] (-U2p[0] V3[t]+U1p[0] V4[t])+U1[t] (U2p[0] V2[t]-U4p[0]
V4[t])) (-m U2[t] V2p[0]+m U4[t] V4p[0]))/((2 U2[t] V2[t]-2 U4[t] V4[t])^3
(((-U2p[0] V2[t]+U4p[0] V4[t]) (-m U2[t] V2p[0]+m U4[t] V4p[0]))/(2 U2[t]
V2[t]-2 U4[t] V4[t])^2-(m (U2p[0] U4[t]-U2[t] U4p[0]) (V2p[0] V4[t]-V2[t]
V4p[0]))/(2 U2[t] V2[t]-2 U4[t] V4[t])^2)^2)-(sq2p2i (U2p[0] U4[t]-U2[t]
U4p[0]) (U2[t] (U4p[0] V1[t]-U3p[0] V2[t])+U4[t] (-U2p[0] V1[t]+U3p[0] V4[t])+U3[t]
(U2p[0] V2[t]-U4p[0] V4[t])) (-m U2[t] V2p[0]+m U4[t] V4p[0]))/((2 U2[t]
V2[t]-2 U4[t] V4[t])^3 (((-U2p[0] V2[t]+U4p[0] V4[t]) (-m U2[t] V2p[0]+m
U4[t] V4p[0]))/(2 U2[t] V2[t]-2 U4[t] V4[t])^2-(m (U2p[0] U4[t]-U2[t] U4p[0])
(V2p[0] V4[t]-V2[t] V4p[0]))/(2 U2[t] V2[t]-2 U4[t] V4[t])^2)^2)-(vep2i veq2i
(U2p[0] U4[t]-U2[t] U4p[0]) (U2[t] (U4p[0] V1[t]-U3p[0] V2[t])+U4[t] (-U2p[0]
V1[t]+U3p[0] V4[t])+U3[t] (U2p[0] V2[t]-U4p[0] V4[t])) (-m U2[t] V2p[0]+m
U4[t] V4p[0]))/((2 U2[t] V2[t]-2 U4[t] V4[t])^3 (((-U2p[0] V2[t]+U4p[0] V4[t])
(-m U2[t] V2p[0]+m U4[t] V4p[0]))/(2 U2[t] V2[t]-2 U4[t] V4[t])^2-(m (U2p[0]
U4[t]-U2[t] U4p[0]) (V2p[0] V4[t]-V2[t] V4p[0]))/(2 U2[t] V2[t]-2 U4[t] V4[t])^2)^2)
- ((U2p[0] U4[t]-U2[t] U4p[0]) (u12[t] u2[t] u4[t] v1[t] v2[t]+u2[t] u21[t]
u4[t] v1[t] v2[t]-u2[t]^2 u41[t] v1[t] v2[t]+u13[t] u2[t]^2 v2[t]^2-u12[t]
u2[t] u3[t] v2[t]^2-u2[t] u21[t] u3[t] v2[t]^2+u2[t]^2 u31[t] v2[t]^2+u2[t]^2

v13[t] v2[t]^2-u2[t]^2 v1[t] v2[t] v23[t]-2 u2[t] u24[t] u4[t] v1[t] v3[t]+2
u22[t] u4[t]^2 v1[t] v3[t]-2 u2[t] u4[t] u42[t] v1[t] v3[t]+2 u2[t]^2 u44[t]
v1[t] v3[t]+u2[t] u24[t] u3[t] v2[t] v3[t]-u2[t]^2 u34[t] v2[t] v3[t]+u2[t]
u23[t] u4[t] v2[t] v3[t]-2 u22[t] u3[t] u4[t] v2[t] v3[t]+u2[t] u32[t] u4[t]
v2[t] v3[t]+u2[t] u3[t] u42[t] v2[t] v3[t]-u2[t]^2 u43[t] v2[t] v3[t]-u2[t]^2
v12[t] v2[t] v3[t]+u2[t] u4[t] v14[t] v2[t] v3[t]-u2[t]^2 v2[t] v21[t] v3[t]+2
u2[t]^2 v1[t] v22[t] v3[t]-2 u2[t] u4[t] v1[t] v24[t] v3[t]+u2[t] u3[t] v2[t]
v24[t] v3[t]+u2[t]^2 v2[t]^2 v31[t]-u2[t]^2 v1[t] v2[t] v32[t]+u2[t] u4[t]
v1[t] v2[t] v34[t]-u2[t] u3[t] v2[t]^2 v34[t]-u12[t] u4[t]^2 v1[t] v4[t]-u21[t]
u4[t]^2 v1[t] v4[t]+u2[t] u4[t] u41[t] v1[t] v4[t]-2 u13[t] u2[t] u4[t] v2[t]
v4[t]+u12[t] u3[t] u4[t] v2[t] v4[t]+u21[t] u3[t] u4[t] v2[t] v4[t]-2 u2[t]
u31[t] u4[t] v2[t] v4[t]+u2[t] u3[t] u41[t] v2[t] v4[t]-2 u2[t] u4[t] v13[t]
v2[t] v4[t]+u2[t] u4[t] v1[t] v23[t] v4[t]+u2[t] u3[t] v2[t] v23[t] v4[t]+u24[t]
u3[t] u4[t] v3[t] v4[t]+u2[t] u34[t] u4[t] v3[t] v4[t]-u23[t] u4[t]^2 v3[t]
v4[t]-u32[t] u4[t]^2 v3[t] v4[t]+u3[t] u4[t] u42[t] v3[t] v4[t]+u2[t] u4[t]
u43[t] v3[t] v4[t]-2 u2[t] u3[t] u44[t] v3[t] v4[t]+u2[t] u4[t] v12[t] v3[t]
v4[t]-u4[t]^2 v14[t] v3[t] v4[t]+u2[t] u4[t] v21[t] v3[t] v4[t]-2 u2[t] u3[t]
v22[t] v3[t] v4[t]+u3[t] u4[t] v24[t] v3[t] v4[t]-2 u2[t] u4[t] v2[t] v31[t]
v4[t]+u2[t] u4[t] v1[t] v32[t] v4[t]+u2[t] u3[t] v2[t] v32[t] v4[t]-u4[t]^2
v1[t] v34[t] v4[t]+u3[t] u4[t] v2[t] v34[t] v4[t]+u13[t] u4[t]^2 v4[t]^2+u31[t]
u4[t]^2 v4[t]^2-u3[t] u4[t] u41[t] v4[t]^2+u4[t]^2 v13[t] v4[t]^2-u3[t] u4[t]
v23[t] v4[t]^2+u4[t]^2 v31[t] v4[t]^2-u3[t] u4[t] v32[t] v4[t]^2-u14[t] (u2[t]
v1[t]-u3[t] v4[t]) (u2[t] v2[t]-u4[t] v4[t])+u2[t] u4[t] v2[t] v3[t] v41[t]-u4[t]^2
v3[t] v4[t] v41[t]-2 u2[t] u4[t] v1[t] v3[t] v42[t]+u2[t] u3[t] v2[t] v3[t]
v42[t]+u3[t] u4[t] v3[t] v4[t] v42[t]+u2[t] u4[t] v1[t] v2[t] v43[t]-u2[t]
u3[t] v2[t]^2 v43[t]-u4[t]^2 v1[t] v4[t] v43[t]+u3[t] u4[t] v2[t] v4[t] v43[t]+2
u4[t] (u4[t] v1[t]-u3[t] v2[t]) v3[t] v44[t]+u1[t] (2 u22[t] v2[t] (-u4[t]
v1[t]+u3[t] v2[t])+u2[t] (u24[t] v1[t] v2[t]+u42[t] v1[t] v2[t]-u23[t] v2[t]^2-u32[t]
v2[t]^2-v14[t] v2[t]^2+v1[t] v2[t] v24[t]-2 u44[t] v1[t] v4[t]+u34[t] v2[t]
v4[t]+u43[t] v2[t] v4[t]+v12[t] v2[t] v4[t]+v2[t] v21[t] v4[t]-2 v1[t] v22[t]
v4[t]-v2[t]^2 v41[t]+v1[t] v2[t] v42[t])+v4[t] (u24[t] (u4[t] v1[t]-2 u3[t]
v2[t])+u4[t] (-(u34[t]+u43[t] +v12[t]+v21[t]) v4[t]+v2[t] (u23[t]+u32[t] +

v14[t] + v41[t]) + v1[t] (u42[t] + v24[t] + v42[t]))-2 u3[t] (-(u44[t]+v22[t])
v4[t]+v2[t] (u42[t] + v24[t] + v42[t])))+2 v2[t] (-u4[t] v1[t]+u3[t] v2[t])
v44[t])) (-m U2[t] V2p[0]+m U4[t] V4p[0]))/(2 (u2[t] v2[t]-u4[t] v4[t])^2
(2 U2[t] V2[t]-2 U4[t] V4[t])^2 (((-U2p[0] V2[t]+U4p[0] V4[t]) (-m U2[t]
V2p[0]+m U4[t] V4p[0]))/(2 U2[t] V2[t]-2 U4[t] V4[t])^2-(m (U2p[0] U4[t]-U2[t]
U4p[0]) (V2p[0] V4[t]-V2[t] V4p[0]))/(2 U2[t] V2[t]-2 U4[t] V4[t])^2)^2)+(sp1p1i
(-m U2[t] V2p[0]+m U4[t] V4p[0])^2)/(4 (2 U2[t] V2[t]-2 U4[t] V4[t])^2 (((-U2p[0]
V2[t]+U4p[0] V4[t]) (-m U2[t] V2p[0]+m U4[t] V4p[0]))/(2 U2[t] V2[t]-2 U4[t]
V4[t])^2-(m (U2p[0] U4[t]-U2[t] U4p[0]) (V2p[0] V4[t]-V2[t] V4p[0]))/(2 U2[t]
V2[t]-2 U4[t] V4[t])^2)^2)+(vep1i^2 (-m U2[t] V2p[0]+m U4[t] V4p[0])^2)/(4
(2 U2[t] V2[t]-2 U4[t] V4[t])^2 (((-U2p[0] V2[t]+U4p[0] V4[t]) (-m U2[t]
V2p[0]+m U4[t] V4p[0]))/(2 U2[t] V2[t]-2 U4[t] V4[t])^2-(m (U2p[0] U4[t]-U2[t]
U4p[0]) (V2p[0] V4[t]-V2[t] V4p[0]))/(2 U2[t] V2[t]-2 U4[t] V4[t])^2)^2)+(sq1p1i
(U2[t] (-U1p[0] V2[t]+U4p[0] V3[t])+U4[t] (-U2p[0] V3[t]+U1p[0] V4[t])+U1[t]
(U2p[0] V2[t]-U4p[0] V4[t])) (-m U2[t] V2p[0]+m U4[t] V4p[0])^2)/((2 U2[t]
V2[t]-2 U4[t] V4[t])^3 (((-U2p[0] V2[t]+U4p[0] V4[t]) (-m U2[t] V2p[0]+m
U4[t] V4p[0]))/(2 U2[t] V2[t]-2 U4[t] V4[t])^2-(m (U2p[0] U4[t]-U2[t] U4p[0])
(V2p[0] V4[t]-V2[t] V4p[0]))/(2 U2[t] V2[t]-2 U4[t] V4[t])^2)^2)+(vep1i veq1i
(U2[t] (-U1p[0] V2[t]+U4p[0] V3[t])+U4[t] (-U2p[0] V3[t]+U1p[0] V4[t])+U1[t]
(U2p[0] V2[t]-U4p[0] V4[t])) (-m U2[t] V2p[0]+m U4[t] V4p[0])^2)/((2 U2[t]
V2[t]-2 U4[t] V4[t])^3 (((-U2p[0] V2[t]+U4p[0] V4[t]) (-m U2[t] V2p[0]+m
U4[t] V4p[0]))/(2 U2[t] V2[t]-2 U4[t] V4[t])^2-(m (U2p[0] U4[t]-U2[t] U4p[0])
(V2p[0] V4[t]-V2[t] V4p[0]))/(2 U2[t] V2[t]-2 U4[t] V4[t])^2)^2)+(sq1q1i
(U2[t] (-U1p[0] V2[t]+U4p[0] V3[t])+U4[t] (-U2p[0] V3[t]+U1p[0] V4[t])+U1[t]
(U2p[0] V2[t]-U4p[0] V4[t]))^2 (-m U2[t] V2p[0]+m U4[t] V4p[0])^2)/((2 U2[t]
V2[t]-2 U4[t] V4[t])^4 (((-U2p[0] V2[t]+U4p[0] V4[t]) (-m U2[t] V2p[0]+m
U4[t] V4p[0]))/(2 U2[t] V2[t]-2 U4[t] V4[t])^2-(m (U2p[0] U4[t]-U2[t] U4p[0])
(V2p[0] V4[t]-V2[t] V4p[0]))/(2 U2[t] V2[t]-2 U4[t] V4[t])^2)^2)+(veq1i^2
(U2[t] (-U1p[0] V2[t]+U4p[0] V3[t])+U4[t] (-U2p[0] V3[t]+U1p[0] V4[t])+U1[t]
(U2p[0] V2[t]-U4p[0] V4[t]))^2 (-m U2[t] V2p[0]+m U4[t] V4p[0])^2)/((2 U2[t]
V2[t]-2 U4[t] V4[t])^4 (((-U2p[0] V2[t]+U4p[0] V4[t]) (-m U2[t] V2p[0]+m
U4[t] V4p[0]))/(2 U2[t] V2[t]-2 U4[t] V4[t])^2-(m (U2p[0] U4[t]-U2[t] U4p[0])

(V2p[0] V4[t]-V2[t] V4p[0]))/(2 U2[t] V2[t]-2 U4[t] V4[t])^2)^2)+(vep1i veq2i
(U2[t] (U4p[0] V1[t]-U3p[0] V2[t])+U4[t] (-U2p[0] V1[t]+U3p[0] V4[t])+U3[t]
(U2p[0] V2[t]-U4p[0] V4[t])) (-m U2[t] V2p[0]+m U4[t] V4p[0])^2)/((2 U2[t]
V2[t]-2 U4[t] V4[t])^3 (((-U2p[0] V2[t]+U4p[0] V4[t]) (-m U2[t] V2p[0]+m
U4[t] V4p[0]))/(2 U2[t] V2[t]-2 U4[t] V4[t])^2-(m (U2p[0] U4[t]-U2[t] U4p[0])
(V2p[0] V4[t]-V2[t] V4p[0]))/(2 U2[t] V2[t]-2 U4[t] V4[t])^2)^2)+(2 veq1i
veq2i (U2[t] (-U1p[0] V2[t]+U4p[0] V3[t])+U4[t] (-U2p[0] V3[t]+U1p[0] V4[t])+U1[t]
(U2p[0] V2[t]-U4p[0] V4[t])) (U2[t] (U4p[0] V1[t]-U3p[0] V2[t])+U4[t] (-U2p[0]
V1[t]+U3p[0] V4[t])+U3[t] (U2p[0] V2[t]-U4p[0] V4[t])) (-m U2[t] V2p[0]+m
U4[t] V4p[0])^2)/((2 U2[t] V2[t]-2 U4[t] V4[t])^4 (((-U2p[0] V2[t]+U4p[0]
V4[t]) (-m U2[t] V2p[0]+m U4[t] V4p[0]))/(2 U2[t] V2[t]-2 U4[t] V4[t])^2-(m
(U2p[0] U4[t]-U2[t] U4p[0]) (V2p[0] V4[t]-V2[t] V4p[0]))/(2 U2[t] V2[t]-2
U4[t] V4[t])^2)^2)+(sq2q2i (U2[t] (U4p[0] V1[t]-U3p[0] V2[t])+U4[t] (-U2p[0]
V1[t]+U3p[0] V4[t])+U3[t] (U2p[0] V2[t]-U4p[0] V4[t]))^2 (-m U2[t] V2p[0]+m
U4[t] V4p[0])^2)/((2 U2[t] V2[t]-2 U4[t] V4[t])^4 (((-U2p[0] V2[t]+U4p[0]
V4[t]) (-m U2[t] V2p[0]+m U4[t] V4p[0]))/(2 U2[t] V2[t]-2 U4[t] V4[t])^2-(m
(U2p[0] U4[t]-U2[t] U4p[0]) (V2p[0] V4[t]-V2[t] V4p[0]))/(2 U2[t] V2[t]-2
U4[t] V4[t])^2)^2)+(veq2i^2 (U2[t] (U4p[0] V1[t]-U3p[0] V2[t])+U4[t] (-U2p[0]
V1[t]+U3p[0] V4[t])+U3[t] (U2p[0] V2[t]-U4p[0] V4[t]))^2 (-m U2[t] V2p[0]+m
U4[t] V4p[0])^2)/((2 U2[t] V2[t]-2 U4[t] V4[t])^4 (((-U2p[0] V2[t]+U4p[0]
V4[t]) (-m U2[t] V2p[0]+m U4[t] V4p[0]))/(2 U2[t] V2[t]-2 U4[t] V4[t])^2-(m
(U2p[0] U4[t]-U2[t] U4p[0]) (V2p[0] V4[t]-V2[t] V4p[0]))/(2 U2[t] V2[t]-2
U4[t] V4[t])^2)^2)+((-u14[t] u2[t]^2 v2[t] v3[t]+u12[t] u2[t] u4[t] v2[t]
v3[t]+u2[t] u21[t] u4[t] v2[t] v3[t]-u2[t]^2 u41[t] v2[t] v3[t]-u2[t]^2 v2[t]
v23[t] v3[t]-u2[t] u24[t] u4[t] v3[t]^2+u22[t] u4[t]^2 v3[t]^2-u2[t] u4[t]
u42[t] v3[t]^2+u2[t]^2 u44[t] v3[t]^2+u2[t]^2 u22[t] v3[t]^2-u2[t] u4[t]
v24[t] v3[t]^2-u2[t]^2 v2[t] v3[t] v32[t]+u2[t]^2 v2[t]^2 v33[t]+u2[t] u4[t]
v2[t] v3[t] v34[t]+u14[t] u2[t] u4[t] v3[t] v4[t]-u12[t] u4[t]^2 v3[t] v4[t]-u21[t]
u4[t]^2 v3[t] v4[t]+u2[t] u4[t] u41[t] v3[t] v4[t]+u2[t] u4[t] v23[t] v3[t]
v4[t]+u2[t] u4[t] v3[t] v32[t] v4[t]-2 u2[t] u4[t] v2[t] v33[t] v4[t]-u4[t]^2
v3[t] v34[t] v4[t]+u4[t]^2 v33[t] v4[t]^2+u11[t] (u2[t] v2[t]-u4[t] v4[t])^2-u2[t]
u4[t] v3[t]^2 v42[t]+u2[t] u4[t] v2[t] v3[t] v43[t]-u4[t]^2 v3[t] v4[t] v43[t]+u4[t]^2

v3[t]^2 v44[t]+u1[t]^2 (u22[t] v2[t]^2-v4[t] (-(u44[t]+v22[t]) v4[t]+v2[t]
(u24[t]+u42[t]+v24[t]+v42[t])) +v2[t]^2 v44[t])+u1[t] (u12[t] v2[t] (-u2[t]
v2[t]+u4[t] v4[t])+u2[t] (-u21[t] v2[t]^2+u24[t] v2[t] v3[t]+u42[t] v2[t]
v3[t]+v2[t] v24[t] v3[t]-v2[t]^2 v34[t]+u14[t] v2[t] v4[t]+u41[t] v2[t] v4[t]+v2[t]
v23[t] v4[t]-2 u44[t] v3[t] v4[t]-2 v22[t] v3[t] v4[t]+v2[t] v32[t] v4[t]+v2[t]
v3[t] v42[t]-v2[t]^2 v43[t])+u4[t] (-2 u22[t] v2[t] v3[t]+v4[t] (-(u14[t]
+ u41[t] + v23[t] +v32[t]) v4[t]+v3[t] (u24[t]+u42[t]+v24[t]+v42[t])+v2[t]
(u21[t]+v34[t]+v43[t]))-2 v2[t] v3[t] v44[t]))) (-m U2[t] V2p[0]+m U4[t]
V4p[0])^2)/(2 (u2[t] v2[t]-u4[t] v4[t])^2 (2 U2[t] V2[t]-2 U4[t] V4[t])^2
(((-U2p[0] V2[t]+U4p[0] V4[t]) (-m U2[t] V2p[0]+m U4[t] V4p[0]))/(2 U2[t]
V2[t]-2 U4[t] V4[t])^2-(m (U2p[0] U4[t]-U2[t] U4p[0]) (V2p[0] V4[t]-V2[t]
V4p[0]))/(2 U2[t] V2[t]-2 U4[t] V4[t])^2)^2) +(m sq2p2i (U2p[0] U4[t]-U2[t]
U4p[0])^2 (U2[t] (-V1p[0] V2[t]+V1[t] V2p[0])+U4[t] (V1p[0] V4[t]-V1[t] V4p[0])+U3[t]
(-V2p[0] V4[t]+V2[t] V4p[0])))/((2 U2[t] V2[t]-2 U4[t] V4[t])^3 (((-U2p[0]
V2[t]+U4p[0] V4[t]) (-m U2[t] V2p[0]+m U4[t] V4p[0]))/(2 U2[t] V2[t]-2 U4[t]
V4[t])^2-(m (U2p[0] U4[t]-U2[t] U4p[0]) (V2p[0] V4[t]-V2[t] V4p[0]))/(2 U2[t]
V2[t]-2 U4[t] V4[t])^2)^2) +(m vep2i veq2i (U2p[0] U4[t]-U2[t] U4p[0])^2
(U2[t] (-V1p[0] V2[t]+V1[t] V2p[0])+U4[t] (V1p[0] V4[t]-V1[t] V4p[0])+U3[t]
(-V2p[0] V4[t]+V2[t] V4p[0])))/((2 U2[t] V2[t]-2 U4[t] V4[t])^3 (((-U2p[0]
V2[t]+U4p[0] V4[t]) (-m U2[t] V2p[0]+m U4[t] V4p[0]))/(2 U2[t] V2[t]-2 U4[t]
V4[t])^2-(m (U2p[0] U4[t]-U2[t] U4p[0]) (V2p[0] V4[t]-V2[t] V4p[0]))/(2 U2[t]
V2[t]-2 U4[t] V4[t])^2)^2)-(m vep1i veq2i (U2p[0] U4[t]-U2[t] U4p[0]) (-m
U2[t] V2p[0]+m U4[t] V4p[0]) (U2[t] (-V1p[0] V2[t]+V1[t] V2p[0])+U4[t] (V1p[0]
V4[t]-V1[t] V4p[0])+U3[t] (-V2p[0] V4[t]+V2[t] V4p[0])))/((2 U2[t] V2[t]-2
U4[t] V4[t])^3 (((-U2p[0] V2[t]+U4p[0] V4[t]) (-m U2[t] V2p[0]+m U4[t] V4p[0]))/(2
U2[t] V2[t]-2 U4[t] V4[t])^2-(m (U2p[0] U4[t]-U2[t] U4p[0]) (V2p[0] V4[t]-V2[t]
V4p[0]))/(2 U2[t] V2[t]-2 U4[t] V4[t])^2)^2) - (2 m veq1i veq2i (U2p[0] U4[t]-U2[t]
U4p[0]) (U2[t] (-U1p[0] V2[t]+U4p[0] V3[t])+U4[t] (-U2p[0] V3[t]+U1p[0] V4[t])+U1[t]
(U2p[0] V2[t]-U4p[0] V4[t])) (-m U2[t] V2p[0]+m U4[t] V4p[0]) (U2[t] (-V1p[0]
V2[t]+V1[t] V2p[0])+U4[t] (V1p[0] V4[t]-V1[t] V4p[0])+U3[t] (-V2p[0] V4[t]+V2[t]
V4p[0])))/((2 U2[t] V2[t]-2 U4[t] V4[t])^4 (((-U2p[0] V2[t]+U4p[0] V4[t])
(-m U2[t] V2p[0]+m U4[t] V4p[0]))/(2 U2[t] V2[t]-2 U4[t] V4[t])^2-(m (U2p[0]

U4[t]-U2[t] U4p[0]) (V2p[0] V4[t]-V2[t] V4p[0]))/(2 U2[t] V2[t]-2 U4[t] V4[t])^2)^2)

- (2 m sq2q2i (U2p[0] U4[t]-U2[t] U4p[0]) (U2[t] (U4p[0] V1[t]-U3p[0] V2[t])+U4[t] (-U2p[0] V1[t]+U3p[0] V4[t])+U3[t] (U2p[0] V2[t]-U4p[0] V4[t])) (-m U2[t] V2p[0]+m U4[t] V4p[0]) (U2[t] (-V1p[0] V2[t]+V1[t] V2p[0])+U4[t] (V1p[0] V4[t]-V1[t] V4p[0])+U3[t] (-V2p[0] V4[t]+V2[t] V4p[0])))/((2 U2[t] V2[t]-2 U4[t] V4[t])^4 (((-U2p[0] V2[t]+U4p[0] V4[t]) (-m U2[t] V2p[0]+m U4[t] V4p[0]))/(2 U2[t] V2[t]-2 U4[t] V4[t])^2-(m (U2p[0] U4[t]-U2[t] U4p[0]) (V2p[0] V4[t]-V2[t] V4p[0]))/(2 U2[t] V2[t]-2 U4[t] V4[t])^2)^2) - (2 m veq2i^2 (U2p[0] U4[t]-U2[t] U4p[0]) (U2[t] (U4p[0] V1[t]-U3p[0] V2[t])+U4[t] (-U2p[0] V1[t]+U3p[0] V4[t])+U3[t] (U2p[0] V2[t]-U4p[0] V4[t])) (-m U2[t] V2p[0]+m U4[t] V4p[0]) (U2[t] (-V1p[0] V2[t]+V1[t] V2p[0])+U4[t] (V1p[0] V4[t]-V1[t] V4p[0])+U3[t] (-V2p[0] V4[t]+V2[t] V4p[0])))/((2 U2[t] V2[t]-2 U4[t] V4[t])^4 (((-U2p[0] V2[t]+U4p[0] V4[t]) (-m U2[t] V2p[0]+m U4[t] V4p[0]))/(2 U2[t] V2[t]-2 U4[t] V4[t])^2-(m (U2p[0] U4[t]-U2[t] U4p[0]) (V2p[0] V4[t]-V2[t] V4p[0]))/(2 U2[t] V2[t]-2 U4[t] V4[t])^2)^2)

+(m^2 sq2q2i (U2p[0] U4[t]-U2[t] U4p[0])^2 (U2[t] (-V1p[0] V2[t]+V1[t] V2p[0])+U4[t] (V1p[0] V4[t]-V1[t] V4p[0])+U3[t] (-V2p[0] V4[t]+V2[t] V4p[0]))^2)/((2 U2[t] V2[t]-2 U4[t] V4[t])^4 (((-U2p[0] V2[t]+U4p[0] V4[t]) (-m U2[t] V2p[0]+m U4[t] V4p[0]))/(2 U2[t] V2[t]-2 U4[t] V4[t])^2-(m (U2p[0] U4[t]-U2[t] U4p[0]) (V2p[0] V4[t]-V2[t] V4p[0]))/(2 U2[t] V2[t]-2 U4[t] V4[t])^2)^2) +(m^2 veq2i^2 (U2p[0] U4[t]-U2[t] U4p[0])^2 (U2[t] (-V1p[0] V2[t]+V1[t] V2p[0])+U4[t] (V1p[0] V4[t]-V1[t] V4p[0])+U3[t] (-V2p[0] V4[t]+V2[t] V4p[0]))^2)/((2 U2[t] V2[t]-2 U4[t] V4[t])^4 (((-U2p[0] V2[t]+U4p[0] V4[t]) (-m U2[t] V2p[0]+m U4[t] V4p[0]))/(2 U2[t] V2[t]-2 U4[t] V4[t])^2-(m (U2p[0] U4[t]-U2[t] U4p[0]) (V2p[0] V4[t]-V2[t] V4p[0]))/(2 U2[t] V2[t]-2 U4[t] V4[t])^2)^2) +(m vep2i veq1i (U2p[0] U4[t]-U2[t] U4p[0])^2 (U2[t] (V2p[0] V3[t]-V2[t] V3p[0])+U1[t] (-V2p[0] V4[t]+V2[t] V4p[0])+U4[t] (V3p[0] V4[t]-V3[t] V4p[0])))/((2 U2[t] V2[t]-2 U4[t] V4[t])^3 (((-U2p[0] V2[t]+U4p[0] V4[t]) (-m U2[t] V2p[0]+m U4[t] V4p[0]))/(2 U2[t] V2[t]-2 U4[t] V4[t])^2-(m (U2p[0] U4[t]-U2[t] U4p[0]) (V2p[0] V4[t]-V2[t] V4p[0]))/(2 U2[t] V2[t]-2 U4[t] V4[t])^2)^2)-(m sq1p1i (U2p[0] U4[t]-U2[t] U4p[0]) (-m U2[t] V2p[0]+m U4[t] V4p[0]) (U2[t] (V2p[0] V3[t]-V2[t] V3p[0])+U1[t] (-V2p[0] V4[t]+V2[t] V4p[0])+U4[t] (V3p[0] V4[t]-V3[t] V4p[0])))/((2 U2[t] V2[t]-2 U4[t] V4[t])^3 (((-U2p[0] V2[t]+U4p[0] V4[t]) (-m U2[t] V2p[0]+m U4[t] V4p[0]))/(2

U2[t] V2[t]-2 U4[t] V4[t])^2-(m (U2p[0] U4[t]-U2[t] U4p[0]) (V2p[0] V4[t]-V2[t] V4p[0]))/(2 U2[t] V2[t]-2 U4[t] V4[t])^2)^2)-(m vep1i veq1i (U2p[0] U4[t]-U2[t] U4p[0]) (-m U2[t] V2p[0]+m U4[t] V4p[0]) (U2[t] (V2p[0] V3[t]-V2[t] V3p[0])+U1[t] (-V2p[0] V4[t]+V2[t] V4p[0])+U4[t] (V3p[0] V4[t]-V3[t] V4p[0])))/((2 U2[t] V2[t]-2 U4[t] V4[t])^3 (((-U2p[0] V2[t]+U4p[0] V4[t]) (-m U2[t] V2p[0]+m U4[t] V4p[0]))/(2 U2[t] V2[t]-2 U4[t] V4[t])^2-(m (U2p[0] U4[t]-U2[t] U4p[0]) (V2p[0] V4[t]-V2[t] V4p[0]))/(2 U2[t] V2[t]-2 U4[t] V4[t])^2)^2) - (2 m sq1q1i (U2p[0] U4[t]-U2[t] U4p[0]) (U2[t] (-U1p[0] V2[t]+U4p[0] V3[t])+U4[t] (-U2p[0] V3[t]+U1p[0] V4[t])+U1[t] (U2p[0] V2[t]-U4p[0] V4[t])) (-m U2[t] V2p[0]+m U4[t] V4p[0]) (U2[t] (V2p[0] V3[t]-V2[t] V3p[0])+U1[t] (-V2p[0] V4[t]+V2[t] V4p[0])+U4[t] (V3p[0] V4[t]-V3[t] V4p[0])))/((2 U2[t] V2[t]-2 U4[t] V4[t])^4 (((-U2p[0] V2[t]+U4p[0] V4[t]) (-m U2[t] V2p[0]+m U4[t] V4p[0]))/(2 U2[t] V2[t]-2 U4[t] V4[t])^2-(m (U2p[0] U4[t]-U2[t] U4p[0]) (V2p[0] V4[t]-V2[t] V4p[0]))/(2 U2[t] V2[t]-2 U4[t] V4[t])^2)^2) - (2 m veq1i^2 (U2p[0] U4[t]-U2[t] U4p[0]) (U2[t] (-U1p[0] V2[t]+U4p[0] V3[t])+U4[t] (-U2p[0] V3[t]+U1p[0] V4[t])+U1[t] (U2p[0] V2[t]-U4p[0] V4[t])) (-m U2[t] V2p[0]+m U4[t] V4p[0]) (U2[t] (V2p[0] V3[t]-V2[t] V3p[0])+U1[t] (-V2p[0] V4[t]+V2[t] V4p[0])+U4[t] (V3p[0] V4[t]-V3[t] V4p[0])))/((2 U2[t] V2[t]-2 U4[t] V4[t])^4 (((-U2p[0] V2[t]+U4p[0] V4[t]) (-m U2[t] V2p[0]+m U4[t] V4p[0]))/(2 U2[t] V2[t]-2 U4[t] V4[t])^2-(m (U2p[0] U4[t]-U2[t] U4p[0]) (V2p[0] V4[t]-V2[t] V4p[0]))/(2 U2[t] V2[t]-2 U4[t] V4[t])^2)^2) - (2 m veq1i veq2i (U2p[0] U4[t]-U2[t] U4p[0]) (U2[t] (U4p[0] V1[t]-U3p[0] V2[t])+U4[t] (-U2p[0] V1[t]+U3p[0] V4[t])+U3[t] (U2p[0] V2[t]-U4p[0] V4[t])) (-m U2[t] V2p[0]+m U4[t] V4p[0]) (U2[t] (V2p[0] V3[t]-V2[t] V3p[0])+U1[t] (-V2p[0] V4[t]+V2[t] V4p[0])+U4[t] (V3p[0] V4[t]-V3[t] V4p[0])))/((2 U2[t] V2[t]-2 U4[t] V4[t])^4 (((-U2p[0] V2[t]+U4p[0] V4[t]) (-m U2[t] V2p[0]+m U4[t] V4p[0]))/(2 U2[t] V2[t]-2 U4[t] V4[t])^2-(m (U2p[0] U4[t]-U2[t] U4p[0]) (V2p[0] V4[t]-V2[t] V4p[0]))/(2 U2[t] V2[t]-2 U4[t] V4[t])^2)^2)+(2 m^2 veq1i veq2i (U2p[0] U4[t]-U2[t] U4p[0])^2 U2[t] (-V1p[0] V2[t]+V1[t] V2p[0])+U4[t] (V1p[0] V4[t]-V1[t] V4p[0])+U3[t] (-V2p[0] V4[t]+V2[t] V4p[0])) (U2[t] (V2p[0] V3[t]-V2[t] V3p[0])+U1[t] (-V2p[0] V4[t]+V2[t] V4p[0])+U4[t] (V3p[0] V4[t]-V3[t] V4p[0])))/((2 U2[t] V2[t]-2 U4[t] V4[t])^4 (((-U2p[0] V2[t]+U4p[0] V4[t]) (-m U2[t] V2p[0]+m U4[t] V4p[0]))/(2 U2[t] V2[t]-2 U4[t] V4[t])^2-(m (U2p[0]

U4[t]-U2[t] U4p[0]) (V2p[0] V4[t]-V2[t] V4p[0])))/(2 U2[t] V2[t]-2 U4[t] V4[t])^2)^2)

+(m^2 sq1q1i (U2p[0] U4[t]-U2[t] U4p[0])^2 (U2[t] (V2p[0] V3[t]-V2[t] V3p[0])+U1[t]

(-V2p[0] V4[t]+V2[t] V4p[0])+U4[t] (V3p[0] V4[t]-V3[t] V4p[0]))^2)/((2 U2[t]

V2[t]-2 U4[t] V4[t])^4 (((-U2p[0] V2[t]+U4p[0] V4[t]) (-m U2[t] V2p[0]+m

U4[t] V4p[0]))/(2 U2[t] V2[t]-2 U4[t] V4[t])^2-(m (U2p[0] U4[t]-U2[t] U4p[0])

(V2p[0] V4[t]-V2[t] V4p[0]))/(2 U2[t] V2[t]-2 U4[t] V4[t])^2)^2) +(m^2 veq1i^2

(U2p[0] U4[t]-U2[t] U4p[0])^2 (U2[t] (V2p[0] V3[t]-V2[t] V3p[0])+U1[t] (-V2p[0]

V4[t]+V2[t] V4p[0])+U4[t] (V3p[0] V4[t]-V3[t] V4p[0]))^2)/((2 U2[t] V2[t]-2

U4[t] V4[t])^4 (((-U2p[0] V2[t]+U4p[0] V4[t]) (-m U2[t] V2p[0]+m U4[t] V4p[0]))/(2

U2[t] V2[t]-2 U4[t] V4[t])^2-(m (U2p[0] U4[t]-U2[t] U4p[0]) (V2p[0] V4[t]-V2[t]

V4p[0]))/(2 U2[t] V2[t]-2 U4[t] V4[t])^2)^2);


sp1p1[t_]:=(4 sq1q1i (U1p[t] U2[t] V2[t]-U1[t] U2p[t] V2[t]+U2p[t] U4[t]

V3[t]-U2[t] U4p[t] V3[t]-U1p[t] U4[t] V4[t]+U1[t] U4p[t] V4[t])^2)/(2 U2[t]

V2[t]-2 U4[t] V4[t])^2+(4 veq1i^2 (U1p[t] U2[t] V2[t]-U1[t] U2p[t] V2[t]+U2p[t]

U4[t] V3[t]-U2[t] U4p[t] V3[t]-U1p[t] U4[t] V4[t]+U1[t] U4p[t] V4[t])^2)/(2

U2[t] V2[t]-2 U4[t] V4[t])^2+(8 veq1i veq2i (U1p[t] U2[t] V2[t]-U1[t] U2p[t]

V2[t]+U2p[t] U4[t] V3[t]-U2[t] U4p[t] V3[t]-U1p[t] U4[t] V4[t]+U1[t] U4p[t]

V4[t]) (U2p[t] U4[t] V1[t]-U2[t] U4p[t] V1[t]-U2p[t] U3[t] V2[t]+U2[t] U3p[t]

V2[t]-U3p[t] U4[t] V4[t]+U3[t] U4p[t] V4[t]))/(2 U2[t] V2[t]-2 U4[t] V4[t])^2+(4

sq2q2i (U2p[t] U4[t] V1[t]-U2[t] U4p[t] V1[t]-U2p[t] U3[t] V2[t]+U2[t] U3p[t]

V2[t]-U3p[t] U4[t] V4[t]+U3[t] U4p[t] V4[t])^2)/(2 U2[t] V2[t]-2 U4[t] V4[t])^2+(4

veq2i^2 (U2p[t] U4[t] V1[t]-U2[t] U4p[t] V1[t]-U2p[t] U3[t] V2[t]+U2[t] U3p[t]

V2[t]-U3p[t] U4[t] V4[t]+U3[t] U4p[t] V4[t])^2)/(2 U2[t] V2[t]-2 U4[t] V4[t])^2+(2

(u22[t] v2[t]^2-v4[t] (-(u44[t]+v22[t]) v4[t]+v2[t] (u24[t]+u42[t]+v24[t]+v42[t]))

+v2[t]^2 v44[t])))/(u2[t] v2[t]-u4[t] v4[t])^2+(sp2p2i (-U2p[t] U4[t]+U2[t]

U4p[t])^2)/(-m U2[t] V2p[0]+m U4[t] V4p[0])^2+(vep2i^2 (-U2p[t] U4[t]+U2[t]

U4p[t])^2)/(-m U2[t] V2p[0]+m U4[t] V4p[0])^2+(2 (-U2p[t] U4[t]+U2[t] U4p[t])^2

(u22[t] (u4[t] v1[t]-u3[t] v2[t])^2+u2[t]^2 (u44[t] v1[t]^2-v2[t] (-(u33[t]+v11[t])

v2[t]+v1[t] (u34[t]+u43[t]+v12[t]+v21[t]))+v1[t]^2 v22[t])+u24[t] u3[t] u4[t]

v1[t] v4[t]-u23[t] u4[t]^2 v1[t] v4[t]-u32[t] u4[t]^2 v1[t] v4[t]+u3[t] u4[t]

u42[t] v1[t] v4[t]-u4[t]^2 v1[t] v14[t] v4[t]-u24[t] u3[t]^2 v2[t] v4[t]+u23[t]

u3[t] u4[t] v2[t] v4[t]+u3[t] u32[t] u4[t] v2[t] v4[t]-u3[t]^2 u42[t] v2[t]
v4[t]+u3[t] u4[t] v14[t] v2[t] v4[t]+u3[t] u4[t] v1[t] v24[t] v4[t]-u3[t]^2
v2[t] v24[t] v4[t]-u3[t] u34[t] u4[t] v4[t]^2+u33[t] u4[t]^2 v4[t]^2-u3[t]
u4[t] u43[t] v4[t]^2+u3[t]^2 u44[t] v4[t]^2+u4[t]^2 v11[t] v4[t]^2-u3[t]
u4[t] v12[t] v4[t]^2-u3[t] u4[t] v21[t] v4[t]^2+u3[t]^2 v22[t] v4[t]^2-u4[t]^2
v1[t] v4[t] v41[t]+u3[t] u4[t] v2[t] v4[t] v41[t]+u3[t] u4[t] v1[t] v4[t]
v42[t]-u3[t]^2 v2[t] v4[t] v42[t]+u2[t] (u24[t] v1[t] (-u4[t] v1[t]+u3[t]
v2[t])+u4[t] (-u42[t] v1[t]^2+u23[t] v1[t] v2[t]+u32[t] v1[t] v2[t]+v1[t]
v14[t] v2[t]-v1[t]^2 v24[t]+u34[t] v1[t] v4[t]+u43[t] v1[t] v4[t]+v1[t] v12[t]
v4[t]-2 u33[t] v2[t] v4[t]-2 v11[t] v2[t] v4[t]+v1[t] v21[t] v4[t]+v1[t]
v2[t] v41[t]-v1[t]^2 v42[t])+u3[t] (u42[t] v1[t] v2[t]-u23[t] v2[t]^2-u32[t]
v2[t]^2-v14[t] v2[t]^2+v1[t] v2[t] v24[t]-2 u44[t] v1[t] v4[t]+u34[t] v2[t]
v4[t]+u43[t] v2[t] v4[t]+v12[t] v2[t] v4[t]+v2[t] v21[t] v4[t]-2 v1[t] v22[t]
v4[t]-v2[t]^2 v41[t]+v1[t] v2[t] v42[t]))+(u4[t] v1[t]-u3[t] v2[t])^2 v44[t]))/((u2[t]
v2[t]-u4[t] v4[t])^2 (-m U2[t] V2p[0]+m U4[t] V4p[0])^2)-(4 vep2i veq1i (-U2p[t]
U4[t]+U2[t] U4p[t]) (U1p[t] U2[t] V2[t]-U1[t] U2p[t] V2[t]+U2p[t] U4[t] V3[t]-U2[t]
U4p[t] V3[t]-U1p[t] U4[t] V4[t]+U1[t] U4p[t] V4[t]))/((2 U2[t] V2[t]-2 U4[t]
V4[t]) (-m U2[t] V2p[0]+m U4[t] V4p[0]))-(4 sq2p2i (-U2p[t] U4[t]+U2[t] U4p[t])
(U2p[t] U4[t] V1[t]-U2[t] U4p[t] V1[t]-U2p[t] U3[t] V2[t]+U2[t] U3p[t] V2[t]-U3p[t]
U4[t] V4[t]+U3[t] U4p[t] V4[t]))/((2 U2[t] V2[t]-2 U4[t] V4[t]) (-m U2[t]
V2p[0]+m U4[t] V4p[0]))-(4 vep2i veq2i (-U2p[t] U4[t]+U2[t] U4p[t]) (U2p[t]
U4[t] V1[t]-U2[t] U4p[t] V1[t]-U2p[t] U3[t] V2[t]+U2[t] U3p[t] V2[t]-U3p[t]
U4[t] V4[t]+U3[t] U4p[t] V4[t]))/((2 U2[t] V2[t]-2 U4[t] V4[t]) (-m U2[t]
V2p[0]+m U4[t] V4p[0]))+(2 (-U2p[t] U4[t]+U2[t] U4p[t]) (2 u22[t] v2[t] (-u4[t]
v1[t]+u3[t] v2[t])+u2[t] (u24[t] v1[t] v2[t]+u42[t] v1[t] v2[t]-u23[t] v2[t]^2-u32[t]
v2[t]^2-v14[t] v2[t]^2+v1[t] v2[t] v24[t]-2 u44[t] v1[t] v4[t]+u34[t] v2[t]
v4[t]+u43[t] v2[t] v4[t]+v12[t] v2[t] v4[t]+v2[t] v21[t] v4[t]-2 v1[t] v22[t]
v4[t]-v2[t]^2 v41[t]+v1[t] v2[t] v42[t])+v4[t] (u24[t] (u4[t] v1[t]-2 u3[t]
v2[t])+u4[t] (-(u34[t]+u43[t] +v12[t]+v21[t]) v4[t]+v2[t] (u23[t]+u32[t] +
v14[t] + v41[t]) +v1[t] (u42[t] + v24[t] + v42[t]))-2 u3[t] (-(u44[t]+v22[t])
v4[t]+v2[t] (u42[t] + v24[t] + v42[t])))+2 v2[t] (-u4[t] v1[t]+u3[t] v2[t])
v44[t]))/((u2[t] v2[t]-u4[t] v4[t])^2 (-m U2[t] V2p[0]+m U4[t] V4p[0]))+(2

vep1i vep2i (-U2p[t] U4[t]+U2[t] U4p[t]) (U2p[t] V2[t]-U4p[t] V4[t]))/((2
U2[t] V2[t]-2 U4[t] V4[t])^2 (((-U2p[0] V2[t]+U4p[0] V4[t]) (-m U2[t] V2p[0]+m
U4[t] V4p[0]))/(2 U2[t] V2[t]-2 U4[t] V4[t])^2-(m (U2p[0] U4[t]-U2[t] U4p[0])
(V2p[0] V4[t]-V2[t] V4p[0]))/(2 U2[t] V2[t]-2 U4[t] V4[t])^2)) +(4 vep2i
veq1i (U2p[0] U4[t]-U2[t] U4p[0]) (U2p[t] V2[t]-U4p[t] V4[t]) (U1p[t] U2[t]
V2[t]-U1[t] U2p[t] V2[t]+U2p[t] U4[t] V3[t]-U2[t] U4p[t] V3[t]-U1p[t] U4[t]
V4[t]+U1[t] U4p[t] V4[t]))/((2 U2[t] V2[t]-2 U4[t] V4[t])^3 (((-U2p[0] V2[t]+U4p[0]
V4[t]) (-m U2[t] V2p[0]+m U4[t] V4p[0]))/(2 U2[t] V2[t]-2 U4[t] V4[t])^2-(m
(U2p[0] U4[t]-U2[t] U4p[0]) (V2p[0] V4[t]-V2[t] V4p[0]))/(2 U2[t] V2[t]-2
U4[t] V4[t])^2)) +(4 sq2p2i (U2p[0] U4[t]-U2[t] U4p[0]) (U2p[t] V2[t]-U4p[t]
V4[t]) (U2p[t] U4[t] V1[t]-U2[t] U4p[t] V1[t]-U2p[t] U3[t] V2[t]+U2[t] U3p[t]
V2[t]-U3p[t] U4[t] V4[t]+U3[t] U4p[t] V4[t]))/((2 U2[t] V2[t]-2 U4[t] V4[t])^3
(((-U2p[0] V2[t]+U4p[0] V4[t]) (-m U2[t] V2p[0]+m U4[t] V4p[0]))/(2 U2[t]
V2[t]-2 U4[t] V4[t])^2-(m (U2p[0] U4[t]-U2[t] U4p[0]) (V2p[0] V4[t]-V2[t]
V4p[0]))/(2 U2[t] V2[t]-2 U4[t] V4[t])^2)) +(4 vep2i veq2i (U2p[0] U4[t]-U2[t]
U4p[0]) (U2p[t] V2[t]-U4p[t] V4[t]) (U2p[t] U4[t] V1[t]-U2[t] U4p[t] V1[t]-U2p[t]
U3[t] V2[t]+U2[t] U3p[t] V2[t]-U3p[t] U4[t] V4[t]+U3[t] U4p[t] V4[t]))/((2
U2[t] V2[t]-2 U4[t] V4[t])^3 (((-U2p[0] V2[t]+U4p[0] V4[t]) (-m U2[t] V2p[0]+m
U4[t] V4p[0]))/(2 U2[t] V2[t]-2 U4[t] V4[t])^2-(m (U2p[0] U4[t]-U2[t] U4p[0])
(V2p[0] V4[t]-V2[t] V4p[0]))/(2 U2[t] V2[t]-2 U4[t] V4[t])^2)) +(4 vep2i
veq1i (-U2p[t] U4[t]+U2[t] U4p[t]) (U2p[t] V2[t]-U4p[t] V4[t]) (U2[t] (-U1p[0]
V2[t]+U4p[0] V3[t])+U4[t] (-U2p[0] V3[t]+U1p[0] V4[t])+U1[t] (U2p[0] V2[t]-U4p[0]
V4[t])))/((2 U2[t] V2[t]-2 U4[t] V4[t])^3 (((-U2p[0] V2[t]+U4p[0] V4[t])
(-m U2[t] V2p[0]+m U4[t] V4p[0]))/(2 U2[t] V2[t]-2 U4[t] V4[t])^2-(m (U2p[0]
U4[t]-U2[t] U4p[0]) (V2p[0] V4[t]-V2[t] V4p[0]))/(2 U2[t] V2[t]-2 U4[t] V4[t])^2))
+(4 sq2p2i (-U2p[t] U4[t]+U2[t] U4p[t]) (U2p[t] V2[t]-U4p[t] V4[t]) (U2[t]
(U4p[0] V1[t]-U3p[0] V2[t])+U4[t] (-U2p[0] V1[t]+U3p[0] V4[t])+U3[t] (U2p[0]
V2[t]-U4p[0] V4[t])))/((2 U2[t] V2[t]-2 U4[t] V4[t])^3 (((-U2p[0] V2[t]+U4p[0]
V4[t]) (-m U2[t] V2p[0]+m U4[t] V4p[0]))/(2 U2[t] V2[t]-2 U4[t] V4[t])^2-(m
(U2p[0] U4[t]-U2[t] U4p[0]) (V2p[0] V4[t]-V2[t] V4p[0]))/(2 U2[t] V2[t]-2
U4[t] V4[t])^2)) +(4 vep2i veq2i (-U2p[t] U4[t]+U2[t] U4p[t]) (U2p[t] V2[t]-U4p[t]
V4[t]) (U2[t] (U4p[0] V1[t]-U3p[0] V2[t])+U4[t] (-U2p[0] V1[t]+U3p[0] V4[t])+U3[t]

(U2p[0] V2[t]-U4p[0] V4[t])))/((2 U2[t] V2[t]-2 U4[t] V4[t])^3 (((-U2p[0]
V2[t]+U4p[0] V4[t]) (-m U2[t] V2p[0]+m U4[t] V4p[0]))/(2 U2[t] V2[t]-2 U4[t]
V4[t])^2-(m (U2p[0] U4[t]-U2[t] U4p[0]) (V2p[0] V4[t]-V2[t] V4p[0]))/(2 U2[t]
V2[t]-2 U4[t] V4[t])^2)) -(2 (U2p[0] U4[t]-U2[t] U4p[0]) (U2p[t] V2[t]-U4p[t]
V4[t]) (2 u22[t] v2[t] (-u4[t] v1[t]+u3[t] v2[t])+u2[t] (u24[t] v1[t] v2[t]+u42[t]
v1[t] v2[t]-u23[t] v2[t]^2-u32[t] v2[t]^2-v14[t] v2[t]^2+v1[t] v2[t] v24[t]-2
u44[t] v1[t] v4[t]+u34[t] v2[t] v4[t]+u43[t] v2[t] v4[t]+v12[t] v2[t] v4[t]+v2[t]
v21[t] v4[t]-2 v1[t] v22[t] v4[t]-v2[t]^2 v41[t]+v1[t] v2[t] v42[t])+v4[t]
(u24[t] (u4[t] v1[t]-2 u3[t] v2[t])+u4[t] (-(u34[t]+u43[t] +v12[t]+v21[t])
v4[t]+v2[t] (u23[t]+u32[t] + v14[t] + v41[t]) +v1[t] (u42[t] + u24[t] + v42[t]))-2
u3[t] (-(u44[t]+v22[t]) v4[t]+v2[t] (u42[t] + u24[t] + v42[t])))+2 v2[t]
(-u4[t] v1[t]+u3[t] v2[t]) v44[t]))/((u2[t] v2[t]-u4[t] v4[t])^2 (2 U2[t]
V2[t]-2 U4[t] V4[t])^2 (((-U2p[0] V2[t]+U4p[0] V4[t]) (-m U2[t] V2p[0]+m
U4[t] V4p[0]))/(2 U2[t] V2[t]-2 U4[t] V4[t])^2-(m (U2p[0] U4[t]-U2[t] U4p[0])
(V2p[0] V4[t]-V2[t] V4p[0]))/(2 U2[t] V2[t]-2 U4[t] V4[t])^2))+(2 (-U2p[t]
U4[t]+U2[t] U4p[t]) (U2p[t] V2[t]-U4p[t] V4[t]) (u12[t] u2[t] u4[t] v1[t]
v2[t]+u2[t] u21[t] u4[t] v1[t] v2[t]-u2[t]^2 u41[t] v1[t] v2[t]+u13[t] u2[t]^2
v2[t]^2-u12[t] u2[t] u3[t] v2[t]^2-u2[t] u21[t] u3[t] v2[t]^2+u2[t]^2 u31[t]
v2[t]^2+u2[t]^2 v13[t] v2[t]^2-u2[t]^2 v1[t] v2[t] v23[t]-2 u2[t] u24[t]
u4[t] v1[t] v3[t]+2 u22[t] u4[t]^2 v1[t] v3[t]-2 u2[t] u4[t] u42[t] v1[t]
v3[t]+2 u2[t]^2 u44[t] v1[t] v3[t]+u2[t] u24[t] u3[t] v2[t] v3[t]-u2[t]^2
u34[t] v2[t] v3[t]+u2[t] u23[t] u4[t] v2[t] v3[t]-2 u22[t] u3[t] u4[t] v2[t]
v3[t]+u2[t] u32[t] u4[t] v2[t] v3[t]+u2[t] u3[t] u42[t] v2[t] v3[t]-u2[t]^2
u43[t] v2[t] v3[t]-u2[t]^2 v12[t] v2[t] v3[t]+u2[t] u4[t] v14[t] v2[t] v3[t]-u2[t]^2
v2[t] v21[t] v3[t]+2 u2[t]^2 v1[t] v22[t] v3[t]-2 u2[t] u4[t] v1[t] v24[t]
v3[t]+u2[t] u3[t] v2[t] v24[t] v3[t]+u2[t]^2 v2[t]^2 v31[t]-u2[t]^2 v1[t]
v2[t] v32[t]+u2[t] u4[t] v1[t] v2[t] v34[t]-u2[t] u3[t] v2[t]^2 v34[t]-u12[t]
u4[t]^2 v1[t] v4[t]-u21[t] u4[t]^2 v1[t] v4[t]+u2[t] u4[t] u41[t] v1[t] v4[t]-2
u13[t] u2[t] u4[t] v2[t] v4[t]+u12[t] u3[t] u4[t] v2[t] v4[t]+u21[t] u3[t]
u4[t] v2[t] v4[t]-2 u2[t] u31[t] u4[t] v2[t] v4[t]+u2[t] u3[t] u41[t] v2[t]
v4[t]-2 u2[t] u4[t] v13[t] v2[t] v4[t]+u2[t] u4[t] v1[t] v23[t] v4[t]+u2[t]
u3[t] v2[t] v23[t] v4[t]+u24[t] u3[t] u4[t] v3[t] v4[t]+u2[t] u34[t] u4[t]

v3[t] v4[t]−u23[t] u4[t]^2 v3[t] v4[t]−u32[t] u4[t]^2 v3[t] v4[t]+u3[t] u4[t]
u42[t] v3[t] v4[t]+u2[t] u4[t] u43[t] v3[t] v4[t]−2 u2[t] u3[t] u44[t] v3[t]
v4[t]+u2[t] u4[t] v12[t] v3[t] v4[t]−u4[t]^2 v14[t] v3[t] v4[t]+u2[t] u4[t]
v21[t] v3[t] v4[t]−2 u2[t] u3[t] v22[t] v3[t] v4[t]+u3[t] u4[t] v24[t] v3[t]
v4[t]−2 u2[t] u4[t] v2[t] v31[t] v4[t]+u2[t] u4[t] v1[t] v32[t] v4[t]+u2[t]
u3[t] v2[t] v32[t] v4[t]−u4[t]^2 v1[t] v34[t] v4[t]+u3[t] u4[t] v2[t] v34[t]
v4[t]+u13[t] u4[t]^2 v4[t]^2+u31[t] u4[t]^2 v4[t]^2−u3[t] u4[t] u41[t] v4[t]^2+u4[t]^2
v13[t] v4[t]^2−u3[t] u4[t] v23[t] v4[t]^2+u4[t]^2 v31[t] v4[t]^2−u3[t] u4[t]
v32[t] v4[t]^2−u14[t] (u2[t] v1[t]−u3[t] v4[t]) (u2[t] v2[t]−u4[t] v4[t])+u2[t]
u4[t] v2[t] v3[t] v41[t]−u4[t]^2 v3[t] v4[t] v41[t]−2 u2[t] u4[t] v1[t] v3[t]
v42[t]+u2[t] u3[t] v2[t] v3[t] v42[t]+u3[t] u4[t] v3[t] v4[t] v42[t]+u2[t]
u4[t] v1[t] v2[t] v43[t]−u2[t] u3[t] v2[t]^2 v43[t]−u4[t]^2 v1[t] v4[t] v43[t]+u3[t]
u4[t] v2[t] v4[t] v43[t]+2 u4[t] (u4[t] v1[t]−u3[t] v2[t]) v3[t] v44[t]+u1[t]
(2 u22[t] v2[t] (−u4[t] v1[t]+u3[t] v2[t])+u2[t] (u24[t] v1[t] v2[t]+u42[t]
v1[t] v2[t]−u23[t] v2[t]^2−u32[t] v2[t]^2−v14[t] v2[t]^2+v1[t] v2[t] v24[t]−2
u44[t] v1[t] v4[t]+u34[t] v2[t] v4[t]+u43[t] v2[t] v4[t]+v12[t] v2[t] v4[t]+v2[t]
v21[t] v4[t]−2 v1[t] v22[t] v4[t]−v2[t]^2 v41[t]+v1[t] v2[t] v42[t])+v4[t]
(u24[t] (u4[t] v1[t]−2 u3[t] v2[t])+u4[t] (−(u34[t]+u43[t] +v12[t]+v21[t])
v4[t]+v2[t] (u23[t]+u32[t] + v14[t] + v41[t]) +v1[t] (u42[t] + v24[t] + v42[t]))−2
u3[t] (−(u44[t]+v22[t]) v4[t]+v2[t] (u42[t] + v24[t] + v42[t])))+2 v2[t]
(−u4[t] v1[t]+u3[t] v2[t]) v44[t])))/((u2[t] v2[t]−u4[t] v4[t])^2 (2 U2[t]
V2[t]−2 U4[t] V4[t])^2 (((−U2p[0] V2[t]+U4p[0] V4[t]) (−m U2[t] V2p[0]+m
U4[t] V4p[0]))/(2 U2[t] V2[t]−2 U4[t] V4[t])^2−(m (U2p[0] U4[t]−U2[t] U4p[0])
(V2p[0] V4[t]−V2[t] V4p[0]))/(2 U2[t] V2[t]−2 U4[t] V4[t])^2)) −(2 sp2p2i
(U2p[0] U4[t]−U2[t] U4p[0]) (−U2p[t] U4[t]+U2[t] U4p[t]) (U2p[t] V2[t]−U4p[t]
V4[t]))/((2 U2[t] V2[t]−2 U4[t] V4[t])^2 (−m U2[t] V2p[0]+m U4[t] V4p[0])
(((−U2p[0] V2[t]+U4p[0] V4[t]) (−m U2[t] V2p[0]+m U4[t] V4p[0]))/(2 U2[t]
V2[t]−2 U4[t] V4[t])^2−(m (U2p[0] U4[t]−U2[t] U4p[0]) (V2p[0] V4[t]−V2[t]
V4p[0]))/(2 U2[t] V2[t]−2 U4[t] V4[t])^2)) −(2 vep2i^2 (U2p[0] U4[t]−U2[t]
U4p[0]) (−U2p[t] U4[t]+U2[t] U4p[t]) (U2p[t] V2[t]−U4p[t] V4[t]))/((2 U2[t]
V2[t]−2 U4[t] V4[t])^2 (−m U2[t] V2p[0]+m U4[t] V4p[0]) (((−U2p[0] V2[t]+U4p[0]
V4[t]) (−m U2[t] V2p[0]+m U4[t] V4p[0]))/(2 U2[t] V2[t]−2 U4[t] V4[t])^2−(m

(U2p[0] U4[t]-U2[t] U4p[0]) (V2p[0] V4[t]-V2[t] V4p[0]))/(2 U2[t] V2[t]-2
U4[t] V4[t])^2)) -(4 (U2p[0] U4[t]-U2[t] U4p[0]) (-U2p[0] U4[t]+U2[t] U4p[t])
(U2p[t] V2[t]-U4p[t] V4[t]) (u22[t] (u4[t] v1[t]-u3[t] v2[t])^2+u2[t]^2 (u44[t]
v1[t]^2-v2[t] (-(u33[t]+v11[t]) v2[t]+v1[t] (u34[t]+u43[t] +v12[t]+v21[t]))+v1[t]^2
v22[t])+u24[t] u3[t] u4[t] v1[t] v4[t]-u23[t] u4[t]^2 v1[t] v4[t]-u32[t]
u4[t]^2 v1[t] v4[t]+u3[t] u4[t] u42[t] v1[t] v4[t]-u4[t]^2 v1[t] v14[t] v4[t]-u24[t]
u3[t]^2 v2[t] v4[t]+u23[t] u3[t] u4[t] v2[t] v4[t]+u3[t] u32[t] u4[t] v2[t]
v4[t]-u3[t]^2 u42[t] v2[t] v4[t]+u3[t] u4[t] v14[t] v2[t] v4[t]+u3[t] u4[t]
v1[t] v24[t] v4[t]-u3[t]^2 v2[t] v24[t] v4[t]-u3[t] u34[t] u4[t] v4[t]^2+u33[t]
u4[t]^2 v4[t]^2-u3[t] u4[t] u43[t] v4[t]^2+u3[t]^2 u44[t] v4[t]^2+u4[t]^2
v11[t] v4[t]^2-u3[t] u4[t] v12[t] v4[t]^2-u3[t] u4[t] v21[t] v4[t]^2+u3[t]^2
v22[t] v4[t]^2-u4[t]^2 v1[t] v4[t] v41[t]+u3[t] u4[t] v2[t] v4[t] v41[t]+u3[t]
u4[t] v1[t] v4[t] v42[t]-u3[t]^2 v2[t] v4[t] v42[t]+u2[t] (u24[t] v1[t] (-u4[t]
v1[t]+u3[t] v2[t])+u4[t] (-u42[t] v1[t]^2+u23[t] v1[t] v2[t]+u32[t] v1[t]
v2[t]+v1[t] v14[t] v2[t]-v1[t]^2 v24[t]+u34[t] v1[t] v4[t]+u43[t] v1[t] v4[t]+v1[t]
v12[t] v4[t]-2 u33[t] v2[t] v4[t]-2 v11[t] v2[t] v4[t]+v1[t] v21[t] v4[t]+v1[t]
v2[t] v41[t]-v1[t]^2 v42[t])+u3[t] (u42[t] v1[t] v2[t]-u23[t] v2[t]^2-u32[t]
v2[t]^2-v14[t] v2[t]^2+v1[t] v2[t] v24[t]-2 u44[t] v1[t] v4[t]+u34[t] v2[t]
v4[t]+u43[t] v2[t] v4[t]+v12[t] v2[t] v4[t]+v2[t] v21[t] v4[t]-2 v1[t] v22[t]
v4[t]-v2[t]^2 v41[t]+v1[t] v2[t] v42[t]))+(u4[t] v1[t]-u3[t] v2[t])^2 v44[t]))/((u2[t]
v2[t]-u4[t] v4[t])^2 (2 U2[t] V2[t]-2 U4[t] V4[t])^2 (-m U2[t] V2p[0]+m U4[t]
V4p[0]) (((-U2p[0] V2[t]+U4p[0] V4[t]) (-m U2[t] V2p[0]+m U4[t] V4p[0]))/(2
U2[t] V2[t]-2 U4[t] V4[t])^2-(m (U2p[0] U4[t]-U2[t] U4p[0]) (V2p[0] V4[t]-V2[t]
V4p[0]))/(2 U2[t] V2[t]-2 U4[t] V4[t])^2)) -(4 sq1p1i (U2p[t] V2[t]-U4p[t]
V4[t]) (U1p[t] U2[t] V2[t]-U1[t] U2p[t] V2[t]+U2p[t] U4[t] V3[t]-U2[t] U4p[t]
V3[t]-U1p[t] U4[t] V4[t]+U1[t] U4p[t] V4[t]) (-m U2[t] V2p[0]+m U4[t] V4p[0]))/((2
U2[t] V2[t]-2 U4[t] V4[t])^3 (((-U2p[0] V2[t]+U4p[0] V4[t]) (-m U2[t] V2p[0]+m
U4[t] V4p[0]))/(2 U2[t] V2[t]-2 U4[t] V4[t])^2-(m (U2p[0] U4[t]-U2[t] U4p[0])
(V2p[0] V4[t]-V2[t] V4p[0]))/(2 U2[t] V2[t]-2 U4[t] V4[t])^2)) -(4 vep1i
veq1i (U2p[t] V2[t]-U4p[t] V4[t]) (U1p[t] U2[t] V2[t]-U1[t] U2p[t] V2[t]+U2p[t]
U4[t] V3[t]-U2[t] U4p[t] V3[t]-U1p[t] U4[t] V4[t]+U1[t] U4p[t] V4[t]) (-m
U2[t] V2p[0]+m U4[t] V4p[0]))/((2 U2[t] V2[t]-2 U4[t] V4[t])^3 (((-U2p[0]

V2[t]+U4p[0] V4[t]) (-m U2[t] V2p[0]+m U4[t] V4p[0]))/(2 U2[t] V2[t]-2 U4[t]
V4[t])^2-(m (U2p[0] U4[t]-U2[t] U4p[0]) (V2p[0] V4[t]-V2[t] V4p[0]))/(2 U2[t]
V2[t]-2 U4[t] V4[t])^2)) -(4 vep1i veq2i (U2p[t] V2[t]-U4p[t] V4[t]) (U2p[t]
U4[t] V1[t]-U2[t] U4p[t] V1[t]-U2[t] U3[t] V2[t]+U2[t] U3p[t] V2[t]-U3p[t]
U4[t] V4[t]+U3[t] U4p[t] V4[t]) (-m U2[t] V2p[0]+m U4[t] V4p[0]))/((2 U2[t]
V2[t]-2 U4[t] V4[t])^3 (((-U2p[0] V2[t]+U4p[0] V4[t]) (-m U2[t] V2p[0]+m
U4[t] V4p[0]))/(2 U2[t] V2[t]-2 U4[t] V4[t])^2-(m (U2p[0] U4[t]-U2[t] U4p[0])
(V2p[0] V4[t]-V2[t] V4p[0]))/(2 U2[t] V2[t]-2 U4[t] V4[t])^2)) -(8 sq1q1i
(U2p[t] V2[t]-U4p[t] V4[t]) (U1p[t] U2[t] V2[t]-U1[t] U2p[t] V2[t]+U2p[t]
U4[t] V3[t]-U2[t] U4p[t] V3[t]-U1p[t] U4[t] V4[t]+U1[t] U4p[t] V4[t]) (U2[t]
(-U1p[0] V2[t]+U4p[0] V3[t])+U4[t] (-U2p[0] V3[t]+U1p[0] V4[t])+U1[t] (U2p[0]
V2[t]-U4p[0] V4[t])) (-m U2[t] V2p[0]+m U4[t] V4p[0]))/((2 U2[t] V2[t]-2
U4[t] V4[t])^4 (((-U2p[0] V2[t]+U4p[0] V4[t]) (-m U2[t] V2p[0]+m U4[t] V4p[0]))/(2
U2[t] V2[t]-2 U4[t] V4[t])^2-(m (U2p[0] U4[t]-U2[t] U4p[0]) (V2p[0] V4[t]-V2[t]
V4p[0]))/(2 U2[t] V2[t]-2 U4[t] V4[t])^2)) -(8 veq1i^2 (U2p[t] V2[t]-U4p[t]
V4[t]) (U1p[t] U2[t] V2[t]-U1[t] U2p[t] V2[t]+U2p[t] U4[t] V3[t]-U2[t] U4p[t]
V3[t]-U1p[t] U4[t] V4[t]+U1[t] U4p[t] V4[t]) (U2[t] (-U1p[0] V2[t]+U4p[0]
V3[t])+U4[t] (-U2p[0] V3[t]+U1p[0] V4[t])+U1[t] (U2p[0] V2[t]-U4p[0] V4[t]))
(-m U2[t] V2p[0]+m U4[t] V4p[0]))/((2 U2[t] V2[t]-2 U4[t] V4[t])^4 (((-U2p[0]
V2[t]+U4p[0] V4[t]) (-m U2[t] V2p[0]+m U4[t] V4p[0]))/(2 U2[t] V2[t]-2 U4[t]
V4[t])^2-(m (U2p[0] U4[t]-U2[t] U4p[0]) (V2p[0] V4[t]-V2[t] V4p[0]))/(2 U2[t]
V2[t]-2 U4[t] V4[t])^2)) -(8 veq1i veq2i (U2p[t] V2[t]-U4p[t] V4[t]) (U2p[t]
U4[t] V1[t]-U2[t] U4p[t] V1[t]-U2[t] U3[t] V2[t]+U2[t] U3p[t] V2[t]-U3p[t]
U4[t] V4[t]+U3[t] U4p[t] V4[t]) (U2[t] (-U1p[0] V2[t]+U4p[0] V3[t])+U4[t]
(-U2p[0] V3[t]+U1p[0] V4[t])+U1[t] (U2p[0] V2[t]-U4p[0] V4[t])) (-m U2[t]
V2p[0]+m U4[t] V4p[0]))/((2 U2[t] V2[t]-2 U4[t] V4[t])^4 (((-U2p[0] V2[t]+U4p[0]
V4[t]) (-m U2[t] V2p[0]+m U4[t] V4p[0]))/(2 U2[t] V2[t]-2 U4[t] V4[t])^2-(m
(U2p[0] U4[t]-U2[t] U4p[0]) (V2p[0] V4[t]-V2[t] V4p[0]))/(2 U2[t] V2[t]-2
U4[t] V4[t])^2)) -(8 veq1i veq2i (U2p[t] V2[t]-U4p[t] V4[t]) (U1p[t] U2[t]
V2[t]-U1[t] U2p[t] V2[t]+U2p[t] U4[t] V3[t]-U2[t] U4p[t] V3[t]-U1p[t] U4[t]
V4[t]+U1[t] U4p[t] V4[t]) (U2[t] (U4p[0] V1[t]-U3p[0] V2[t])+U4[t] (-U2p[0]
V1[t]+U3p[0] V4[t])+U3[t] (U2p[0] V2[t]-U4p[0] V4[t])) (-m U2[t] V2p[0]+m

31

U4[t] V4p[0]))/((2 U2[t] V2[t]-2 U4[t] V4[t])^4 (((-U2p[0] V2[t]+U4p[0] V4[t])

(-m U2[t] V2p[0]+m U4[t] V4p[0]))/(2 U2[t] V2[t]-2 U4[t] V4[t])^2-(m (U2p[0]

U4[t]-U2[t] U4p[0]) (V2p[0] V4[t]-V2[t] V4p[0]))/(2 U2[t] V2[t]-2 U4[t] V4[t])^2))

-(8 sq2q2i (U2p[t] V2[t]-U4p[t] V4[t]) (U2p[t] U4[t] V1[t]-U2[t] U4p[t] V1[t]-U2p[t]

U3[t] V2[t]+U2[t] U3p[t] V2[t]-U3p[t] U4[t] V4[t]+U3[t] U4p[t] V4[t]) (U2[t]

(U4p[0] V1[t]-U3p[0] V2[t])+U4[t] (-U2p[0] V1[t]+U3p[0] V4[t])+U3[t] (U2p[0]

V2[t]-U4p[0] V4[t])) (-m U2[t] V2p[0]+m U4[t] V4p[0]))/((2 U2[t] V2[t]-2

U4[t] V4[t])^4 (((-U2p[0] V2[t]+U4p[0] V4[t]) (-m U2[t] V2p[0]+m U4[t] V4p[0]))/(2

U2[t] V2[t]-2 U4[t] V4[t])^2-(m (U2p[0] U4[t]-U2[t] U4p[0]) (V2p[0] V4[t]-V2[t]

V4p[0]))/(2 U2[t] V2[t]-2 U4[t] V4[t])^2)) -(8 veq2i^2 (U2p[t] V2[t]-U4p[t]

V4[t]) (U2p[t] U4[t] V1[t]-U2[t] U4p[t] V1[t]-U2p[t] U3[t] V2[t]+U2[t] U3p[t]

V2[t]-U3p[t] U4[t] V4[t]+U3[t] U4p[t] V4[t]) (U2[t] (U4p[0] V1[t]-U3p[0]

V2[t])+U4[t] (-U2p[0] V1[t]+U3p[0] V4[t])+U3[t] (U2p[0] V2[t]-U4p[0] V4[t]))

(-m U2[t] V2p[0]+m U4[t] V4p[0]))/((2 U2[t] V2[t]-2 U4[t] V4[t])^4 (((-U2p[0]

V2[t]+U4p[0] V4[t]) (-m U2[t] V2p[0]+m U4[t] V4p[0]))/(2 U2[t] V2[t]-2 U4[t]

V4[t])^2-(m (U2p[0] U4[t]-U2[t] U4p[0]) (V2p[0] V4[t]-V2[t] V4p[0]))/(2 U2[t]

V2[t]-2 U4[t] V4[t])^2)) -(2 (U2p[t] V2[t]-U4p[t] V4[t]) (-2 u1[t] u22[t]

v2[t]^2+2 u22[t] u4[t] v2[t] v3[t]+2 u1[t] u24[t] v2[t] v4[t]-u21[t] u4[t]

v2[t] v4[t]+2 u1[t] u42[t] v2[t] v4[t]+2 u1[t] v2[t] v24[t] v4[t]-u24[t]

u4[t] v3[t] v4[t]-u4[t] u42[t] v3[t] v4[t]-u4[t] v24[t] v3[t] v4[t]-u4[t]

v2[t] v34[t] v4[t]+u14[t] u4[t] v4[t]^2+u4[t] u41[t] v4[t]^2-2 u1[t] u44[t]

v4[t]^2-2 u1[t] v22[t] v4[t]^2+u4[t] v23[t] v4[t]^2+u4[t] v32[t] v4[t]^2+u12[t]

v2[t] (u2[t] v2[t]-u4[t] v4[t])+2 u1[t] v2[t] v4[t] v42[t]-u4[t] v3[t] v4[t]

v42[t]-u4[t] v2[t] v4[t] v43[t]+u2[t] (u21[t] v2[t]^2-u24[t] v2[t] v3[t]-u42[t]

v2[t] v3[t]-v2[t] v24[t] v3[t]+v2[t]^2 v34[t]-u14[t] v2[t] v4[t]-u41[t] v2[t]

v4[t]-v2[t] v23[t] v4[t]+2 u44[t] v3[t] v4[t]+2 v22[t] v3[t] v4[t]-v2[t]

v32[t] v4[t]-v2[t] v3[t] v42[t]+v2[t]^2 v43[t])+2 v2[t] (-u1[t] v2[t]+u4[t]

v3[t]) v44[t]) (-m U2[t] V2p[0]+m U4[t] V4p[0]))/((u2[t] v2[t]-u4[t] v4[t])^2

(2 U2[t] V2[t]-2 U4[t] V4[t])^2 (((-U2p[0] V2[t]+U4p[0] V4[t]) (-m U2[t]

V2p[0]+m U4[t] V4p[0]))/(2 U2[t] V2[t]-2 U4[t] V4[t])^2-(m (U2p[0] U4[t]-U2[t]

U4p[0]) (V2p[0] V4[t]-V2[t] V4p[0]))/(2 U2[t] V2[t]-2 U4[t] V4[t])^2)) +(4

m sq1p1i (-U2p[t] U4[t]+U2[t] U4p[t]) (U1p[t] U2[t] V2[t]-U1[t] U2p[t] V2[t]+U2p[t]

U4[t] V3[t]-U2[t] U4p[t] V3[t]-U1p[t] U4[t] V4[t]+U1[t] U4p[t] V4[t]) (V2p[0]
V4[t]-V2[t] V4p[0]))/((2 U2[t] V2[t]-2 U4[t] V4[t])^3 (((-U2p[0] V2[t]+U4p[0]
V4[t]) (-m U2[t] V2p[0]+m U4[t] V4p[0]))/(2 U2[t] V2[t]-2 U4[t] V4[t])^2-(m
(U2p[0] U4[t]-U2[t] U4p[0]) (V2p[0] V4[t]-V2[t] V4p[0]))/(2 U2[t] V2[t]-2
U4[t] V4[t])^2)) +(4 m vep1i veq1i (-U2p[t] U4[t]+U2[t] U4p[t]) (U1p[t] U2[t]
V2[t]-U1[t] U2p[t] V2[t]+U2p[t] U4[t] V3[t]-U2[t] U4p[t] V3[t]-U1p[t] U4[t]
V4[t]+U1[t] U4p[t] V4[t]) (V2p[0] V4[t]-V2[t] V4p[0]))/((2 U2[t] V2[t]-2
U4[t] V4[t])^3 (((-U2p[0] V2[t]+U4p[0] V4[t]) (-m U2[t] V2p[0]+m U4[t] V4p[0]))/(2
U2[t] V2[t]-2 U4[t] V4[t])^2-(m (U2p[0] U4[t]-U2[t] U4p[0]) (V2p[0] V4[t]-V2[t]
V4p[0]))/(2 U2[t] V2[t]-2 U4[t] V4[t])^2)) +(4 m vep1i veq2i (-U2p[t] U4[t]+U2[t]
U4p[t]) (U2p[t] U4[t] V1[t]-U2[t] U4p[t] V1[t]-U2p[t] U3[t] V2[t]+U2[t] U3p[t]
V2[t]-U3p[t] U4[t] V4[t]+U3[t] U4p[t] V4[t]) (V2p[0] V4[t]-V2[t] V4p[0]))/((2
U2[t] V2[t]-2 U4[t] V4[t])^3 (((-U2p[0] V2[t]+U4p[0] V4[t]) (-m U2[t] V2p[0]+m
U4[t] V4p[0]))/(2 U2[t] V2[t]-2 U4[t] V4[t])^2-(m (U2p[0] U4[t]-U2[t] U4p[0])
(V2p[0] V4[t]-V2[t] V4p[0]))/(2 U2[t] V2[t]-2 U4[t] V4[t])^2)) +(8 m sq1q1i
(-U2p[t] U4[t]+U2[t] U4p[t]) (U1p[t] U2[t] V2[t]-U1[t] U2p[t] V2[t]+U2p[t]
U4[t] V3[t]-U2[t] U4p[t] V3[t]-U1p[t] U4[t] V4[t]+U1[t] U4p[t] V4[t]) (U2[t]
(-U1p[0] V2[t]+U4p[0] V3[t])+U4[t] (-U2p[0] V3[t]+U1p[0] V4[t])+U1[t] (U2p[0]
V2[t]-U4p[0] V4[t])) (V2p[0] V4[t]-V2[t] V4p[0]))/((2 U2[t] V2[t]-2 U4[t]
V4[t])^4 (((-U2p[0] V2[t]+U4p[0] V4[t]) (-m U2[t] V2p[0]+m U4[t] V4p[0]))/(2
U2[t] V2[t]-2 U4[t] V4[t])^2-(m (U2p[0] U4[t]-U2[t] U4p[0]) (V2p[0] V4[t]-V2[t]
V4p[0]))/(2 U2[t] V2[t]-2 U4[t] V4[t])^2)) +(8 m veq1i^2 (-U2p[t] U4[t]+U2[t]
U4p[t]) (U1p[t] U2[t] V2[t]-U1[t] U2p[t] V2[t]+U2p[t] U4[t] V3[t]-U2[t] U4p[t]
V3[t]-U1p[t] U4[t] V4[t]+U1[t] U4p[t] V4[t]) (U2[t] (-U1p[0] V2[t]+U4p[0]
V3[t])+U4[t] (-U2p[0] V3[t]+U1p[0] V4[t])+U1[t] (U2p[0] V2[t]-U4p[0] V4[t]))
(V2p[0] V4[t]-V2[t] V4p[0]))/((2 U2[t] V2[t]-2 U4[t] V4[t])^4 (((-U2p[0]
V2[t]+U4p[0] V4[t]) (-m U2[t] V2p[0]+m U4[t] V4p[0]))/(2 U2[t] V2[t]-2 U4[t]
V4[t])^2-(m (U2p[0] U4[t]-U2[t] U4p[0]) (V2p[0] V4[t]-V2[t] V4p[0]))/(2 U2[t]
V2[t]-2 U4[t] V4[t])^2)) +(8 m veq1i veq2i (-U2p[t] U4[t]+U2[t] U4p[t]) (U2p[t]
U4[t] V1[t]-U2[t] U4p[t] V1[t]-U2p[t] U3[t] V2[t]+U2[t] U3p[t] V2[t]-U3p[t]
U4[t] V4[t]+U3[t] U4p[t] V4[t]) (U2[t] (-U1p[0] V2[t]+U4p[0] V3[t])+U4[t]
(-U2p[0] V3[t]+U1p[0] V4[t])+U1[t] (U2p[0] V2[t]-U4p[0] V4[t])) (V2p[0] V4[t]-V2[t]

V4p[0]))/((2 U2[t] V2[t]-2 U4[t] V4[t])^4 (((-U2p[0] V2[t]+U4p[0] V4[t])

(-m U2[t] V2p[0]+m U4[t] V4p[0]))/(2 U2[t] V2[t]-2 U4[t] V4[t])^2-(m (U2p[0]

U4[t]-U2[t] U4p[0]) (V2p[0] V4[t]-V2[t] V4p[0]))/(2 U2[t] V2[t]-2 U4[t] V4[t])^2))

+(8 m veq1i veq2i (-U2p[t] U4[t]+U2[t] U4p[t]) (U1p[t] U2[t] V2[t]-U1[t]

U2p[t] V2[t]+U2p[t] U4[t] V3[t]-U2[t] U4p[t] V3[t]-U1p[t] U4[t] V4[t]+U1[t]

U4p[t] V4[t]) (U2[t] (U4p[0] V1[t]-U3p[0] V2[t])+U4[t] (-U2p[0] V1[t]+U3p[0]

V4[t])+U3[t] (U2p[0] V2[t]-U4p[0] V4[t])) (V2p[0] V4[t]-V2[t] V4p[0]))/((2

U2[t] V2[t]-2 U4[t] V4[t])^4 (((-U2p[0] V2[t]+U4p[0] V4[t]) (-m U2[t] V2p[0]+m

U4[t] V4p[0]))/(2 U2[t] V2[t]-2 U4[t] V4[t])^2-(m (U2p[0] U4[t]-U2[t] U4p[0])

(V2p[0] V4[t]-V2[t] V4p[0]))/(2 U2[t] V2[t]-2 U4[t] V4[t])^2)) +(8 m sq2q2i

(-U2p[t] U4[t]+U2[t] U4p[t]) (U2p[t] U4[t] V1[t]-U2[t] U4p[t] V1[t]-U2p[t]

U3[t] V2[t]+U2[t] U3p[t] V2[t]-U3p[t] U4[t] V4[t]+U3[t] U4p[t] V4[t]) (U2[t]

(U4p[0] V1[t]-U3p[0] V2[t])+U4[t] (-U2p[0] V1[t]+U3p[0] V4[t])+U3[t] (U2p[0]

V2[t]-U4p[0] V4[t])) (V2p[0] V4[t]-V2[t] V4p[0]))/((2 U2[t] V2[t]-2 U4[t]

V4[t])^4 (((-U2p[0] V2[t]+U4p[0] V4[t]) (-m U2[t] V2p[0]+m U4[t] V4p[0]))/(2

U2[t] V2[t]-2 U4[t] V4[t])^2-(m (U2p[0] U4[t]-U2[t] U4p[0]) (V2p[0] V4[t]-V2[t]

V4p[0]))/(2 U2[t] V2[t]-2 U4[t] V4[t])^2)) +(8 m veq2i^2 (-U2p[t] U4[t]+U2[t]

U4p[t]) (U2p[t] U4[t] V1[t]-U2[t] U4p[t] V1[t]-U2p[t] U3[t] V2[t]+U2[t] U3p[t]

V2[t]-U3p[t] U4[t] V4[t]+U3[t] U4p[t] V4[t]) (U2[t] (U4p[0] V1[t]-U3p[0]

V2[t])+U4[t] (-U2p[0] V1[t]+U3p[0] V4[t])+U3[t] (U2p[0] V2[t]-U4p[0] V4[t]))

(V2p[0] V4[t]-V2[t] V4p[0]))/((2 U2[t] V2[t]-2 U4[t] V4[t])^4 (((-U2p[0]

V2[t]+U4p[0] V4[t]) (-m U2[t] V2p[0]+m U4[t] V4p[0]))/(2 U2[t] V2[t]-2 U4[t]

V4[t])^2-(m (U2p[0] U4[t]-U2[t] U4p[0]) (V2p[0] V4[t]-V2[t] V4p[0]))/(2 U2[t]

V2[t]-2 U4[t] V4[t])^2))+(2 m (-U2p[t] U4[t]+U2[t] U4p[t]) (-2 u1[t] u22[t]

v2[t]^2+2 u22[t] u4[t] v2[t] v3[t]+2 u1[t] u24[t] v2[t] v4[t]-u21[t] u4[t]

v2[t] v4[t]+2 u1[t] u42[t] v2[t] v4[t]+2 u1[t] v2[t] v24[t] v4[t]-u24[t]

u4[t] v3[t] v4[t]-u4[t] u42[t] v3[t] v4[t]-u4[t] v24[t] v3[t] v4[t]-u4[t]

v2[t] v34[t] v4[t]+u14[t] u4[t] v4[t]^2+u4[t] u41[t] v4[t]^2-2 u1[t] u44[t]

v4[t]^2-2 u1[t] v22[t] v4[t]^2+u4[t] v23[t] v4[t]^2+u4[t] v32[t] v4[t]^2+u12[t]

v2[t] (u2[t] v2[t]-u4[t] v4[t])+2 u1[t] v2[t] v4[t] v42[t]-u4[t] v3[t] v4[t]

v42[t]-u4[t] v2[t] v4[t] v43[t]+u2[t] (u21[t] v2[t]^2-u24[t] v2[t] v3[t]-u42[t]

v2[t] v3[t]-v2[t] v24[t] v3[t]+v2[t]^2 v34[t]-u14[t] v2[t] v4[t]-u41[t] v2[t]

v4[t]-v2[t] v23[t] v4[t]+2 u44[t] v3[t] v4[t]+2 v22[t] v3[t] v4[t]-v2[t]

v32[t] v4[t]-v2[t] v3[t] v42[t]+v2[t]^2 v43[t])+2 v2[t] (-u1[t] v2[t]+u4[t]

v3[t]) v44[t]) (V2p[0] V4[t]-V2[t] V4p[0]))/((u2[t] v2[t]-u4[t] v4[t])^2

(2 U2[t] V2[t]-2 U4[t] V4[t])^2 (((-U2p[0] V2[t]+U4p[0] V4[t]) (-m U2[t]

V2p[0]+m U4[t] V4p[0]))/(2 U2[t] V2[t]-2 U4[t] V4[t])^2-(m (U2p[0] U4[t]-U2[t]

U4p[0]) (V2p[0] V4[t]-V2[t] V4p[0]))/(2 U2[t] V2[t]-2 U4[t] V4[t])^2))+(2

m sp2p2i (U2p[0] U4[t]-U2[t] U4p[0]) (-U2p[t] U4[t]+U2[t] U4p[t])^2 (V2p[0]

V4[t]-V2[t] V4p[0]))/((2 U2[t] V2[t]-2 U4[t] V4[t])^2 (-m U2[t] V2p[0]+m

U4[t] V4p[0])^2 (((-U2p[0] V2[t]+U4p[0] V4[t]) (-m U2[t] V2p[0]+m U4[t] V4p[0]))/(2

U2[t] V2[t]-2 U4[t] V4[t])^2-(m (U2p[0] U4[t]-U2[t] U4p[0]) (V2p[0] V4[t]-V2[t]

V4p[0]))/(2 U2[t] V2[t]-2 U4[t] V4[t])^2))+(2 m vep2i^2 (U2p[0] U4[t]-U2[t]

U4p[0]) (-U2p[t] U4[t]+U2[t] U4p[t])^2 (V2p[0] V4[t]-V2[t] V4p[0]))/((2 U2[t]

V2[t]-2 U4[t] V4[t])^2 (-m U2[t] V2p[0]+m U4[t] V4p[0])^2 (((-U2p[0] V2[t]+U4p[0]

V4[t]) (-m U2[t] V2p[0]+m U4[t] V4p[0]))/(2 U2[t] V2[t]-2 U4[t] V4[t])^2-(m

(U2p[0] U4[t]-U2[t] U4p[0]) (V2p[0] V4[t]-V2[t] V4p[0]))/(2 U2[t] V2[t]-2

U4[t] V4[t])^2)) +(4 m (U2p[0] U4[t]-U2[t] U4p[0]) (-U2p[t] U4[t]+U2[t] U4p[t])^2

(u22[t] (u4[t] v1[t]-u3[t] v2[t])^2+u2[t]^2 (u44[t] v1[t]^2-v2[t] (-(u33[t]+v11[t])

v2[t]+v1[t] (u34[t]+u43[t] +v12[t]+v21[t]))+v1[t]^2 v22[t])+u24[t] u3[t]

u4[t] v1[t] v4[t]-u23[t] u4[t]^2 v1[t] v4[t]-u32[t] u4[t]^2 v1[t] v4[t]+u3[t]

u4[t] u42[t] v1[t] v4[t]-u4[t]^2 v1[t] v14[t] v4[t]-u24[t] u3[t]^2 v2[t]

v4[t]+u23[t] u3[t] u4[t] v2[t] v4[t]+u3[t] u32[t] u4[t] v2[t] v4[t]-u3[t]^2

u42[t] v2[t] v4[t]+u3[t] u4[t] v14[t] v2[t] v4[t]+u3[t] u4[t] v1[t] v24[t]

v4[t]-u3[t]^2 v2[t] v24[t] v4[t]-u3[t] u34[t] u4[t] v4[t]^2+u33[t] u4[t]^2

v4[t]^2-u3[t] u4[t] u43[t] v4[t]^2+u3[t]^2 u44[t] v4[t]^2+u4[t]^2 v11[t]

v4[t]^2-u3[t] u4[t] v12[t] v4[t]^2-u3[t] u4[t] v21[t] v4[t]^2+u3[t]^2 v22[t]

v4[t]^2-u4[t]^2 v1[t] v4[t] v41[t]+u3[t] u4[t] v2[t] v4[t] v41[t]+u3[t] u4[t]

v1[t] v4[t] v42[t]-u3[t]^2 v2[t] v4[t] v42[t]+u2[t] (u24[t] v1[t] (-u4[t]

v1[t]+u3[t] v2[t])+u4[t] (-u42[t] v1[t]^2+u23[t] v1[t] v2[t]+u32[t] v1[t]

v2[t]+v1[t] v14[t] v2[t]-v1[t]^2 v24[t]+u34[t] v1[t] v4[t]+u43[t] v1[t] v4[t]+v1[t]

v12[t] v4[t]-2 u33[t] v2[t] v4[t]-2 v11[t] v2[t] v4[t]+v1[t] v21[t] v4[t]+v1[t]

v2[t] v41[t]-v1[t]^2 v42[t])+u3[t] (u42[t] v1[t] v2[t]-u23[t] v2[t]^2-u32[t]

v2[t]^2-v14[t] v2[t]^2+v1[t] v2[t] v24[t]-2 u44[t] v1[t] v4[t]+u34[t] v2[t]

v4[t]+u43[t] v2[t] v4[t]+v12[t] v2[t] v4[t]+v2[t] v21[t] v4[t]-2 v1[t] v22[t]
v4[t]-v2[t]^2 v41[t]+v1[t] v2[t] v42[t]))+(u4[t] v1[t]-u3[t] v2[t])^2 v44[t])
(V2p[0] V4[t]-V2[t] V4p[0]))/((u2[t] v2[t]-u4[t] v4[t])^2 (2 U2[t] V2[t]-2
U4[t] V4[t])^2 (-m U2[t] V2p[0]+m U4[t] V4p[0])^2 (((-U2p[0] V2[t]+U4p[0]
V4[t]) (-m U2[t] V2p[0]+m U4[t] V4p[0]))/(2 U2[t] V2[t]-2 U4[t] V4[t])^2-(m
(U2p[0] U4[t]-U2[t] U4p[0]) (V2p[0] V4[t]-V2[t] V4p[0]))/(2 U2[t] V2[t]-2
U4[t] V4[t])^2)) -(2 m vep1i vep2i (-U2p[t] U4[t]+U2[t] U4p[t])^2 (V2p[0]
V4[t]-V2[t] V4p[0]))/((2 U2[t] V2[t]-2 U4[t] V4[t])^2 (-m U2[t] V2p[0]+m
U4[t] V4p[0]) (((-U2p[0] V2[t]+U4p[0] V4[t]) (-m U2[t] V2p[0]+m U4[t] V4p[0]))/(2
U2[t] V2[t]-2 U4[t] V4[t])^2-(m (U2p[0] U4[t]-U2[t] U4p[0]) (V2p[0] V4[t]-V2[t]
V4p[0]))/(2 U2[t] V2[t]-2 U4[t] V4[t])^2)) -(4 m vep2i veq1i (U2p[0] U4[t]-U2[t]
U4p[0]) (-U2p[t] U4[t]+U2[t] U4p[t]) (U1p[t] U2[t] V2[t]-U1[t] U2p[t] V2[t]+U2p[t]
U4[t] V3[t]-U2[t] U4p[t] V3[t]-U1p[t] U4[t] V4[t]+U1[t] U4p[t] V4[t]) (V2p[0]
V4[t]-V2[t] V4p[0]))/((2 U2[t] V2[t]-2 U4[t] V4[t])^3 (-m U2[t] V2p[0]+m
U4[t] V4p[0]) (((-U2p[0] V2[t]+U4p[0] V4[t]) (-m U2[t] V2p[0]+m U4[t] V4p[0]))/(2
U2[t] V2[t]-2 U4[t] V4[t])^2-(m (U2p[0] U4[t]-U2[t] U4p[0]) (V2p[0] V4[t]-V2[t]
V4p[0]))/(2 U2[t] V2[t]-2 U4[t] V4[t])^2)) -(4 m sq2p2i (U2p[0] U4[t]-U2[t]
U4p[0]) (-U2p[t] U4[t]+U2[t] U4p[t]) (U2p[t] U4[t] V1[t]-U2[t] U4p[t] V1[t]-U2p[t]
U3[t] V2[t]+U2[t] U3p[t] V2[t]-U3p[t] U4[t] V4[t]+U3[t] U4p[t] V4[t]) (V2p[0]
V4[t]-V2[t] V4p[0]))/((2 U2[t] V2[t]-2 U4[t] V4[t])^3 (-m U2[t] V2p[0]+m
U4[t] V4p[0]) (((-U2p[0] V2[t]+U4p[0] V4[t]) (-m U2[t] V2p[0]+m U4[t] V4p[0]))/(2
U2[t] V2[t]-2 U4[t] V4[t])^2-(m (U2p[0] U4[t]-U2[t] U4p[0]) (V2p[0] V4[t]-V2[t]
V4p[0]))/(2 U2[t] V2[t]-2 U4[t] V4[t])^2)) -(4 m vep2i veq2i (U2p[0] U4[t]-U2[t]
U4p[0]) (-U2p[t] U4[t]+U2[t] U4p[t]) (U2p[t] U4[t] V1[t]-U2[t] U4p[t] V1[t]-U2p[t]
U3[t] V2[t]+U2[t] U3p[t] V2[t]-U3p[t] U4[t] V4[t]+U3[t] U4p[t] V4[t]) (V2p[0]
V4[t]-V2[t] V4p[0]))/((2 U2[t] V2[t]-2 U4[t] V4[t])^3 (-m U2[t] V2p[0]+m
U4[t] V4p[0]) (((-U2p[0] V2[t]+U4p[0] V4[t]) (-m U2[t] V2p[0]+m U4[t] V4p[0]))/(2
U2[t] V2[t]-2 U4[t] V4[t])^2-(m (U2p[0] U4[t]-U2[t] U4p[0]) (V2p[0] V4[t]-V2[t]
V4p[0]))/(2 U2[t] V2[t]-2 U4[t] V4[t])^2)) -(4 m vep2i veq1i (-U2p[t] U4[t]+U2[t]
U4p[t])^2 (U2[t] (-U1p[0] V2[t]+U4p[0] V3[t])+U4[t] (-U2p[0] V3[t]+U1p[0]
V4[t])+U1[t] (U2p[0] V2[t]-U4p[0] V4[t])) (V2p[0] V4[t]-V2[t] V4p[0]))/((2
U2[t] V2[t]-2 U4[t] V4[t])^3 (-m U2[t] V2p[0]+m U4[t] V4p[0]) (((-U2p[0]

36

V2[t]+U4p[0] V4[t]) (-m U2[t] V2p[0]+m U4[t] V4p[0]))/(2 U2[t] V2[t]-2 U4[t]
V4[t])^2-(m (U2p[0] U4[t]-U2[t] U4p[0]) (V2p[0] V4[t]-V2[t] V4p[0]))/(2 U2[t]
V2[t]-2 U4[t] V4[t])^2)) -(4 m sq2p2i (-U2p[t] U4[t]+U2[t] U4p[t])^2 (U2[t]
(U4p[0] V1[t]-U3p[0] V2[t])+U4[t] (-U2p[0] V1[t]+U3p[0] V4[t])+U3[t] (U2p[0]
V2[t]-U4p[0] V4[t])) (V2p[0] V4[t]-V2[t] V4p[0]))/((2 U2[t] V2[t]-2 U4[t]
V4[t])^3 (-m U2[t] V2p[0]+m U4[t] V4p[0]) ((((-U2p[0] V2[t]+U4p[0] V4[t])
(-m U2[t] V2p[0]+m U4[t] V4p[0]))/(2 U2[t] V2[t]-2 U4[t] V4[t])^2-(m (U2p[0]
U4[t]-U2[t] U4p[0]) (V2p[0] V4[t]-V2[t] V4p[0]))/(2 U2[t] V2[t]-2 U4[t] V4[t])^2))
-(4 m vep2i veq2i (-U2p[t] U4[t]+U2[t] U4p[t])^2 (U2[t] (U4p[0] V1[t]-U3p[0]
V2[t])+U4[t] (-U2p[0] V1[t]+U3p[0] V4[t])+U3[t] (U2p[0] V2[t]-U4p[0] V4[t]))
(V2p[0] V4[t]-V2[t] V4p[0]))/((2 U2[t] V2[t]-2 U4[t] V4[t])^3 (-m U2[t] V2p[0]+m
U4[t] V4p[0]) ((((-U2p[0] V2[t]+U4p[0] V4[t]) (-m U2[t] V2p[0]+m U4[t] V4p[0]))/(2
U2[t] V2[t]-2 U4[t] V4[t])^2-(m (U2p[0] U4[t]-U2[t] U4p[0]) (V2p[0] V4[t]-V2[t]
V4p[0]))/(2 U2[t] V2[t]-2 U4[t] V4[t])^2))+(2 m (U2p[0] U4[t]-U2[t] U4p[0])
(-U2p[t] U4[t]+U2[t] U4p[t]) (2 u22[t] v2[t] (-u4[t] v1[t]+u3[t] v2[t])+u2[t]
(u24[t] v1[t] v2[t]+u42[t] v1[t] v2[t]-u23[t] v2[t]^2-u32[t] v2[t]^2-v14[t]
v2[t]^2+v1[t] v2[t] v24[t]-2 u44[t] v1[t] v4[t]+u34[t] v2[t] v4[t]+u43[t]
v2[t] v4[t]+v12[t] v2[t] v4[t]+v2[t] v21[t] v4[t]-2 v1[t] v22[t] v4[t]-v2[t]^2
v41[t]+v1[t] v2[t] v42[t])+v4[t] (u24[t] (u4[t] v1[t]-2 u3[t] v2[t])+u4[t]
(-(u34[t]+u43[t] +v12[t]+v21[t]) v4[t]+v2[t] (u23[t]+u32[t]+ v14[t] + v41[t])
+v1[t] (u42[t] + u24[t] + v42[t]))-2 u3[t] (-(u44[t]+v22[t]) v4[t]+v2[t]
(u42[t] + u24[t] + v42[t])))+2 v2[t] (-u4[t] v1[t]+u3[t] v2[t]) v44[t]) (V2p[0]
V4[t]-V2[t] V4p[0]))/((u2[t] v2[t]-u4[t] v4[t])^2 (2 U2[t] V2[t]-2 U4[t]
V4[t])^2 (-m U2[t] V2p[0]+m U4[t] V4p[0]) ((((-U2p[0] V2[t]+U4p[0] V4[t])
(-m U2[t] V2p[0]+m U4[t] V4p[0]))/(2 U2[t] V2[t]-2 U4[t] V4[t])^2-(m (U2p[0]
U4[t]-U2[t] U4p[0]) (V2p[0] V4[t]-V2[t] V4p[0]))/(2 U2[t] V2[t]-2 U4[t] V4[t])^2))
-(2 m (-U2p[t] U4[t]+U2[t] U4p[t])^2 (u12[t] u2[t] u4[t] v1[t] v2[t]+u2[t]
u21[t] u4[t] v1[t] v2[t]-u2[t]^2 u41[t] v1[t] v2[t]+u13[t] u2[t]^2 v2[t]^2-u12[t]
u2[t] u3[t] v2[t]^2-u2[t] u21[t] u3[t] v2[t]^2+u2[t]^2 u31[t] v2[t]^2+u2[t]^2
v13[t] v2[t]^2-u2[t]^2 v1[t] v2[t] v23[t]-2 u2[t] u24[t] u4[t] v1[t] v3[t]+2
u22[t] u4[t]^2 v1[t] v3[t]-2 u2[t] u4[t] u42[t] v1[t] v3[t]+2 u2[t]^2 u44[t]
v1[t] v3[t]+u2[t] u24[t] u3[t] v2[t] v3[t]-u2[t]^2 u34[t] v2[t] v3[t]+u2[t]

37

u23[t] u4[t] v2[t] v3[t]-2 u22[t] u3[t] u4[t] v2[t] v3[t]+u2[t] u32[t] u4[t]
v2[t] v3[t]+u2[t] u3[t] u42[t] v2[t] v3[t]-u2[t]^2 u43[t] v2[t] v3[t]-u2[t]^2
v12[t] v2[t] v3[t]+u2[t] u4[t] v14[t] v2[t] v3[t]-u2[t]^2 v2[t] v21[t] v3[t]+2
u2[t]^2 v1[t] v22[t] v3[t]-2 u2[t] u4[t] v1[t] v24[t] v3[t]+u2[t] u3[t] v2[t]
v24[t] v3[t]+u2[t]^2 v2[t]^2 v31[t]-u2[t]^2 v1[t] v2[t] v32[t]+u2[t] u4[t]
v1[t] v2[t] v34[t]-u2[t] u3[t] v2[t]^2 v34[t]-u12[t] u4[t]^2 v1[t] v4[t]-u21[t]
u4[t]^2 v1[t] v4[t]+u2[t] u4[t] u41[t] v1[t] v4[t]-2 u13[t] u2[t] u4[t] v2[t]
v4[t]+u12[t] u3[t] u4[t] v2[t] v4[t]+u21[t] u3[t] u4[t] v2[t] v4[t]-2 u2[t]
u31[t] u4[t] v2[t] v4[t]+u2[t] u3[t] u41[t] v2[t] v4[t]-2 u2[t] u4[t] v13[t]
v2[t] v4[t]+u2[t] u4[t] v1[t] v23[t] v4[t]+u2[t] u3[t] v2[t] v23[t] v4[t]+u24[t]
u3[t] u4[t] v3[t] v4[t]+u2[t] u34[t] u4[t] v3[t] v4[t]-u23[t] u4[t]^2 v3[t]
v4[t]-u32[t] u4[t]^2 v3[t] v4[t]+u3[t] u4[t] u42[t] v3[t] v4[t]+u2[t] u4[t]
u43[t] v3[t] v4[t]-2 u2[t] u3[t] u44[t] v3[t] v4[t]+u2[t] u4[t] v12[t] v3[t]
v4[t]-u4[t]^2 v14[t] v3[t] v4[t]+u2[t] u4[t] v21[t] v3[t] v4[t]-2 u2[t] u3[t]
v22[t] v3[t] v4[t]+u3[t] u4[t] v24[t] v3[t] v4[t]-2 u2[t] u4[t] v2[t] v31[t]
v4[t]+u2[t] u4[t] v1[t] v32[t] v4[t]+u2[t] u3[t] v2[t] v32[t] v4[t]-u4[t]^2
v1[t] v34[t] v4[t]+u3[t] u4[t] v2[t] v34[t] v4[t]+u13[t] u4[t]^2 v4[t]^2+u31[t]
u4[t]^2 v4[t]^2-u3[t] u4[t] u41[t] v4[t]^2+u4[t]^2 v13[t] v4[t]^2-u3[t] u4[t]
v23[t] v4[t]^2+u4[t]^2 v31[t] v4[t]^2-u3[t] u4[t] v32[t] v4[t]^2-u14[t] (u2[t]
v1[t]-u3[t] v4[t]) (u2[t] v2[t]-u4[t] v4[t])+u2[t] u4[t] v2[t] v3[t] v41[t]-u4[t]^2
v3[t] v4[t] v41[t]-2 u2[t] u4[t] v1[t] v3[t] v42[t]+u2[t] u3[t] v2[t] v3[t]
v42[t]+u3[t] u4[t] v3[t] v4[t] v42[t]+u2[t] u4[t] v1[t] v2[t] v43[t]-u2[t]
u3[t] v2[t]^2 v43[t]-u4[t]^2 v1[t] v4[t] v43[t]+u3[t] u4[t] v2[t] v4[t] v43[t]+2
u4[t] (u4[t] v1[t]-u3[t] v2[t]) v3[t] v44[t]+u1[t] (2 u22[t] v2[t] (-u4[t]
v1[t]+u3[t] v2[t])+u2[t] (u24[t] v1[t] v2[t]+u42[t] v1[t] v2[t]-u23[t] v2[t]^2-u32[t]
v2[t]^2-v14[t] v2[t]^2+v1[t] v2[t] v24[t]-2 u44[t] v1[t] v4[t]+u34[t] v2[t]
v4[t]+u43[t] v2[t] v4[t]+v12[t] v2[t] v4[t]+v2[t] v21[t] v4[t]-2 v1[t] v22[t]
v4[t]-v2[t]^2 v41[t]+v1[t] v2[t] v42[t])+v4[t] (u24[t] (u4[t] v1[t]-2 u3[t]
v2[t])+u4[t] (-(u34[t]+u43[t] +v12[t]+v21[t]) v4[t]+v2[t] (u23[t]+u32[t] +
v14[t] + v41[t]) +v1[t] (u42[t] + v24[t] + v42[t]))-2 u3[t] (-(u44[t]+v22[t])
v4[t]+v2[t] (u42[t] + v24[t] + v42[t])))+2 v2[t] (-u4[t] v1[t]+u3[t] v2[t])
v44[t])) (V2p[0] V4[t]-V2[t] V4p[0]))/((u2[t] v2[t]-u4[t] v4[t])^2 (2 U2[t]

```
V2[t]-2 U4[t] V4[t])^2 (-m U2[t] V2p[0]+m U4[t] V4p[0]) (((-U2p[0] V2[t]+U4p[0]
V4[t]) (-m U2[t] V2p[0]+m U4[t] V4p[0]))/(2 U2[t] V2[t]-2 U4[t] V4[t])^2-(m
(U2p[0] U4[t]-U2[t] U4p[0]) (V2p[0] V4[t]-V2[t] V4p[0]))/(2 U2[t] V2[t]-2
U4[t] V4[t])^2)) +(4 sq1q1[t] (2 U2[t] V2[t]-2 U4[t] V4[t])^2 (-(((U2p[t]
V2[t]-U4p[t] V4[t]) (-m U2[t] V2p[0]+m U4[t] V4p[0]))/(2 U2[t] V2[t]-2 U4[t]
V4[t])^2)+(m (-U2p[t] U4[t]+U2[t] U4p[t]) (V2p[0] V4[t]-V2[t] V4p[0]))/(2
U2[t] V2[t]-2 U4[t] V4[t])^2)^2) /(-m U2[t] V2p[0]+m U4[t] V4p[0])^2+(4 m
sq2p2i (-U2p[t] U4[t]+U2[t] U4p[t])^2 (U2[t] (-V1p[0] V2[t]+V1[t] V2p[0])+U4[t]
(V1p[0] V4[t]-V1[t] V4p[0])+U3[t] (-V2p[0] V4[t]+V2[t] V4p[0])))/((2 U2[t]
V2[t]-2 U4[t] V4[t]) (-m U2[t] V2p[0]+m U4[t] V4p[0])^2)+(4 m vep2i veq2i
(-U2p[t] U4[t]+U2[t] U4p[t])^2 (U2[t] (-V1p[0] V2[t]+V1[t] V2p[0])+U4[t]
(V1p[0] V4[t]-V1[t] V4p[0])+U3[t] (-V2p[0] V4[t]+V2[t] V4p[0])))/((2 U2[t]
V2[t]-2 U4[t] V4[t]) (-m U2[t] V2p[0]+m U4[t] V4p[0])^2)-(8 m veq1i veq2i
(-U2p[t] U4[t]+U2[t] U4p[t]) (U1p[t] U2[t] V2[t]-U1[t] U2p[t] V2[t]+U2p[t]
U4[t] V3[t]-U2[t] U4p[t] V3[t]-U1p[t] U4[t] V4[t]+U1[t] U4p[t] V4[t]) (U2[t]
(-V1p[0] V2[t]+V1[t] V2p[0])+U4[t] (V1p[0] V4[t]-V1[t] V4p[0])+U3[t] (-V2p[0]
V4[t]+V2[t] V4p[0])))/((2 U2[t] V2[t]-2 U4[t] V4[t])^2 (-m U2[t] V2p[0]+m
U4[t] V4p[0]))-(8 m sq2q2i (-U2p[t] U4[t]+U2[t] U4p[t]) (U2p[t] U4[t] V1[t]-U2[t]
U4p[t] V1[t]-U2p[t] U3[t] V2[t]+U2[t] U3p[t] V2[t]-U3p[t] U4[t] V4[t]+U3[t]
U4p[t] V4[t]) (U2[t] (-V1p[0] V2[t]+V1[t] V2p[0])+U4[t] (V1p[0] V4[t]-V1[t]
V4p[0])+U3[t] (-V2p[0] V4[t]+V2[t] V4p[0])))/((2 U2[t] V2[t]-2 U4[t] V4[t])^2
(-m U2[t] V2p[0]+m U4[t] V4p[0]))-(8 m veq2i^2 (-U2p[t] U4[t]+U2[t] U4p[t])
(U2p[t] U4[t] V1[t]-U2[t] U4p[t] V1[t]-U2p[t] U3[t] V2[t]+U2[t] U3p[t] V2[t]-U3p[t]
U4[t] V4[t]+U3[t] U4p[t] V4[t]) (U2[t] (-V1p[0] V2[t]+V1[t] V2p[0])+U4[t]
(V1p[0] V4[t]-V1[t] V4p[0])+U3[t] (-V2p[0] V4[t]+V2[t] V4p[0])))/((2 U2[t]
V2[t]-2 U4[t] V4[t])^2 (-m U2[t] V2p[0]+m U4[t] V4p[0]))+(4 m vep1i veq2i
(-U2p[t] U4[t]+U2[t] U4p[t]) (U2p[t] V2[t]-U4p[t] V4[t]) (U2[t] (-V1p[0]
V2[t]+V1[t] V2p[0])+U4[t] (V1p[0] V4[t]-V1[t] V4p[0])+U3[t] (-V2p[0] V4[t]+V2[t]
V4p[0])))/((2 U2[t] V2[t]-2 U4[t] V4[t])^3 (((-U2p[0] V2[t]+U4p[0] V4[t])
(-m U2[t] V2p[0]+m U4[t] V4p[0]))/(2 U2[t] V2[t]-2 U4[t] V4[t])^2-(m (U2p[0]
U4[t]-U2[t] U4p[0]) (V2p[0] V4[t]-V2[t] V4p[0]))/(2 U2[t] V2[t]-2 U4[t] V4[t])^2))
+(8 m veq1i veq2i (U2p[0] U4[t]-U2[t] U4p[0]) (U2p[t] V2[t]-U4p[t] V4[t])
```

$$\begin{aligned}
&(U1p[t]\ U2[t]\ V2[t]-U1[t]\ U2p[t]\ V2[t]+U2p[t]\ U4[t]\ V3[t]-U2[t]\ U4p[t]\ V3[t]-U1p[t]\\
&U4[t]\ V4[t]+U1[t]\ U4p[t]\ V4[t])\ (U2[t]\ (-V1p[0]\ V2[t]+V1[t]\ V2p[0])+U4[t]\\
&(V1p[0]\ V4[t]-V1[t]\ V4p[0])+U3[t]\ (-V2p[0]\ V4[t]+V2[t]\ V4p[0])))/((2\ U2[t]\\
&V2[t]-2\ U4[t]\ V4[t])^{\wedge}4\ (((-U2p[0]\ V2[t]+U4p[0]\ V4[t])\ (-m\ U2[t]\ V2p[0]+m\\
&U4[t]\ V4p[0]))/(2\ U2[t]\ V2[t]-2\ U4[t]\ V4[t])^{\wedge}2-(m\ (U2p[0]\ U4[t]-U2[t]\ U4p[0])\\
&(V2p[0]\ V4[t]-V2[t]\ V4p[0]))/(2\ U2[t]\ V2[t]-2\ U4[t]\ V4[t])^{\wedge}2))\ +(8\ m\ sq2q2i\\
&(U2p[0]\ U4[t]-U2[t]\ U4p[0])\ (U2p[t]\ V2[t]-U4p[t]\ V4[t])\ (U2p[t]\ U4[t]\ V1[t]-U2[t]\\
&U4p[t]\ V1[t]-U2p[t]\ U3[t]\ V2[t]+U2[t]\ U3p[t]\ V2[t]-U3p[t]\ U4[t]\ V4[t]+U3[t]\\
&U4p[t]\ V4[t])\ (U2[t]\ (-V1p[0]\ V2[t]+V1[t]\ V2p[0])+U4[t]\ (V1p[0]\ V4[t]-V1[t]\\
&V4p[0])+U3[t]\ (-V2p[0]\ V4[t]+V2[t]\ V4p[0])))/((2\ U2[t]\ V2[t]-2\ U4[t]\ V4[t])^{\wedge}4\\
&(((-U2p[0]\ V2[t]+U4p[0]\ V4[t])\ (-m\ U2[t]\ V2p[0]+m\ U4[t]\ V4p[0]))/(2\ U2[t]\\
&V2[t]-2\ U4[t]\ V4[t])^{\wedge}2-(m\ (U2p[0]\ U4[t]-U2[t]\ U4p[0])\ (V2p[0]\ V4[t]-V2[t]\\
&V4p[0]))/(2\ U2[t]\ V2[t]-2\ U4[t]\ V4[t])^{\wedge}2))\ +(8\ m\ veq2i^{\wedge}2\ (U2p[0]\ U4[t]-U2[t]\\
&U4p[0])\ (U2p[t]\ V2[t]-U4p[t]\ V4[t])\ (U2p[t]\ U4[t]\ V1[t]-U2[t]\ U4p[t]\ V1[t]-U2p[t]\\
&U3[t]\ V2[t]+U2[t]\ U3p[t]\ V2[t]-U3p[t]\ U4[t]\ V4[t]+U3[t]\ U4p[t]\ V4[t])\ (U2[t]\\
&(-V1p[0]\ V2[t]+V1[t]\ V2p[0])+U4[t]\ (V1p[0]\ V4[t]-V1[t]\ V4p[0])+U3[t]\ (-V2p[0]\\
&V4[t]+V2[t]\ V4p[0])))/((2\ U2[t]\ V2[t]-2\ U4[t]\ V4[t])^{\wedge}4\ (((-U2p[0]\ V2[t]+U4p[0]\\
&V4[t])\ (-m\ U2[t]\ V2p[0]+m\ U4[t]\ V4p[0]))/(2\ U2[t]\ V2[t]-2\ U4[t]\ V4[t])^{\wedge}2-(m\\
&(U2p[0]\ U4[t]-U2[t]\ U4p[0])\ (V2p[0]\ V4[t]-V2[t]\ V4p[0]))/(2\ U2[t]\ V2[t]-2\\
&U4[t]\ V4[t])^{\wedge}2))\ +(8\ m\ veq1i\ veq2i\ (-U2p[0]\ U4[t]+U2[t]\ U4p[t])\ (U2p[t]\ V2[t]-U4p[t]\\
&V4[t])\ (U2[t]\ (-U1p[0]\ V2[t]+U4p[0]\ V3[t])+U4[t]\ (-U2p[0]\ V3[t]+U1p[0]\ V4[t])+U1[t]\\
&(U2p[0]\ V2[t]-U4p[0]\ V4[t]))\ (U2[t]\ (-V1p[0]\ V2[t]+V1[t]\ V2p[0])+U4[t]\ (V1p[0]\\
&V4[t]-V1[t]\ V4p[0])+U3[t]\ (-V2p[0]\ V4[t]+V2[t]\ V4p[0])))/((2\ U2[t]\ V2[t]-2\\
&U4[t]\ V4[t])^{\wedge}4\ (((-U2p[0]\ V2[t]+U4p[0]\ V4[t])\ (-m\ U2[t]\ V2p[0]+m\ U4[t]\ V4p[0]))/(2\\
&U2[t]\ V2[t]-2\ U4[t]\ V4[t])^{\wedge}2-(m\ (U2p[0]\ U4[t]-U2[t]\ U4p[0])\ (V2p[0]\ V4[t]-V2[t]\\
&V4p[0]))/(2\ U2[t]\ V2[t]-2\ U4[t]\ V4[t])^{\wedge}2))\ +(8\ m\ sq2q2i\ (-U2p[t]\ U4[t]+U2[t]\\
&U4p[t])\ (U2p[t]\ V2[t]-U4p[t]\ V4[t])\ (U2[t]\ (U4p[0]\ V1[t]-U3p[0]\ V2[t])+U4[t]\\
&(-U2p[0]\ V1[t]+U3p[0]\ V4[t])+U3[t]\ (U2p[0]\ V2[t]-U4p[0]\ V4[t]))\ (U2[t]\ (-V1p[0]\\
&V2[t]+V1[t]\ V2p[0])+U4[t]\ (V1p[0]\ V4[t]-V1[t]\ V4p[0])+U3[t]\ (-V2p[0]\ V4[t]+V2[t]\\
&V4p[0])))/((2\ U2[t]\ V2[t]-2\ U4[t]\ V4[t])^{\wedge}4\ (((-U2p[0]\ V2[t]+U4p[0]\ V4[t])\\
&(-m\ U2[t]\ V2p[0]+m\ U4[t]\ V4p[0]))/(2\ U2[t]\ V2[t]-2\ U4[t]\ V4[t])^{\wedge}2-(m\ (U2p[0]\\
&U4[t]-U2[t]\ U4p[0])\ (V2p[0]\ V4[t]-V2[t]\ V4p[0]))/(2\ U2[t]\ V2[t]-2\ U4[t]\ V4[t])^{\wedge}2))
\end{aligned}$$

$+$(8 m veq2i^2 (-U2p[t] U4[t]+U2[t] U4p[t]) (U2p[t] V2[t]-U4p[t] V4[t]) (U2[t] (U4p[0] V1[t]-U3p[0] V2[t])+U4[t] (-U2p[0] V1[t]+U3p[0] V4[t])+U3[t] (U2p[0] V2[t]-U4p[0] V4[t])) (U2[t] (-V1p[0] V2[t]+V1[t] V2p[0])+U4[t] (V1p[0] V4[t]-V1[t] V4p[0])+U3[t] (-V2p[0] V4[t]+V2[t] V4p[0])))/((2 U2[t] V2[t]-2 U4[t] V4[t])^4 (((-U2p[0] V2[t]+U4p[0] V4[t]) (-m U2[t] V2p[0]+m U4[t] V4p[0]))/(2 U2[t] V2[t]-2 U4[t] V4[t])^2-(m (U2p[0] U4[t]-U2[t] U4p[0]) (V2p[0] V4[t]-V2[t] V4p[0]))/(2 U2[t] V2[t]-2 U4[t] V4[t])^2)) -(8 m sq2p2i (U2p[0] U4[t]-U2[t] U4p[0]) (-U2p[t] U4[t]+U2[t] U4p[t]) (U2p[t] V2[t]-U4p[t] V4[t]) (U2[t] (-V1p[0] V2[t]+V1[t] V2p[0])+U4[t] (V1p[0] V4[t]-V1[t] V4p[0])+U3[t] (-V2p[0] V4[t]+V2[t] V4p[0])))/((2 U2[t] V2[t]-2 U4[t] V4[t])^3 (-m U2[t] V2p[0]+m U4[t] V4p[0]) (((-U2p[0] V2[t]+U4p[0] V4[t]) (-m U2[t] V2p[0]+m U4[t] V4p[0]))/(2 U2[t] V2[t]-2 U4[t] V4[t])^2-(m (U2p[0] U4[t]-U2[t] U4p[0]) (V2p[0] V4[t]-V2[t] V4p[0]))/(2 U2[t] V2[t]-2 U4[t] V4[t])^2)) -(8 m vep2i veq2i (U2p[0] U4[t]-U2[t] U4p[0]) (-U2p[t] U4[t]+U2[t] U4p[t]) (U2p[t] V2[t]-U4p[t] V4[t]) (U2[t] (-V1p[0] V2[t]+V1[t] V2p[0])+U4[t] (V1p[0] V4[t]-V1[t] V4p[0])+U3[t] (-V2p[0] V4[t]+V2[t] V4p[0])))/((2 U2[t] V2[t]-2 U4[t] V4[t])^3 (-m U2[t] V2p[0]+m U4[t] V4p[0]) (((-U2p[0] V2[t]+U4p[0] V4[t]) (-m U2[t] V2p[0]+m U4[t] V4p[0]))/(2 U2[t] V2[t]-2 U4[t] V4[t])^2-(m (U2p[0] U4[t]-U2[t] U4p[0]) (V2p[0] V4[t]-V2[t] V4p[0]))/(2 U2[t] V2[t]-2 U4[t] V4[t])^2)) +(8 m^2 sq2p2i (U2p[0] U4[t]-U2[t] U4p[0]) (-U2p[t] U4[t]+U2[t] U4p[t])^2 (V2p[0] V4[t]-V2[t] V4p[0]) (U2[t] (-V1p[0] V2[t]+V1[t] V2p[0])+U4[t] (V1p[0] V4[t]-V1[t] V4p[0])+U3[t] (-V2p[0] V4[t]+V2[t] V4p[0])))/((2 U2[t] V2[t]-2 U4[t] V4[t])^3 (-m U2[t] V2p[0]+m U4[t] V4p[0])^2 (((-U2p[0] V2[t]+U4p[0] V4[t]) (-m U2[t] V2p[0]+m U4[t] V4p[0]))/(2 U2[t] V2[t]-2 U4[t] V4[t])^2-(m (U2p[0] U4[t]-U2[t] U4p[0]) (V2p[0] V4[t]-V2[t] V4p[0]))/(2 U2[t] V2[t]-2 U4[t] V4[t])^2)) +(8 m^2 vep2i veq2i (U2p[0] U4[t]-U2[t] U4p[0]) (-U2p[t] U4[t]+U2[t] U4p[t])^2 (V2p[0] V4[t]-V2[t] V4p[0]) (U2[t] (-V1p[0] V2[t]+V1[t] V2p[0])+U4[t] (V1p[0] V4[t]-V1[t] V4p[0])+U3[t] (-V2p[0] V4[t]+V2[t] V4p[0])))/((2 U2[t] V2[t]-2 U4[t] V4[t])^3 (-m U2[t] V2p[0]+m U4[t] V4p[0])^2 (((-U2p[0] V2[t]+U4p[0] V4[t]) (-m U2[t] V2p[0]+m U4[t] V4p[0]))/(2 U2[t] V2[t]-2 U4[t] V4[t])^2-(m (U2p[0] U4[t]-U2[t] U4p[0]) (V2p[0] V4[t]-V2[t] V4p[0]))/(2 U2[t] V2[t]-2 U4[t] V4[t])^2)) -(4 m^2 vep1i veq2i (-U2p[t] U4[t]+U2[t] U4p[t])^2 (V2p[0] V4[t]-V2[t] V4p[0]) (U2[t] (-V1p[0] V2[t]+V1[t] V2p[0])+U4[t]

(V1p[0] V4[t]-V1[t] V4p[0])+U3[t] (-V2p[0] V4[t]+V2[t] V4p[0])))/((2 U2[t]
V2[t]-2 U4[t] V4[t])^3 (-m U2[t] V2p[0]+m U4[t] V4p[0]) (((-U2p[0] V2[t]+U4p[0]
V4[t]) (-m U2[t] V2p[0]+m U4[t] V4p[0]))/(2 U2[t] V2[t]-2 U4[t] V4[t])^2-(m
(U2p[0] U4[t]-U2[t] U4p[0]) (V2p[0] V4[t]-V2[t] V4p[0]))/(2 U2[t] V2[t]-2
U4[t] V4[t])^2)) -(8 m^2 veq1i veq2i (U2p[0] U4[t]-U2[t] U4p[0]) (-U2p[t]
U4[t]+U2[t] U4p[t]) (U1p[t] U2[t] V2[t]-U1[t] U2p[t] V2[t]+U2p[t] U4[t] V3[t]-U2[t]
U4p[t] V3[t]-U1p[t] U4[t] V4[t]+U1[t] U4p[t] V4[t]) (V2p[0] V4[t]-V2[t] V4p[0])
(U2[t] (-V1p[0] V2[t]+V1[t] V2p[0])+U4[t] (V1p[0] V4[t]-V1[t] V4p[0])+U3[t]
(-V2p[0] V4[t]+V2[t] V4p[0])))/((2 U2[t] V2[t]-2 U4[t] V4[t])^4 (-m U2[t]
V2p[0]+m U4[t] V4p[0]) (((-U2p[0] V2[t]+U4p[0] V4[t]) (-m U2[t] V2p[0]+m
U4[t] V4p[0]))/(2 U2[t] V2[t]-2 U4[t] V4[t])^2-(m (U2p[0] U4[t]-U2[t] U4p[0])
(V2p[0] V4[t]-V2[t] V4p[0]))/(2 U2[t] V2[t]-2 U4[t] V4[t])^2)) -(8 m^2 sq2q2i
(U2p[0] U4[t]-U2[t] U4p[0]) (-U2p[t] U4[t]+U2[t] U4p[t]) (U2p[t] U4[t] V1[t]-U2[t]
U4p[t] V1[t]-U2p[t] U3[t] V2[t]+U2[t] U3p[t] V2[t]-U3p[t] U4[t] V4[t]+U3[t]
U4p[t] V4[t]) (V2p[0] V4[t]-V2[t] V4p[0]) (U2[t] (-V1p[0] V2[t]+V1[t] V2p[0])+U4[t]
(V1p[0] V4[t]-V1[t] V4p[0])+U3[t] (-V2p[0] V4[t]+V2[t] V4p[0])))/((2 U2[t]
V2[t]-2 U4[t] V4[t])^4 (-m U2[t] V2p[0]+m U4[t] V4p[0]) (((-U2p[0] V2[t]+U4p[0]
V4[t]) (-m U2[t] V2p[0]+m U4[t] V4p[0]))/(2 U2[t] V2[t]-2 U4[t] V4[t])^2-(m
(U2p[0] U4[t]-U2[t] U4p[0]) (V2p[0] V4[t]-V2[t] V4p[0]))/(2 U2[t] V2[t]-2
U4[t] V4[t])^2)) -(8 m^2 veq2i^2 (U2p[0] U4[t]-U2[t] U4p[0]) (-U2p[t] U4[t]+U2[t]
U4p[t]) (U2p[t] U4[t] V1[t]-U2[t] U4p[t] V1[t]-U2p[t] U3[t] V2[t]+U2[t] U3p[t]
V2[t]-U3p[t] U4[t] V4[t]+U3[t] U4p[t] V4[t]) (V2p[0] V4[t]-V2[t] V4p[0])
(U2[t] (-V1p[0] V2[t]+V1[t] V2p[0])+U4[t] (V1p[0] V4[t]-V1[t] V4p[0])+U3[t]
(-V2p[0] V4[t]+V2[t] V4p[0])))/((2 U2[t] V2[t]-2 U4[t] V4[t])^4 (-m U2[t]
V2p[0]+m U4[t] V4p[0]) (((-U2p[0] V2[t]+U4p[0] V4[t]) (-m U2[t] V2p[0]+m
U4[t] V4p[0]))/(2 U2[t] V2[t]-2 U4[t] V4[t])^2-(m (U2p[0] U4[t]-U2[t] U4p[0])
(V2p[0] V4[t]-V2[t] V4p[0]))/(2 U2[t] V2[t]-2 U4[t] V4[t])^2)) -(8 m^2 veq1i
veq2i (-U2p[t] U4[t]+U2[t] U4p[t])^2 (U2[t] (-U1p[0] V2[t]+U4p[0] V3[t])+U4[t]
(-U2p[0] V3[t]+U1p[0] V4[t])+U1[t] (U2p[0] V2[t]-U4p[0] V4[t])) (V2p[0] V4[t]-V2[t]
V4p[0]) (U2[t] (-V1p[0] V2[t]+V1[t] V2p[0])+U4[t] (V1p[0] V4[t]-V1[t] V4p[0])+U3[t]
(-V2p[0] V4[t]+V2[t] V4p[0])))/((2 U2[t] V2[t]-2 U4[t] V4[t])^4 (-m U2[t]
V2p[0]+m U4[t] V4p[0]) (((-U2p[0] V2[t]+U4p[0] V4[t]) (-m U2[t] V2p[0]+m

$$U4[t] \ V4p[0]))/(2 \ U2[t] \ V2[t]-2 \ U4[t] \ V4[t])^{\wedge}2-(m \ (U2p[0] \ U4[t]-U2[t] \ U4p[0])$$

$$(V2p[0] \ V4[t]-V2[t] \ V4p[0]))/(2 \ U2[t] \ V2[t]-2 \ U4[t] \ V4[t])^{\wedge}2)) \ -(8 \ m^{\wedge}2 \ sq2q2i$$

$$(-U2p[t] \ U4[t]+U2[t] \ U4p[t])^{\wedge}2 \ (U2[t] \ (U4p[0] \ V1[t]-U3p[0] \ V2[t])+U4[t] \ (-U2p[0]$$

$$V1[t]+U3p[0] \ V4[t])+U3[t] \ (U2p[0] \ V2[t]-U4p[0] \ V4[t])) \ (V2p[0] \ V4[t]-V2[t]$$

$$V4p[0]) \ (U2[t] \ (-V1p[0] \ V2[t]+V1[t] \ V2p[0])+U4[t] \ (V1p[0] \ V4[t]-V1[t] \ V4p[0])+U3[t]$$

$$(-V2p[0] \ V4[t]+V2[t] \ V4p[0])))/((2 \ U2[t] \ V2[t]-2 \ U4[t] \ V4[t])^{\wedge}4 \ (-m \ U2[t]$$

$$V2p[0]+m \ U4[t] \ V4p[0]) \ (((-U2p[0] \ V2[t]+U4p[0] \ V4[t]) \ (-m \ U2[t] \ V2p[0]+m$$

$$U4[t] \ V4p[0]))/(2 \ U2[t] \ V2[t]-2 \ U4[t] \ V4[t])^{\wedge}2-(m \ (U2p[0] \ U4[t]-U2[t] \ U4p[0])$$

$$(V2p[0] \ V4[t]-V2[t] \ V4p[0]))/(2 \ U2[t] \ V2[t]-2 \ U4[t] \ V4[t])^{\wedge}2)) \ -(8 \ m^{\wedge}2 \ veq2i^{\wedge}2$$

$$(-U2p[t] \ U4[t]+U2[t] \ U4p[t])^{\wedge}2 \ (U2[t] \ (U4p[0] \ V1[t]-U3p[0] \ V2[t])+U4[t] \ (-U2p[0]$$

$$V1[t]+U3p[0] \ V4[t])+U3[t] \ (U2p[0] \ V2[t]-U4p[0] \ V4[t])) \ (V2p[0] \ V4[t]-V2[t]$$

$$V4p[0]) \ (U2[t] \ (-V1p[0] \ V2[t]+V1[t] \ V2p[0])+U4[t] \ (V1p[0] \ V4[t]-V1[t] \ V4p[0])+U3[t]$$

$$(-V2p[0] \ V4[t]+V2[t] \ V4p[0])))/((2 \ U2[t] \ V2[t]-2 \ U4[t] \ V4[t])^{\wedge}4 \ (-m \ U2[t]$$

$$V2p[0]+m \ U4[t] \ V4p[0]) \ (((-U2p[0] \ V2[t]+U4p[0] \ V4[t]) \ (-m \ U2[t] \ V2p[0]+m$$

$$U4[t] \ V4p[0]))/(2 \ U2[t] \ V2[t]-2 \ U4[t] \ V4[t])^{\wedge}2-(m \ (U2p[0] \ U4[t]-U2[t] \ U4p[0])$$

$$(V2p[0] \ V4[t]-V2[t] \ V4p[0]))/(2 \ U2[t] \ V2[t]-2 \ U4[t] \ V4[t])^{\wedge}2)) \ +(4 \ m^{\wedge}2 \ sq2q2i$$

$$(-U2p[t] \ U4[t]+U2[t] \ U4p[t])^{\wedge}2 \ (U2[t] \ (-V1p[0] \ V2[t]+V1[t] \ V2p[0])+U4[t]$$

$$(V1p[0] \ V4[t]-V1[t] \ V4p[0])+U3[t] \ (-V2p[0] \ V4[t]+V2[t] \ V4p[0]))^{\wedge}2)/((2 \ U2[t]$$

$$V2[t]-2 \ U4[t] \ V4[t])^{\wedge}2 \ (-m \ U2[t] \ V2p[0]+m \ U4[t] \ V4p[0])^{\wedge}2)+(4 \ m^{\wedge}2 \ veq2i^{\wedge}2$$

$$(-U2p[t] \ U4[t]+U2[t] \ U4p[t])^{\wedge}2 \ (U2[t] \ (-V1p[0] \ V2[t]+V1[t] \ V2p[0])+U4[t]$$

$$(V1p[0] \ V4[t]-V1[t] \ V4p[0])+U3[t] \ (-V2p[0] \ V4[t]+V2[t] \ V4p[0]))^{\wedge}2)/((2 \ U2[t]$$

$$V2[t]-2 \ U4[t] \ V4[t])^{\wedge}2 \ (-m \ U2[t] \ V2p[0]+m \ U4[t] \ V4p[0])^{\wedge}2)-(8 \ m^{\wedge}2 \ sq2q2i$$

$$(U2p[0] \ U4[t]-U2[t] \ U4p[0]) \ (-U2p[t] \ U4[t]+U2[t] \ U4p[t]) \ (U2p[t] \ V2[t]-U4p[t]$$

$$V4[t]) \ (U2[t] \ (-V1p[0] \ V2[t]+V1[t] \ V2p[0])+U4[t] \ (V1p[0] \ V4[t]-V1[t] \ V4p[0])+U3[t]$$

$$(-V2p[0] \ V4[t]+V2[t] \ V4p[0]))^{\wedge}2)/((2 \ U2[t] \ V2[t]-2 \ U4[t] \ V4[t])^{\wedge}4 \ (-m \ U2[t]$$

$$V2p[0]+m \ U4[t] \ V4p[0]) \ (((-U2p[0] \ V2[t]+U4p[0] \ V4[t]) \ (-m \ U2[t] \ V2p[0]+m$$

$$U4[t] \ V4p[0]))/(2 \ U2[t] \ V2[t]-2 \ U4[t] \ V4[t])^{\wedge}2-(m \ (U2p[0] \ U4[t]-U2[t] \ U4p[0])$$

$$(V2p[0] \ V4[t]-V2[t] \ V4p[0]))/(2 \ U2[t] \ V2[t]-2 \ U4[t] \ V4[t])^{\wedge}2)) \ -(8 \ m^{\wedge}2 \ veq2i^{\wedge}2$$

$$(U2p[0] \ U4[t]-U2[t] \ U4p[0]) \ (-U2p[t] \ U4[t]+U2[t] \ U4p[t]) \ (U2p[t] \ V2[t]-U4p[t]$$

$$V4[t]) \ (U2[t] \ (-V1p[0] \ V2[t]+V1[t] \ V2p[0])+U4[t] \ (V1p[0] \ V4[t]-V1[t] \ V4p[0])+U3[t]$$

$$(-V2p[0] \ V4[t]+V2[t] \ V4p[0]))^{\wedge}2)/((2 \ U2[t] \ V2[t]-2 \ U4[t] \ V4[t])^{\wedge}4 \ (-m \ U2[t]$$

$$V2p[0]+m \ U4[t] \ V4p[0]) \ (((-U2p[0] \ V2[t]+U4p[0] \ V4[t]) \ (-m \ U2[t] \ V2p[0]+m$$

U4[t] V4p[0]))/(2 U2[t] V2[t]-2 U4[t] V4[t])^2-(m (U2p[0] U4[t]-U2[t] U4p[0]) (V2p[0] V4[t]-V2[t] V4p[0]))/(2 U2[t] V2[t]-2 U4[t] V4[t])^2)) +(8 m^3 sq2q2i (U2p[0] U4[t]-U2[t] U4p[0]) (-U2p[0] U4[t]+U2[t] U4p[t])^2 (V2p[0] V4[t]-V2[t] V4p[0]) (U2[t] (-V1p[0] V2[t]+V1[t] V2p[0])+U4[t] (V1p[0] V4[t]-V1[t] V4p[0])+U3[t] (-V2p[0] V4[t]+V2[t] V4p[0]))^2)/((2 U2[t] V2[t]-2 U4[t] V4[t])^4 (-m U2[t] V2p[0]+m U4[t] V4p[0])^2 (((-U2p[0] V2[t]+U4p[0] V4[t]) (-m U2[t] V2p[0]+m U4[t] V4p[0]))/(2 U2[t] V2[t]-2 U4[t] V4[t])^2-(m (U2p[0] U4[t]-U2[t] U4p[0]) (V2p[0] V4[t]-V2[t] V4p[0]))/(2 U2[t] V2[t]-2 U4[t] V4[t])^2)) +(8 m^3 veq2i^2 (U2p[0] U4[t]-U2[t] U4p[0]) (-U2p[0] U4[t]+U2[t] U4p[t])^2 (V2p[0] V4[t]-V2[t] V4p[0]) (U2[t] (-V1p[0] V2[t]+V1[t] V2p[0])+U4[t] (V1p[0] V4[t]-V1[t] V4p[0])+U3[t] (-V2p[0] V4[t]+V2[t] V4p[0]))^2)/((2 U2[t] V2[t]-2 U4[t] V4[t])^4 (-m U2[t] V2p[0]+m U4[t] V4p[0])^2 (((-U2p[0] V2[t]+U4p[0] V4[t]) (-m U2[t] V2p[0]+m U4[t] V4p[0]))/(2 U2[t] V2[t]-2 U4[t] V4[t])^2-(m (U2p[0] U4[t]-U2[t] U4p[0]) (V2p[0] V4[t]-V2[t] V4p[0]))/(2 U2[t] V2[t]-2 U4[t] V4[t])^2)) +(4 m vep2i veq1i (-U2p[t] U4[t]+U2[t] U4p[t])^2 (U2[t] (V2p[0] V3[t]-V2[t] V3p[0])+U1[t] (-V2p[0] V4[t]+V2[t] V4p[0])+U4[t] (V3p[0] V4[t]-V3[t] V4p[0])))/((2 U2[t] V2[t]-2 U4[t] V4[t]) (-m U2[t] V2p[0]+m U4[t] V4p[0])^2)-(8 m sq1q1i (-U2p[t] U4[t]+U2[t] U4p[t]) (U1p[t] U2[t] V2[t]-U1[t] U2p[t] V2[t]+U2p[t] U4[t] V3[t]-U2[t] U4p[t] V3[t]-U1p[t] U4[t] V4[t]+U1[t] U4p[t] V4[t]) (U2[t] (V2p[0] V3[t]-V2[t] V3p[0])+U1[t] (-V2p[0] V4[t]+V2[t] V4p[0])+U4[t] (V3p[0] V4[t]-V3[t] V4p[0])))/((2 U2[t] V2[t]-2 U4[t] V4[t])^2 (-m U2[t] V2p[0]+m U4[t] V4p[0]))-(8 m veq1i^2 (-U2p[t] U4[t]+U2[t] U4p[t]) (U1p[t] U2[t] V2[t]-U1[t] U2p[t] V2[t]+U2p[t] U4[t] V3[t]-U2[t] U4p[t] V3[t]-U1p[t] U4[t] V4[t]+U1[t] U4p[t] V4[t]) (U2[t] (V2p[0] V3[t]-V2[t] V3p[0])+U1[t] (-V2p[0] V4[t]+V2[t] V4p[0])+U4[t] (V3p[0] V4[t]-V3[t] V4p[0])))/((2 U2[t] V2[t]-2 U4[t] V4[t])^2 (-m U2[t] V2p[0]+m U4[t] V4p[0]))-(8 m veq1i veq2i (-U2p[t] U4[t]+U2[t] U4p[t]) (U2p[t] U4[t] V1[t]-U2[t] U4p[t] V1[t]-U2p[t] U3[t] V2[t]+U2[t] U3p[t] V2[t]-U3p[t] U4[t] V4[t]+U3[t] U4p[t] V4[t]) (U2[t] (V2p[0] V3[t]-V2[t] V3p[0])+U1[t] (-V2p[0] V4[t]+V2[t] V4p[0])+U4[t] (V3p[0] V4[t]-V3[t] V4p[0])))/((2 U2[t] V2[t]-2 U4[t] V4[t])^2 (-m U2[t] V2p[0]+m U4[t] V4p[0]))+(4 m sq1p1i (-U2p[t] U4[t]+U2[t] U4p[t]) (U2p[t] V2[t]-U4p[t] V4[t]) (U2[t] (V2p[0] V3[t]-V2[t] V3p[0])+U1[t] (-V2p[0] V4[t]+V2[t] V4p[0])+U4[t] (V3p[0] V4[t]-V3[t] V4p[0])))/((2 U2[t]

V2[t]-2 U4[t] V4[t])^3 (((-U2p[0] V2[t]+U4p[0] V4[t]) (-m U2[t] V2p[0]+m
U4[t] V4p[0]))/(2 U2[t] V2[t]-2 U4[t] V4[t])^2-(m (U2p[0] U4[t]-U2[t] U4p[0])
(V2p[0] V4[t]-V2[t] V4p[0]))/(2 U2[t] V2[t]-2 U4[t] V4[t])^2)) +(4 m vep1i
veq1i (-U2p[t] U4[t]+U2[t] U4p[t]) (U2p[t] V2[t]-U4p[t] V4[t]) (U2[t] (V2p[0]
V3[t]-V2[t] V3p[0])+U1[t] (-V2p[0] V4[t]+V2[t] V4p[0])+U4[t] (V3p[0] V4[t]-V3[t]
V4p[0])))/((2 U2[t] V2[t]-2 U4[t] V4[t])^3 (((-U2p[0] V2[t]+U4p[0] V4[t])
(-m U2[t] V2p[0]+m U4[t] V4p[0]))/(2 U2[t] V2[t]-2 U4[t] V4[t])^2-(m (U2p[0]
U4[t]-U2[t] U4p[0]) (V2p[0] V4[t]-V2[t] V4p[0]))/(2 U2[t] V2[t]-2 U4[t] V4[t])^2))
+(8 m sq1q1i (U2p[0] U4[t]-U2[t] U4p[0]) (U2p[t] V2[t]-U4p[t] V4[t]) (U1p[t]
U2[t] V2[t]-U1[t] U2p[t] V2[t]+U2p[t] U4[t] V3[t]-U2[t] U4p[t] V3[t]-U1p[t]
U4[t] V4[t]+U1[t] U4p[t] V4[t]) (U2[t] (V2p[0] V3[t]-V2[t] V3p[0])+U1[t]
(-V2p[0] V4[t]+V2[t] V4p[0])+U4[t] (V3p[0] V4[t]-V3[t] V4p[0])))/((2 U2[t]
V2[t]-2 U4[t] V4[t])^4 (((-U2p[0] V2[t]+U4p[0] V4[t]) (-m U2[t] V2p[0]+m
U4[t] V4p[0]))/(2 U2[t] V2[t]-2 U4[t] V4[t])^2-(m (U2p[0] U4[t]-U2[t] U4p[0])
(V2p[0] V4[t]-V2[t] V4p[0]))/(2 U2[t] V2[t]-2 U4[t] V4[t])^2)) +(8 m veq1i^2
(U2p[0] U4[t]-U2[t] U4p[0]) (U2p[t] V2[t]-U4p[t] V4[t]) (U1p[t] U2[t] V2[t]-U1[t]
U2p[t] V2[t]+U2p[t] U4[t] V3[t]-U2[t] U4p[t] V3[t]-U1p[t] U4[t] V4[t]+U1[t]
U4p[t] V4[t]) (U2[t] (V2p[0] V3[t]-V2[t] V3p[0])+U1[t] (-V2p[0] V4[t]+V2[t]
V4p[0])+U4[t] (V3p[0] V4[t]-V3[t] V4p[0])))/((2 U2[t] V2[t]-2 U4[t] V4[t])^4
(((-U2p[0] V2[t]+U4p[0] V4[t]) (-m U2[t] V2p[0]+m U4[t] V4p[0]))/(2 U2[t]
V2[t]-2 U4[t] V4[t])^2-(m (U2p[0] U4[t]-U2[t] U4p[0]) (V2p[0] V4[t]-V2[t]
V4p[0]))/(2 U2[t] V2[t]-2 U4[t] V4[t])^2)) +(8 m veq1i veq2i (U2p[0] U4[t]-U2[t]
U4p[0]) (U2p[t] V2[t]-U4p[t] V4[t]) (U2p[t] U4[t] V1[t]-U2[t] U4p[t] V1[t]-U2p[t]
U3[t] V2[t]+U2[t] U3p[t] V2[t]-U3p[t] U4[t] V4[t]+U3[t] U4p[t] V4[t]) (U2[t]
(V2p[0] V3[t]-V2[t] V3p[0])+U1[t] (-V2p[0] V4[t]+V2[t] V4p[0])+U4[t] (V3p[0]
V4[t]-V3[t] V4p[0])))/((2 U2[t] V2[t]-2 U4[t] V4[t])^4 (((-U2p[0] V2[t]+U4p[0]
V4[t]) (-m U2[t] V2p[0]+m U4[t] V4p[0]))/(2 U2[t] V2[t]-2 U4[t] V4[t])^2-(m
(U2p[0] U4[t]-U2[t] U4p[0]) (V2p[0] V4[t]-V2[t] V4p[0]))/(2 U2[t] V2[t]-2
U4[t] V4[t])^2)) +(8 m sq1q1i (-U2p[t] U4[t]+U2[t] U4p[t]) (U2p[t] V2[t]-U4p[t]
V4[t]) (U2[t] (-U1p[0] V2[t]+U4p[0] V3[t])+U4[t] (-U2p[0] V3[t]+U1p[0] V4[t])+U1[t]
(U2p[0] V2[t]-U4p[0] V4[t])) (U2[t] (V2p[0] V3[t]-V2[t] V3p[0])+U1[t] (-V2p[0]
V4[t]+V2[t] V4p[0])+U4[t] (V3p[0] V4[t]-V3[t] V4p[0])))/((2 U2[t] V2[t]-2

U4[t] V4[t])^4 (((-U2p[0] V2[t]+U4p[0] V4[t]) (-m U2[t] V2p[0]+m U4[t] V4p[0]))/(2
U2[t] V2[t]-2 U4[t] V4[t])^2-(m (U2p[0] U4[t]-U2[t] U4p[0]) (V2p[0] V4[t]-V2[t]
V4p[0]))/(2 U2[t] V2[t]-2 U4[t] V4[t])^2)) +(8 m veq1i^2 (-U2p[0] U4[t]+U2[t]
U4p[t]) (U2p[t] V2[t]-U4p[t] V4[t]) (U2[t] (-U1p[0] V2[t]+U4p[0] V3[t])+U4[t]
(-U2p[0] V3[t]+U1p[0] V4[t])+U1[t] (U2p[0] V2[t]-U4p[0] V4[t])) (U2[t] (V2p[0]
V3[t]-V2[t] V3p[0])+U1[t] (-V2p[0] V4[t]+V2[t] V4p[0])+U4[t] (V3p[0] V4[t]-V3[t]
V4p[0])))/((2 U2[t] V2[t]-2 U4[t] V4[t])^4 (((-U2p[0] V2[t]+U4p[0] V4[t])
(-m U2[t] V2p[0]+m U4[t] V4p[0]))/(2 U2[t] V2[t]-2 U4[t] V4[t])^2-(m (U2p[0]
U4[t]-U2[t] U4p[0]) (V2p[0] V4[t]-V2[t] V4p[0]))/(2 U2[t] V2[t]-2 U4[t] V4[t])^2))
+(8 m veq1i veq2i (-U2p[t] U4[t]+U2[t] U4p[t]) (U2p[t] V2[t]-U4p[t] V4[t])
(U2[t] (U4p[0] V1[t]-U3p[0] V2[t])+U4[t] (-U2p[0] V1[t]+U3p[0] V4[t])+U3[t]
(U2p[0] V2[t]-U4p[0] V4[t])) (U2[t] (V2p[0] V3[t]-V2[t] V3p[0])+U1[t] (-V2p[0]
V4[t]+V2[t] V4p[0])+U4[t] (V3p[0] V4[t]-V3[t] V4p[0])))/((2 U2[t] V2[t]-2
U4[t] V4[t])^4 (((-U2p[0] V2[t]+U4p[0] V4[t]) (-m U2[t] V2p[0]+m U4[t] V4p[0]))/(2
U2[t] V2[t]-2 U4[t] V4[t])^2-(m (U2p[0] U4[t]-U2[t] U4p[0]) (V2p[0] V4[t]-V2[t]
V4p[0]))/(2 U2[t] V2[t]-2 U4[t] V4[t])^2)) -(8 m vep2i veq1i (U2p[0] U4[t]-U2[t]
U4p[0]) (-U2p[t] U4[t]+U2[t] U4p[t]) (U2p[t] V2[t]-U4p[t] V4[t]) (U2[t] (V2p[0]
V3[t]-V2[t] V3p[0])+U1[t] (-V2p[0] V4[t]+V2[t] V4p[0])+U4[t] (V3p[0] V4[t]-V3[t]
V4p[0])))/((2 U2[t] V2[t]-2 U4[t] V4[t])^3 (-m U2[t] V2p[0]+m U4[t] V4p[0])
(((-U2p[0] V2[t]+U4p[0] V4[t]) (-m U2[t] V2p[0]+m U4[t] V4p[0]))/(2 U2[t]
V2[t]-2 U4[t] V4[t])^2-(m (U2p[0] U4[t]-U2[t] U4p[0]) (V2p[0] V4[t]-V2[t]
V4p[0]))/(2 U2[t] V2[t]-2 U4[t] V4[t])^2)) +(8 m^2 vep2i veq1i (U2p[0] U4[t]-U2[t]
U4p[0]) (-U2p[t] U4[t]+U2[t] U4p[t])^2 (V2p[0] V4[t]-V2[t] V4p[0]) (U2[t]
(V2p[0] V3[t]-V2[t] V3p[0])+U1[t] (-V2p[0] V4[t]+V2[t] V4p[0])+U4[t] (V3p[0]
V4[t]-V3[t] V4p[0])))/((2 U2[t] V2[t]-2 U4[t] V4[t])^3 (-m U2[t] V2p[0]+m
U4[t] V4p[0])^2 (((-U2p[0] V2[t]+U4p[0] V4[t]) (-m U2[t] V2p[0]+m U4[t] V4p[0]))/(2
U2[t] V2[t]-2 U4[t] V4[t])^2-(m (U2p[0] U4[t]-U2[t] U4p[0]) (V2p[0] V4[t]-V2[t]
V4p[0]))/(2 U2[t] V2[t]-2 U4[t] V4[t])^2)) -(4 m^2 sq1p1i (-U2p[t] U4[t]+U2[t]
U4p[t])^2 (V2p[0] V4[t]-V2[t] V4p[0]) (U2[t] (V2p[0] V3[t]-V2[t] V3p[0])+U1[t]
(-V2p[0] V4[t]+V2[t] V4p[0])+U4[t] (V3p[0] V4[t]-V3[t] V4p[0])))/((2 U2[t]
V2[t]-2 U4[t] V4[t])^3 (-m U2[t] V2p[0]+m U4[t] V4p[0]) (((-U2p[0] V2[t]+U4p[0]
V4[t]) (-m U2[t] V2p[0]+m U4[t] V4p[0]))/(2 U2[t] V2[t]-2 U4[t] V4[t])^2-(m

(U2p[0] U4[t]-U2[t] U4p[0]) (V2p[0] V4[t]-V2[t] V4p[0]))/(2 U2[t] V2[t]-2
U4[t] V4[t])^2)) -(4 m^2 vep1i veq1i (-U2p[0] U4[t]+U2[t] U4p[t])^2 (V2p[0]
V4[t]-V2[t] V4p[0]) (U2[t] (V2p[0] V3[t]-V2[t] V3p[0])+U1[t] (-V2p[0] V4[t]+V2[t]
V4p[0])+U4[t] (V3p[0] V4[t]-V3[t] V4p[0])))/((2 U2[t] V2[t]-2 U4[t] V4[t])^3
(-m U2[t] V2p[0]+m U4[t] V4p[0]) (((-U2p[0] V2[t]+U4p[0] V4[t]) (-m U2[t]
V2p[0]+m U4[t] V4p[0]))/(2 U2[t] V2[t]-2 U4[t] V4[t])^2-(m (U2p[0] U4[t]-U2[t]
U4p[0]) (V2p[0] V4[t]-V2[t] V4p[0]))/(2 U2[t] V2[t]-2 U4[t] V4[t])^2)) -(8
m^2 sq1q1i (U2p[0] U4[t]-U2[t] U4p[0]) (-U2p[t] U4[t]+U2[t] U4p[t]) (U1p[t]
U2[t] V2[t]-U1[t] U2p[t] V2[t]+U2p[t] U4[t] V3[t]-U2[t] U4p[t] V3[t]-U1p[t]
U4[t] V4[t]+U1[t] U4p[t] V4[t]) (V2p[0] V4[t]-V2[t] V4p[0]) (U2[t] (V2p[0]
V3[t]-V2[t] V3p[0])+U1[t] (-V2p[0] V4[t]+V2[t] V4p[0])+U4[t] (V3p[0] V4[t]-V3[t]
V4p[0])))/((2 U2[t] V2[t]-2 U4[t] V4[t])^4 (-m U2[t] V2p[0]+m U4[t] V4p[0])
(((-U2p[0] V2[t]+U4p[0] V4[t]) (-m U2[t] V2p[0]+m U4[t] V4p[0]))/(2 U2[t]
V2[t]-2 U4[t] V4[t])^2-(m (U2p[0] U4[t]-U2[t] U4p[0]) (V2p[0] V4[t]-V2[t]
V4p[0]))/(2 U2[t] V2[t]-2 U4[t] V4[t])^2)) -(8 m^2 veq1i^2 (U2p[0] U4[t]-U2[t]
U4p[0]) (-U2p[t] U4[t]+U2[t] U4p[t]) (U1p[t] U2[t] V2[t]-U1[t] U2p[t] V2[t]+U2p[t]
U4[t] V3[t]-U2[t] U4p[t] V3[t]-U1p[t] U4[t] V4[t]+U1[t] U4p[t] V4[t]) (V2p[0]
V4[t]-V2[t] V4p[0]) (U2[t] (V2p[0] V3[t]-V2[t] V3p[0])+U1[t] (-V2p[0] V4[t]+V2[t]
V4p[0])+U4[t] (V3p[0] V4[t]-V3[t] V4p[0]))))/((2 U2[t] V2[t]-2 U4[t] V4[t])^4
(-m U2[t] V2p[0]+m U4[t] V4p[0]) (((-U2p[0] V2[t]+U4p[0] V4[t]) (-m U2[t]
V2p[0]+m U4[t] V4p[0]))/(2 U2[t] V2[t]-2 U4[t] V4[t])^2-(m (U2p[0] U4[t]-U2[t]
U4p[0]) (V2p[0] V4[t]-V2[t] V4p[0]))/(2 U2[t] V2[t]-2 U4[t] V4[t])^2)) -(8 m^2 veq1i veq2i (U2p[0] U4[t]-U2[t] U4p[0]) (-U2p[t] U4[t]+U2[t] U4p[t])
(U2p[t] U4[t] V1[t]-U2[t] U4p[t] V1[t]-U2p[t] U3[t] V2[t]+U2[t] U3p[t] V2[t]-U3p[t]
U4[t] V4[t]+U3[t] U4p[t] V4[t]) (V2p[0] V4[t]-V2[t] V4p[0]) (U2[t] (V2p[0]
V3[t]-V2[t] V3p[0])+U1[t] (-V2p[0] V4[t]+V2[t] V4p[0])+U4[t] (V3p[0] V4[t]-V3[t]
V4p[0])))/((2 U2[t] V2[t]-2 U4[t] V4[t])^4 (-m U2[t] V2p[0]+m U4[t] V4p[0])
(((-U2p[0] V2[t]+U4p[0] V4[t]) (-m U2[t] V2p[0]+m U4[t] V4p[0]))/(2 U2[t]
V2[t]-2 U4[t] V4[t])^2-(m (U2p[0] U4[t]-U2[t] U4p[0]) (V2p[0] V4[t]-V2[t]
V4p[0]))/(2 U2[t] V2[t]-2 U4[t] V4[t])^2)) -(8 m^2 sq1q1i (-U2p[t] U4[t]+U2[t]
U4p[t])^2 (U2[t] (-U1p[0] V2[t]+U4p[0] V3[t])+U4[t] (-U2p[0] V3[t]+U1p[0]
V4[t])+U1[t] (U2p[0] V2[t]-U4p[0] V4[t])) (V2p[0] V4[t]-V2[t] V4p[0]) (U2[t]

(V2p[0] V3[t]-V2[t] V3p[0])+U1[t] (-V2p[0] V4[t]+V2[t] V4p[0])+U4[t] (V3p[0] V4[t]-V3[t] V4p[0])))/((2 U2[t] V2[t]-2 U4[t] V4[t])^4 (-m U2[t] V2p[0]+m U4[t] V4p[0]) (((-U2p[0] V2[t]+U4p[0] V4[t]) (-m U2[t] V2p[0]+m U4[t] V4p[0]))/(2 U2[t] V2[t]-2 U4[t] V4[t])^2-(m (U2p[0] U4[t]-U2[t] U4p[0]) (V2p[0] V4[t]-V2[t] V4p[0]))/(2 U2[t] V2[t]-2 U4[t] V4[t])^2)) -(8 m^2 veq1i^2 (-U2p[t] U4[t]+U2[t] U4p[t])^2 (U2[t] (-U1p[0] V2[t]+U4p[0] V3[t])+U4[t] (-U2p[0] V3[t]+U1p[0] V4[t])+U1[t] (U2p[0] V2[t]-U4p[0] V4[t])) (V2p[0] V4[t]-V2[t] V4p[0]) (U2[t] (V2p[0] V3[t]-V2[t] V3p[0])+U1[t] (-V2p[0] V4[t]+V2[t] V4p[0])+U4[t] (V3p[0] V4[t]-V3[t] V4p[0])))/((2 U2[t] V2[t]-2 U4[t] V4[t])^4 (-m U2[t] V2p[0]+m U4[t] V4p[0]) (((-U2p[0] V2[t]+U4p[0] V4[t]) (-m U2[t] V2p[0]+m U4[t] V4p[0]))/(2 U2[t] V2[t]-2 U4[t] V4[t])^2-(m (U2p[0] U4[t]-U2[t] U4p[0]) (V2p[0] V4[t]-V2[t] V4p[0]))/(2 U2[t] V2[t]-2 U4[t] V4[t])^2)) -(8 m^2 veq1i veq2i (-U2p[t] U4[t]+U2[t] U4p[t])^2 (U2[t] (U4p[0] V1[t]-U3p[0] V2[t])+U4[t] (-U2p[0] V1[t]+U3p[0] V4[t])+U3[t] (U2p[0] V2[t]-U4p[0] V4[t])) (V2p[0] V4[t]-V2[t] V4p[0]) (U2[t] (V2p[0] V3[t]-V2[t] V3p[0])+U1[t] (-V2p[0] V4[t]+V2[t] V4p[0])+U4[t] (V3p[0] V4[t]-V3[t] V4p[0])))/((2 U2[t] V2[t]-2 U4[t] V4[t])^4 (-m U2[t] V2p[0]+m U4[t] V4p[0]) (((-U2p[0] V2[t]+U4p[0] V4[t]) (-m U2[t] V2p[0]+m U4[t] V4p[0]))/(2 U2[t] V2[t]-2 U4[t] V4[t])^2-(m (U2p[0] U4[t]-U2[t] U4p[0]) (V2p[0] V4[t]-V2[t] V4p[0]))/(2 U2[t] V2[t]-2 U4[t] V4[t])^2)) +(8 m^2 veq1i veq2i (-U2p[t] U4[t]+U2[t] U4p[t])^2 (U2[t] (-V1p[0] V2[t]+V1[t] V2p[0])+U4[t] (V1p[0] V4[t]-V1[t] V4p[0])+U3[t] (-V2p[0] V4[t]+V2[t] V4p[0])) (U2[t] (V2p[0] V3[t]-V2[t] V3p[0])+U1[t] (-V2p[0] V4[t]+V2[t] V4p[0])+U4[t] (V3p[0] V4[t]-V3[t] V4p[0])))/((2 U2[t] V2[t]-2 U4[t] V4[t])^2 (-m U2[t] V2p[0]+m U4[t] V4p[0])^2)-(16 m^2 veq1i veq2i (U2p[0] U4[t]-U2[t] U4p[0]) (-U2p[t] U4[t]+U2[t] U4p[t]) (U2p[t] V2[t]-U4p[t] V4[t]) (U2[t] (-V1p[0] V2[t]+V1[t] V2p[0])+U4[t] (V1p[0] V4[t]-V1[t] V4p[0])+U3[t] (-V2p[0] V4[t]+V2[t] V4p[0])) (U2[t] (V2p[0] V3[t]-V2[t] V3p[0])+U1[t] (-V2p[0] V4[t]+V2[t] V4p[0])+U4[t] (V3p[0] V4[t]-V3[t] V4p[0])))/((2 U2[t] V2[t]-2 U4[t] V4[t])^4 (-m U2[t] V2p[0]+m U4[t] V4p[0]) (((-U2p[0] V2[t]+U4p[0] V4[t]) (-m U2[t] V2p[0]+m U4[t] V4p[0]))/(2 U2[t] V2[t]-2 U4[t] V4[t])^2-(m (U2p[0] U4[t]-U2[t] U4p[0]) (V2p[0] V4[t]-V2[t] V4p[0]))/(2 U2[t] V2[t]-2 U4[t] V4[t])^2)) +(16 m^3 veq1i veq2i (U2p[0] U4[t]-U2[t] U4p[0]) (-U2p[t] U4[t]+U2[t] U4p[t])^2 (V2p[0] V4[t]-V2[t] V4p[0]) (U2[t] (-V1p[0] V2[t]+V1[t] V2p[0])+U4[t] (V1p[0]

V4[t]-V1[t] V4p[0])+U3[t] (-V2p[0] V4[t]+V2[t] V4p[0])) (U2[t] (V2p[0] V3[t]-V2[t] V3p[0])+U1[t] (-V2p[0] V4[t]+V2[t] V4p[0])+U4[t] (V3p[0] V4[t]-V3[t] V4p[0])))/((2 U2[t] V2[t]-2 U4[t] V4[t])^4 (-m U2[t] V2p[0]+m U4[t] V4p[0])^2 (((-U2p[0] V2[t]+U4p[0] V4[t]) (-m U2[t] V2p[0]+m U4[t] V4p[0]))/(2 U2[t] V2[t]-2 U4[t] V4[t])^2-(m (U2p[0] U4[t]-U2[t] U4p[0]) (V2p[0] V4[t]-V2[t] V4p[0]))/(2 U2[t] V2[t]-2 U4[t] V4[t])^2)) +(4 m^2 sq1q1i (-U2p[t] U4[t]+U2[t] U4p[t])^2 (U2[t] (V2p[0] V3[t]-V2[t] V3p[0])+U1[t] (-V2p[0] V4[t]+V2[t] V4p[0])+U4[t] (V3p[0] V4[t]-V3[t] V4p[0]))^2)/((2 U2[t] V2[t]-2 U4[t] V4[t])^2 (-m U2[t] V2p[0]+m U4[t] V4p[0])^2)+(4 m^2 veq1i^2 (-U2p[t] U4[t]+U2[t] U4p[t])^2 (U2[t] (V2p[0] V3[t]-V2[t] V3p[0])+U1[t] (-V2p[0] V4[t]+V2[t] V4p[0])+U4[t] (V3p[0] V4[t]-V3[t] V4p[0]))^2)/((2 U2[t] V2[t]-2 U4[t] V4[t])^2 (-m U2[t] V2p[0]+m U4[t] V4p[0])^2)-(8 m^2 sq1q1i (U2p[0] U4[t]-U2[t] U4p[0]) (-U2p[t] U4[t]+U2[t] U4p[t]) (U2p[t] V2[t]-U4p[t] V4[t]) (U2[t] (V2p[0] V3[t]-V2[t] V3p[0])+U1[t] (-V2p[0] V4[t]+V2[t] V4p[0])+U4[t] (V3p[0] V4[t]-V3[t] V4p[0]))^2)/((2 U2[t] V2[t]-2 U4[t] V4[t])^4 (-m U2[t] V2p[0]+m U4[t] V4p[0]) (((-U2p[0] V2[t]+U4p[0] V4[t]) (-m U2[t] V2p[0]+m U4[t] V4p[0]))/(2 U2[t] V2[t]-2 U4[t] V4[t])^2-(m (U2p[0] U4[t]-U2[t] U4p[0]) (V2p[0] V4[t]-V2[t] V4p[0]))/(2 U2[t] V2[t]-2 U4[t] V4[t])^2)) -(8 m^2 veq1i^2 (U2p[0] U4[t]-U2[t] U4p[0]) (-U2p[t] U4[t]+U2[t] U4p[t]) (U2p[t] V2[t]-U4p[t] V4[t]) (U2[t] (V2p[0] V3[t]-V2[t] V3p[0])+U1[t] (-V2p[0] V4[t]+V2[t] V4p[0])+U4[t] (V3p[0] V4[t]-V3[t] V4p[0]))^2)/((2 U2[t] V2[t]-2 U4[t] V4[t])^4 (-m U2[t] V2p[0]+m U4[t] V4p[0]) (((-U2p[0] V2[t]+U4p[0] V4[t]) (-m U2[t] V2p[0]+m U4[t] V4p[0]))/(2 U2[t] V2[t]-2 U4[t] V4[t])^2-(m (U2p[0] U4[t]-U2[t] U4p[0]) (V2p[0] V4[t]-V2[t] V4p[0]))/(2 U2[t] V2[t]-2 U4[t] V4[t])^2)) +(8 m^3 sq1q1i (U2p[0] U4[t]-U2[t] U4p[0]) (-U2p[t] U4[t]+U2[t] U4p[t])^2 (V2p[0] V4[t]-V2[t] V4p[0]) (U2[t] (V2p[0] V3[t]-V2[t] V3p[0])+U1[t] (-V2p[0] V4[t]+V2[t] V4p[0])+U4[t] (V3p[0] V4[t]-V3[t] V4p[0]))^2)/((2 U2[t] V2[t]-2 U4[t] V4[t])^4 (-m U2[t] V2p[0]+m U4[t] V4p[0])^2 (((-U2p[0] V2[t]+U4p[0] V4[t]) (-m U2[t] V2p[0]+m U4[t] V4p[0]))/(2 U2[t] V2[t]-2 U4[t] V4[t])^2-(m (U2p[0] U4[t]-U2[t] U4p[0]) (V2p[0] V4[t]-V2[t] V4p[0]))/(2 U2[t] V2[t]-2 U4[t] V4[t])^2)) +(8 m^3 veq1i^2 (U2p[0] U4[t]-U2[t] U4p[0]) (-U2p[t] U4[t]+U2[t] U4p[t])^2 (V2p[0] V4[t]-V2[t] V4p[0]) (U2[t] (V2p[0] V3[t]-V2[t] V3p[0])+U1[t] (-V2p[0] V4[t]+V2[t] V4p[0])+U4[t] (V3p[0] V4[t]-V3[t] V4p[0]))^2)/((2 U2[t] V2[t]-2 U4[t] V4[t])^4

```
(-m U2[t] V2p[0]+m U4[t] V4p[0])^2 (((-U2p[0] V2[t]+U4p[0] V4[t]) (-m U2[t]
V2p[0]+m U4[t] V4p[0]))/(2 U2[t] V2[t]-2 U4[t] V4[t])^2-(m (U2p[0] U4[t]-U2[t]
U4p[0]) (V2p[0] V4[t]-V2[t] V4p[0]))/(2 U2[t] V2[t]-2 U4[t] V4[t])^2));
```

---

[1] A. F. Estrada and L. A. Pachon, New J. Phys. **17**, 033038 (2015).

[2] R. Schmidt, A. Negretti, J. Ankerhold, T. Calarco, and J. T. Stockburger, Phys. Rev. Lett. **107**, 130404 (2011).

[3] D. Kirk, *Optimal Control Theory: An Introduction*, Dover Books on Electrical Engineering (Dover Publications, 2012).


}

\end{document}